\documentclass[12pt]{article}
\usepackage{amsmath}
\usepackage{graphicx}
\usepackage{enumerate}
\usepackage{natbib}
\usepackage{array}
\usepackage{algorithmic}
\usepackage[plain]{algorithm}

\newcommand*{\doublerule}{\hrule width \hsize height 0.5pt \kern 0.5mm \hrule width \hsize height 0.5pt}
\newcommand*{\singlerule}{\hrule width \hsize height 0.5pt}
\usepackage[colorlinks = true,
linkcolor = red,
urlcolor  = blue,
filecolor={red},
citecolor = blue,
anchorcolor = blue]{hyperref}
\usepackage{xurl}
\newcommand{\blind}{1}
\usepackage{mathtools}
\usepackage{etoolbox}
\usepackage[margin=1in]{geometry}
\usepackage{longtable,booktabs,array}
\usepackage{xcolor}
\allowdisplaybreaks
\DeclareMathOperator*{\argmin}{arg\,min}
\usepackage{caption}
\usepackage{graphicx,latexsym,amssymb,amsmath}
\usepackage{subcaption}
\usepackage{bbm, bm}

\def\R{{\mathbb{R}}}

\def\P{{\mathbb{P}}}
\def\E{{\mathbb{E}}}
\newcommand{\vertiii}[1]{{\left\vert\kern-0.4ex\left\vert\kern-0.4ex\left\vert #1 
		\right\vert\kern-0.4ex\right\vert\kern-0.4ex\right\vert}}
\usepackage{newfloat}
\newtheorem{Assumption}{Assumption}
\newtheorem{Theorem}{Theorem}

\usepackage{xr}
	\def\spacingset#1{\renewcommand{\baselinestretch}%
		{#1}\small\normalsize} \spacingset{1}
	
\usepackage[normalem]{ulem}

\begin{document}
	

	
	\if1\blind
	{
		\title{\bf Testing for sparse idiosyncratic components in factor-augmented regression models}
		\author{Jad Beyhum\hspace{.2cm}\\
			Department of Economics, KU Leuven, Belgium\\
			and \\
			Jonas Striaukas \\
			Department of Finance, Copenhagen Business School, Denmark}
		\maketitle
	} \fi
	
	\if0\blind
	{
		\bigskip
		\bigskip
		\bigskip
		\begin{center}
			{\LARGE\bf Testing for sparse idiosyncratic components in factor-augmented regression models}
		\end{center}
		\medskip
	} \fi
	
	\bigskip
	\begin{abstract}
		We propose a novel bootstrap test of a dense model, namely factor regression, against a sparse plus dense alternative augmenting model with sparse idiosyncratic components. The asymptotic properties of the test are established under time series dependence and polynomial tails. We outline a data-driven rule to select the tuning parameter and prove its theoretical validity. In simulation experiments, our procedure exhibits high power against sparse alternatives and low power against dense deviations from the null. Moreover, we apply our test to various datasets in macroeconomics and finance and often reject the null. This suggests the presence of sparsity --- on top of a dense model --- in commonly studied economic applications. The R package \texttt{\textquotesingle FAS\textquotesingle}  implements our approach. 
	\end{abstract}
	
	\noindent%
	\vfill
	
	\newpage
	\spacingset{1.5} 

\section{Introduction}
	
In this paper, we investigate a factor-augmented sparse regression model. Our analysis involves an observed sample of $T$ real-valued outcomes $y_1,\dots,y_T$, and high-dimensional regressors $x_1,\dots,x_T\in\R^p$, which are interconnected as follows:
\begin{equation}\label{model}
	\begin{aligned}
		y_t&=f_t^\top \gamma^*+u_t^\top\beta^*+\varepsilon_t,\\
		x_t&=Bf_t+u_t,\quad t=1\dots,T.
	\end{aligned}
\end{equation}
Here, $\varepsilon_t\in \R$ represents a random error, $u_t$ is a $p$-dimensional random vector of idiosyncratic shocks, $f_t$ is a $K$-dimensional random vector of factors, and $B$ is a $p\times K$ (nonrandom) matrix of loadings. The parameters of interest are $\gamma^*\in \R^K$ and $\beta^*\in \R^p$ and the right-hand side of \eqref{model} is unobserved. We consider the case where the number $p$ of regressors is large with respect to the sample size $T$ and a sparsity condition on the high-dimensional parameter vector $\beta^*$ is imposed. In the asymptotic regimes we study, $T$ goes to infinity, $p$ is allowed to grow with $T$ while $K$ remains fixed. The model formulation in equation \eqref{model} effectively merges two popular approaches in handling high-dimensional datasets: factor regression (\cite{stock2002forecasting,bai2006confidence}) and sparse high-dimensional regression (\cite{tibshirani1996regression,bickel2009simultaneous}).
Such a model allows the outcome to be related to the regressors through both common and idiosyncratic shocks and may better explain the data than factor regression or sparse regression alone (see \cite{fan2023latent,fan2021bridging}, which introduce and study model \eqref{model}). As noted in \cite{fan2021bridging}, this type of structure has many applications in forecasting, causal inference and to describe correlation networks. Note that, as in \cite{stock2002forecasting,bai2006confidence,fan2021bridging}, we could augment the model \eqref{model} with additional regressors $w_t$ entering the first equation of \eqref{model} but not the second one. This case is discussed in the Appendix Section \ref{app.modelw}.
	
We develop a test for the hypothesis:
\begin{equation}\label{Htest}
	H_0:\beta^*=0\quad \text{against}\quad H_1:\beta^*\ne 0 \text{ is sparse},
\end{equation}
where our theory outlines the set of sparse alternatives against which our test has power. Our specification test sheds light on the data-generating process by allowing us to determine if the underlying model is dense (as is the factor regression model) or sparse plus dense, as is the factor-augmented sparse regression model. This determination will then tell us if the relation between the regressors and the outcome is only driven by common shocks (factor regression) or if idiosyncratic shocks also play a role (factor-augmented sparse regression). The question of the adequacy of sparse or dense representations has recently garnered significant attention (see, e.g., \cite{giannone2021economic,kolesar2023fragility}). However, existing studies mostly focus on the differences between sparse and dense models and have found that dense models are often more adequate. In contrast, we compare a dense model with a sparse plus dense alternative.

In this paper, we propose a new bootstrap test for \eqref{Htest}. Our test's principle is to compare two estimators of $\sum_{t=1}^T u_t\varepsilon_t$. The first estimator is only consistent under the null, while the second estimator relies on the LASSO, and is, therefore, consistent under sparse alternatives. Our proposed test does not require estimating covariance matrices and is easy to implement. Following \cite{lederer2021estimating}, we outline a data-driven rule to select the tuning parameter of the LASSO estimator and prove its theoretical validity. We establish the validity of the test within a theoretical framework that accommodates scenarios where the number of variables, denoted by $p$, can significantly exceed $T$, and the explanatory variables exhibit strong mixing and possess polynomial tails. We use simulations to evaluate the finite sample properties of our procedure. Our test controls size and exhibits good power against sparse alternatives even when $p$ greatly exceeds $T$ or the data are heavy-tailed and serially correlated. A potential limitation of our approach might be that our approach also rejects when $\beta^*$ is nonzero but has a dense structure. To assess this issue, we conduct simulations with dense $\beta^*$. We find that our test exhibits very low power against such alternatives (in absolute terms and also relatively to sparse alternatives with the same signal-to-noise ratio). Hence, our test exhibits some robustness against dense alternatives. This result is intuitive: the LASSO estimator on which our test relies sets to zero very small coefficients pertaining to dense alternatives. Finally, we apply our test to several commonly studied datasets in macroeconomics and finance and often reject the null. This suggests that sparsity can help describe economic data once a dense component (here, modeled through the factors) is included in the model. This result complements the recent studies \cite{giannone2021economic,kolesar2023fragility}, which concluded that dense representations were often more appropriate for economic data.  The R package \texttt{\textquotesingle FAS\textquotesingle}  implements our approach.\\

\noindent 	\textbf{Related literature.} Our paper relates to the growing literature on factor-augmented sparse models, see \cite{hansen2019factor,fan2023do, fan2023latent, fan2021bridging, vogt2022cce, beyhum2023factor,barigozzi2023fnets} among others. In particular, \cite{fan2021bridging} proposes a general framework for factor-augmented sparse models. As noted by a reviewer, our test can be used to check that the sparse idiosyncratic components are jointly significant after the third step of the methodology by \cite{fan2021bridging}. Note that \cite{fan2021bridging}'s model includes lags of idiosyncratic terms and additional regressors. In the Appendix, we explain how to extend our test to the case with additional regressors. Section \ref{OA-subsec.lag} of the Online Appendix outlines an adapted test with a lag. We do not formally prove that our test works in these cases, but we conjecture that the asymptotic properties extend to these more general models. Simulations reported in Section \ref{OA-subsec.add.sim} of the Online Appendix corroborate this presumption. We also note that the factor-augmented sparse model studied in the present paper is a generalization of an earlier model studied in \cite{fosten2017model,fosten2017confidence} which augments the factor regression of \cite{stock2002forecasting} with a low-dimensional set of idiosyncratic terms.
	
\cite{fan2023latent} recently introduced the Factor-Adjusted deBiased Test (FabTest) for evaluating \eqref{Htest}. However, the FabTest exhibits several limitations. The test relies on a desparsified LASSO estimator based on model \eqref{model}. To achieve desparsification, \cite{fan2023latent} utilized the nodewise LASSO method proposed by \cite{zhang2014confidence} and \cite{van2014asymptotically} for estimating the precision matrix of the idiosyncratic shocks. However, this approach introduces $p$ additional tuning parameters, in addition to the one used in the original LASSO regression. Although the tuning parameters are selected through cross-validation in practice, \cite{fan2023latent} did not provide a theoretical justification for this selection procedure. Besides, inferential theory for LASSO-type regressions is not well understood when the tuning parameter is selected by cross-validation. Moreover, the test's performance may deteriorate due to errors associated with the nodewise LASSO estimates, and it incurs a heavy computational cost. Another limitation of the FabTest is its reliance on estimating the variance of $\varepsilon_t$, which can lead to imprecise results where variance estimation is challenging. Additionally, \cite{fan2023latent} only established the validity of the FabTest for i.i.d. sub-Gaussian data (see Section 2 in \cite{fan2023latent}). An alternative is to use the partial covariance test developed by \cite{fan2021bridging} and applied in \cite{fan2023do}. This test's principle is to estimate the covariance matrix of the idiosyncratic terms (including the idiosyncratic term of $y_t$, that is, $u_t^\top\beta^*+\varepsilon_t$) and then use a Gaussian bootstrap under the null to obtain the critical value of the test statistic. In contrast to our test, this partial covariance test does not make use of the LASSO estimator and requires estimating a high-dimensional covariance matrix, which is challenging.
	
Finally, we would like to note that this paper contributes to various other strands of literature. First, it connects to recent literature considering testing for high-dimensional parameters. There exists several approaches, see \cite{fan2015power,zhu2018linear,chernozhukov2019inference,lederer2021estimating,he2023most} and references therein. Our strategy draws inspiration from \cite{lederer2021estimating}, a recent paper that introduces a bootstrap procedure for selecting the penalty parameter of LASSO in a standard sparse linear regression. They employ this procedure to test the null hypothesis that a specific high-dimensional parameter equals zero. We adapt their approach to the case with unobserved factors, time series dependence and polynomial tails, which poses a challenge beyond the scope of the results in \cite{lederer2021estimating}. Second, our work is related to the literature on inference on parameters of additional low-dimensional regressors in the factor regression model of \cite{stock2002forecasting}, see \cite{bai2006confidence,gonccalves2014bootstrapping,gonccalves2020bootstrapping}. Third, our work connects with the literature on specification tests for models involving unobserved factors. Many papers test for the validity of the assumption that loadings are time-independent in the approximate factor model itself — the second equation in \eqref{model} — (\cite{breitung2011testing, chen2014detecting, han2015tests, yamamoto2015testing, su2017time, su2020testing, baltagi2021estimating, xu2022testing, fu2023testing}), while \cite{corradi2014testing} tests for time-independence of all coefficients in the factor regression model of \cite{stock2002forecasting}. Our approach complements this literature by proposing a specification test of the factor regression model under a different alternative, namely the factor-augmented sparse regression model.  \\

\noindent\textbf{Outline.} The paper is organized as follows. In Section \ref{sec.test}, we outline our testing procedure. Its asymptotic properties are studied in Section \ref{sec.theory}. Then, Section \ref{sec.sim} contains Monte Carlo simulations. The empirical applications can be found in Section \ref{sec.emp}. Section \ref{sec.ccl} concludes. The Appendix contains an adapted testing procedure for a model with additional regressors and lags. The Online Appendix includes all proofs and additional discussions and simulation and empirical results.\\
	
\noindent\textbf{Notation.} For an integer $N\in \mathbb{N}$, let $[N]=\{1,\dots, N\}$. The transpose of a $n_1 \times n_2$ matrix $A$ is written $A^{\top}$. Its $k^{th}$ singular value is $\sigma_k(A)$. Let us also define the Euclidean norm $\left\| A\right\|_2^2=\sum_{i=1}^{n_1} \sum_{j=1}^{n_2}A_{ij}^2$ and the sup-norm $\|A\|_\infty=\max\limits_{i\in[n_1], j\in[n_2]}|A_{ij}|$. The quantity $n_1 \vee n_2$ is the maximum of $n_1$ and $n_2$, $n_1 \wedge  n_2$ is the minimum of $n_1$ and $n_2$. For $N\in \mathbb{N}$, $I_N$ is the identity matrix of size $N\times N$. For a real-valued random variable $Z$ and $g>0$, we let $\vertiii{Z}=\E[|Z|^g]^\frac1g$. For a $d$-dimensional random vector $Z$, we define $\vertiii{Z}_g=\sup\limits_{u\in\R^d: \ \|u\|_2\le 1}\vertiii{u^\top Z}_g$.
	
\section{The test}\label{sec.test}
\subsection{Testing procedure}

In this subsection, we explain our testing procedure, which is then summarized in algorithmic form in subsection \ref{subsec.computation}. To facilitate understanding, we rewrite the model in matrix form as follows:
\begin{equation*}
\begin{aligned}
Y&=F \gamma^*+U\beta^*+\mathcal{E},\\
X&=FB^\top+U,
\end{aligned}
\end{equation*}
where $Y=(y_1,\dots,y_T)^\top$, $F=(f_1,\dots, f_T)^\top$ is a $T\times K$ matrix, $U=(u_1,\dots,u_T)^\top$ and $X=(x_1,\dots,x_T)^\top$ are $T\times p$ matrices and $\mathcal{E}=(\varepsilon_1,\dots,\varepsilon_T)^\top.$

It is important to note that, under the null hypothesis $H_0$, we have $U^\top (Y-F\gamma^*)=U^\top\mathcal{E}$. This observation suggests a testing procedure that involves computing an estimate $2T^{-1}\left\|U^\top (Y-F\gamma^*)\right\|_{\infty}$ and comparing it with the (estimated) quantiles of $2T^{-1}\left\|U^\top \mathcal{E}\right\|_{\infty}$.\footnote{We have a factor 2 in front of $T^{-1}\left\|U^\top (Y-F\gamma^*)\right\|_{\infty}$ and $T^{-1}\left\|U^\top \mathcal{E}\right\|_{\infty}$ because $2T^{-1}\left\|U^\top \mathcal{E}\right\|_{\infty}$ is the effective noise of the problem, a natural concept in the literature on the LASSO, see \cite{lederer2021estimating}.}

We can  estimate $U^\top (Y-F\gamma^*)$ by principal components analysis. First, let $\widehat{K}$ be one of the many estimators of the number of factors $K$ available in the literature (see for instance \cite{bai2002determining,onatski2010determining,ahn2013eigenvalue,bai2019rank,fan2022estimating}). As in \cite{fan2023latent}, we let the columns of $\widehat{F}/\sqrt{T}$ be the eigenvectors corresponding to the leading $\widehat{K}$ eigenvalues of $XX^\top$ and $\widehat{B}= X^\top \widehat{F}(\widehat{F}^\top \widehat{F})^{-1}=T^{-1} X^\top \widehat{F} .$
Then, we project the data on the orthogonal of the vector space generated by the estimated factors. Let $\widehat{P}=T^{-1} \widehat{F}\left(\widehat{F}^\top \widehat{F}/T\right)^{-1}\widehat{F}^\top=T^{-1} \widehat{F}\widehat{F}^\top$ be the projector on the vector space generated by the columns of $\widehat{F}$. A natural estimate for $U$ is $\widehat{U}=X-\widehat{F}\widehat{B}^\top =\left(I_T-\widehat{P}\right) X$.  Similarly, we let $\widetilde{Y}=\left(I_T-\widehat{P}\right)Y$ be an estimate of $Y-F\gamma^*$. The final estimate of $2T^{-1}\left\|U^\top (Y-F\gamma^*)\right\|_{\infty}$ is our test statistic \begin{equation}\label{test_stat_510}2T^{-1}\left\|\widehat{U}^\top \widetilde{Y}\right\|_{\infty}.\end{equation}

Next, to estimate the quantiles of the distribution of $2T^{-1}\left\|U^\top \mathcal{E}\right\|_{\infty},$ we need an estimate of $\mathcal{E}$. We obtain it through the following LASSO estimator:
\begin{equation}\label{eq:lasso}
\widehat{\beta}_\lambda = \argmin_{\beta\in\R^p}\frac{1}{T}\left\|\widetilde{Y}-\widehat{U}\beta\right\|_2^2+\lambda\|\beta\|_1,
\end{equation}
where $\lambda>0$ is a penalty parameter, the choice of which will be fully data-driven in both theory and practice. For $t\in [T]$, we denote by $\widetilde{y}_t$ the $t^{th}$ element of $\widetilde{Y}$ and $\widehat{u}_t$ as the $T\times 1$ vector corresponding to the $t^{th}$ row of $\widehat{U}$.
For a given $\lambda>0$, let $\widehat{\varepsilon}_{\lambda,t}=\widetilde{y}_t -\widehat{u}_t^\top\widehat{\beta}_{\lambda},\ t\in[T]$ be the estimate of $\varepsilon_t$. For a fixed $\alpha \in(0,1)$, we can then estimate $q_\alpha$, the $(1-\alpha)$ quantile of the distribution of $2T^{-1}\|U^\top \mathcal{E}\|_{\infty}$, by the Gaussian multiplier bootstrap. Let $e=(e_1,\dots,e_T)$ be a standard normal random vector independent of the data $(X,Y)$ and define the criterion
$$\widehat{Q}(\lambda,e)=\left\|\frac2T \sum_{t=1}^T\widehat{u}_t\widehat{\varepsilon}_{\lambda,t} e_t\right\|_\infty.$$
The estimate $\widehat{q}_\alpha(\lambda)$ of $q_\alpha$ is then the $(1-\alpha)$-quantile of the distribution of $\widehat{Q}(\lambda,e)$ given $X$ and $Y$. Formally, $\widehat{q}_\alpha(\lambda)=\inf\left\{q:\P_e(\widehat{Q}(\lambda,e)\le q)\ge 1-\alpha\right\}$, where $\P_e(\cdot)=\P(\cdot|X,Y)$.

The only remaining element is the procedure to select $\lambda$. We adapt the approach of \cite{lederer2021estimating} to our setting. Our choice of $\lambda$ is
\begin{equation}\label{choicelambda} \widehat{\lambda}_\alpha =\inf\{\lambda>0\ :\  \widehat{q}_\alpha(\lambda')\le \lambda' \text{ for all } \lambda'\ge \lambda\}.\end{equation}
We explain in Section \ref{subsec.computation} how to compute $\widehat{\lambda}_\alpha$ in practice. The infimum in \eqref{choicelambda} exists because for all $\lambda \ge \bar{\lambda}=2T^{-1}\left\|\widehat{U}^\top\widetilde{Y}\right\|_\infty$, it holds that $\widehat{\beta}_\lambda =\widehat{\beta}_{\bar \lambda}=0.$ Moreover, since $\widehat{U}\widehat{\beta}_\lambda$ is a continuous function of $\lambda$, $\widehat{q}_\alpha(\lambda)$ is also continuous in $\lambda$ and the infimum is attained at a point $\widehat{\lambda}_\alpha>0$ such that $q_\alpha\left(\widehat{\lambda}_\alpha\right)= \widehat{\lambda}_\alpha$. Let us recall briefly the heuristics behind the choice of $\lambda$ and refer the reader to \cite{lederer2021estimating} for more details. First, note that when $\lambda$ is close to $q_{\alpha}$, standard convergence bounds for the LASSO suggest that $\widehat{\beta}_\lambda$ is a precise estimate of $\beta^*$, so that $\widehat{\varepsilon}_{\lambda,t}$ is a good estimate of $\varepsilon_t$ and, in turn, $\widehat{q}_\alpha(\lambda)$ is close to $q_\alpha$. Second, when $\lambda$ becomes (much) larger than $q_\alpha$, the error $\widehat{\varepsilon}_{\lambda,t}-\varepsilon_t$ becomes large and dependent of $\widehat{u}_t$, which in turn increases $\widehat{q}_\alpha(\lambda)$ and leads it to be larger than $q_{\alpha}$. We then let our estimator of $q_\alpha$ be $\widehat{\lambda}_\alpha=\widehat{q}_\alpha\left(\widehat{\lambda}_\alpha\right)$.

The test rejects $H_0$ at the level $\alpha$ when our test statistic is given in \eqref{test_stat_510} is larger than the estimate $\widehat{\lambda}_\alpha$ of $q_\alpha$. Therefore, our testing procedure is free of tuning parameters stemming from the LASSO regression in equation \eqref{eq:lasso}.
	
	\subsection{Computation}\label{subsec.computation}
	Algorithm \ref{algo} below explains how to conduct the test in practice. Let us discuss Step 4 of Algorithm \ref{algo} in detail. It approximates $\widehat{\lambda}_\alpha$ as defined in \eqref{choicelambda}. It is advisable to set the grid size $M$ and the number of bootstrap samples $L$ to be as large as possible. As mentioned in \cite{lederer2021estimating}, one can speed up Step 4b. by computing the LASSO with a warm start along the penalty parameter path (see, e.g., \cite{friedman2007pathwise}), i.e., for each decreasing $\lambda$, the new coefficient estimate is computed by using the previous (i.e., the one that was computed with a larger value of $\lambda$) as a starting value. Furthermore, Step 4c. can be accelerated through parallelization techniques. In our implementation, we use both suggestions, which greatly speed up the computations. We also note that to compute the $p$-value of the test, it suffices to conduct it on a grid of values of $\alpha$ and let the $p$-value be equal to the largest value of $\alpha$ in this grid such that the test of level $\alpha$ rejects $H_0$.

	\begin{algorithm}
		\singlerule\vspace{0.1cm}
		\begin{algorithmic}
			\STATE \textbf{1.} Estimate $\widehat{K}$ by one of the available estimators of the number of factors.
			\STATE \textbf{2.} Let the columns of $\widehat{F}/\sqrt{T}$ be the eigenvectors corresponding to the leading $\widehat{K}$ eigenvalues of $XX^\top$.
			\STATE \textbf{3.} Compute $\widehat{U}=\left(I_T-\widehat{P}\right) X$ and $\widetilde{Y}=\left(I_T-\widehat{P}\right)Y$, where $\widehat{P}=T^{-1} \widehat{F}\widehat{F}^\top$.
			\STATE \textbf{4.} Calculate an approximation $\widehat{\lambda}_{\alpha,emp}$ of $\widehat{\lambda}_\alpha$ as follows:
			\STATE \begin{itemize}
				\item[\textbf{4a.}] Specify a grid $0<\lambda_1<\dots<\lambda_M<\bar{\lambda}$, with $\bar{\lambda}=2T^{-1}\left\|\widehat{U}^\top\widetilde{Y}\right\|_\infty$.
				\item[\textbf{4b.}] For $m\in[M]$ compute $\left\{\widehat{Q}\left(\lambda_m,e^{(\ell)}\right):\ \ell\in[L]\right\}$ for $L$ draws of $e\sim\mathcal{N}(0,I_T)$ and the corresponding empirical $(1-\alpha)$-quantile $\widehat{q}_{\alpha,emp}(\lambda_m)$ from them.
				\item[\textbf{4c.}] Let $\widehat{\lambda}_{\alpha,emp}=\widehat{q}_{\alpha,emp}(\lambda_{\widehat{m}})$,  with $\widehat{m}=\min\{m\in[M]:\ \widehat{q}_{\alpha,emp}(\lambda_{m'})\le \lambda_{m'}\text{ for all }m'\ge m\}.$
			\end{itemize}
			\STATE \textbf{5.} Reject $H_0$ when $2T^{-1}\left\|\widehat{U}^\top \widetilde{Y}\right\|_\infty>\widehat{\lambda}_{\alpha,emp}$.
		\end{algorithmic}
		\doublerule
		\caption{Conducting a test of level $\alpha\in(0,1)$.}
		\label{algo}
	\end{algorithm}

	\noindent 
	
	\section{Asymptotic theory}\label{sec.theory} In this section, we provide the asymptotic properties of the test in a theoretical framework allowing for time series dependence in the factors and the idiosyncratic shocks and polynomial tails. We place ourselves in an asymptotic regime where $T$ goes to infinity and $p$ goes to infinity as a function of $T$. The number of factors $K$ is fixed with $T$. It would be possible to let it grow, see, for instance, \cite{beyhum2022factor}. The distributions of the factors $f_{t}$ and the error terms $\varepsilon_t$ do not depend on $T$, while the distribution of the other variables are allowed to vary with $T$. All the constants we introduce are universal in the sense that they do not vary with the sample size. Our assumptions are similar to that of \cite{fan2021bridging} but significantly weaker than that of \cite{fan2023latent}, which imposes that the variables are i.i.d. sub-Gaussian.
	
	We introduce further notation. The loading $b_{jk}$ corresponds to the $j^{th}$ element of the $k^{th}$ column of $B$. Let also $b_j=(b_{j1},\dots,b_{jK})^\top$. For $t\in[T]$, let 
	$$z_t= \left(u_t^\top,f_t^\top,\varepsilon_t,\frac{1}{\sqrt{p}}\sum_{\ell=1}^pu_{t\ell}b_{\ell}^\top, \frac{1}{\sqrt{p}}\sum_{\ell=1}^pu_{t\ell}f_{t}^\top, \frac{1}{\sqrt{p}}\sum_{\ell=1}^pu_{t\ell}\varepsilon_t\right)^\top.$$ Finally, define $\Sigma=\E[u_tu_t^\top]$.

	We make the following assumptions. 
	\begin{Assumption}\label{as.nb}
		The estimator $\widehat{K}$ is such that $\P(\widehat{K}=K)\to 1$. 
	\end{Assumption}
	\begin{Assumption}\label{as.moments} It holds that
		\begin{enumerate}[\textup{(}i\textup{)}]
			\item\label{f4i} For all $t\in[T]$, $\E[f_t]=0,\ \E[f_tf_t^\top]=I_K$ and $B^\top B$ is diagonal;
			\item \label{f4ii}  All the eigenvalues of the $K\times K$ matrix $p^{-1} B^\top B$ are bounded away from $0$ and $\infty$ as $p\to \infty$;
			\item\label{f4iii} $\|\Sigma -BB^\top\|_2=O(1)$;
			\item\label{f4iv} $\|B\|_\infty=O(1)$.
		\end{enumerate}
	\end{Assumption}
	\begin{Assumption}\label{as.tail}
		The following holds:
		\begin{enumerate}[\textup{(}i\textup{)}]  
			\item\label{taili} The process $\left\{z_t\right\}_{t}$ is weakly stationary. Moreover, it holds that $$\E[u_{tj}]=\E[u_{tj}f_{tk}]=\E\left[f_{tk}\left(\sum_{\ell=1}^pu_{t\ell}b_{\ell h}\right)\right]=0,$$ \text{ for all } $t\in[T], j\in[p], k,h\in[K]$.
			\item\label{tailii} There exist $\kappa_1,\kappa_2>0$ such that,  for all $t\in[T]$, $\sigma_p(\E\left[\varepsilon_t^2u_tu_t^\top\right])>\kappa_1$, $\left\|\E\left[\varepsilon_t^2u_tu_t^\top\right]\right\|_\infty<\kappa_2$, $\sigma_p(\Sigma)>\kappa_1$ and $\|\Sigma\|_\infty<\kappa_2$;
			\item\label{tailiii} There exist $q \ge 8 $, $C_1>0$ and $\zeta>0$, such that for $s,t\in[T]$, we have 
			\begin{align*}
				\vertiii{z_t}_{q+\zeta}&<C_1;\\
				\vertiii{p^{-1/2}\left(u_s^\top u_{t}-\E\left[u_s^\top u_t\right]\right)}_{q}&<C_1;
			\end{align*}
			\item\label{tailiv} $\{u_t\varepsilon_t\}_t$ is uncorrelated across $t$, and, for all $t\in[T], j\in[p], k\in[K]$,
			$$\E[u_{tj}\varepsilon_t]=\E[f_{tk}\varepsilon_t]=\E\left[\varepsilon_t\sum_{\ell=1}^pu_{t\ell}b_{\ell k}\right]=0;$$ 
			\item\label{tailv} $\max_{j\in[p],k\in[K]}\left|T^{-1} \sum_{t=1}^T\E\left[u_{tj}\left(\sum_{\ell=1}^pu_{t\ell}b_{\ell k}\right)\right]\right| =O(1)$.
		\end{enumerate}
	\end{Assumption}
	
	\begin{Assumption}\label{as.mixing}
		Let $\tilde{\alpha}$ denote the strong mixing coefficients of $\left\{z_t\right\}_{t}$. There exist $C_2,c>0$ such that $c>\left[ \left(\frac{h+\xi}{\xi}\right)\left(\frac{h}{2}-1\right)\right]\vee \left(\frac{2}{1-\frac{2}{h}}\right)$ and $\kappa =\left(\frac{\frac{1}{2} +\frac{h}{4(h+1)}}{c+\frac{h}{2(h+1)}}\right)<\frac12$, where $h=\frac{q}{2}$ and $\xi=\frac{\zeta}{2}$, and,  for all  $t\in \mathbb{Z}_+$, we have 
		$\tilde{\alpha} (t)\le C_2t^{-c}.$
	\end{Assumption}
	
	Assumption \ref{as.nb} means that the estimator $\widehat{K}$ of the number of factors $K$ is consistent. Examples of $\widehat{K}$ and sufficient conditions for its consistency can be found in \cite{bai2002determining,onatski2010determining,ahn2013eigenvalue,bai2019rank,fan2022estimating}. Assumption \ref{as.moments} is the same as Assumption 3 in \cite{fan2021bridging}. Its conditions \eqref{f4i} and \eqref{f4ii} constitute a strong factor assumption (\cite{bai2003inferential}). Assumption \ref{as.tail} restricts the moments and the tail behavior of the variables. Assumption \ref{as.tail} \eqref{taili},\eqref{tailii},\eqref{tailiv} contain conditions on the moments of the different variables similar to that of the literature (\cite{bai2006confidence,fan2013large}). We assume that the variables in Assumption \ref{as.tail} \eqref{tailiii} have polynomial tails with common parameter $q+\zeta$. It would be possible to have different tail parameters for each variable but we avoid doing so to simplify our presentation. As in \cite{fan2021bridging} the number of finite moments is at least $8$. In the similar context of inference on factor regression models, Assumption E.2. in  \cite{bai2006confidence} and Assumption 7 in \cite{gonccalves2014bootstrapping} impose conditions analogous to the restriction that $\{u_t\varepsilon_t\}_t$ is uncorrelated across $t$ in Assumption \ref{as.tail} \eqref{tailiv}. A sufficient condition for the latter restriction is that $\{\varepsilon_t\}_t$ is uncorrelated across $t$ and independent of $\{u_t\}_t$. Note that this assumption could be avoided by using a block bootstrap method, but this would complicate the test. It may also not be justified because (most of) the serial correlation in the data may be picked up by the factors $f_t$ and not by the error term $\varepsilon_t$. Next, Assumption \ref{as.tail} \eqref{tailv} is a restriction on the cross-section correlation of the idiosyncratic shocks. This condition holds, for instance, if $\{u_t\}_t$ and $\{b_\ell\}_\ell$ are independent and $\max_{j\in[p]}\sum_{\ell=1}^p|\Sigma_{j\ell}| =O(1)$. Assumption \ref{as.mixing} means that the process $\{z_t\}_t$ has strong mixing coefficients decaying polynomially, which is a restriction on the time-series dependence of the variables. A similar assumption is made in \cite{fan2021bridging}. Note that Assumption \ref{as.mixing}  restricts the full distribution of the process $\{z_t\}_t$, while Assumption \ref{as.tail} \eqref{tailiv} just imposes a condition on a particular serial correlation.
	
	Let us introduce $\varphi^*= \gamma^*-B^\top \beta^*$ . To interpret $\varphi^*$, note that the first equation of \eqref{model} can be rewritten $y_t=f_t^\top\varphi^*+x_t^\top\beta^*+\varepsilon_t$, which becomes a usual high-dimensional sparse regression model when $\varphi^*=0$.
	The next assumption concerns the relative growth rates of $T,p,\|\varphi^*\|_2$ and $\|\beta^*\|_1$. 
	\begin{Assumption}\label{as.Rates}
		The following holds:
		\begin{enumerate}[\textup{(}i\textup{)}]  
			\item\label{ratei} $p=O(T^r),$ where $r<\left(\frac{h}{4} -\frac12\right)\wedge \left[(h+1)(c\kappa -\frac12)\right]$;
			\item\label{rateii} $\frac{\log(T\vee p)^{5}}{\sqrt{T}}\|\beta^*\|_1=o(1);$
			\item\label{rateiii} $\log(T\vee p)^{5/2}\frac{\sqrt{T}}{T\wedge p}(\|\varphi^*\|_2\vee 1)=o(1).$
		\end{enumerate}
	\end{Assumption}
	Condition (i) restricts the rate at which $p$ can grow with respect to $T$. We allow it to grow at a polynomial rate, which is more restrictive than the standard restriction for the LASSO (exponential rate). We need to be more restrictive because of the presence of polynomial tails and time series dependence. Assumption \ref{as.Rates} \eqref{rateii}  contains sparsity restrictions on the alternative hypotheses. When $\|\beta^*\|_\infty=O(1)$, condition \eqref{rateii} corresponds, up to logarithmic factors, to the standard consistency condition for the LASSO with bounded regressors and errors with sub-Gaussian tails that is $\sqrt{\log(p)/T}(s_0\vee 1)=o(1)$, where $s_0$ is the number of nonzero coefficients of $\beta^*$. We can impose a similar sparsity condition as in the standard LASSO literature with sub-Gaussian errors, despite dealing with polynomial tails and serial correlation, because we utilize the high-dimensional central limit theorem for polynomial-tailed time series from \cite{fan2021bridging}. Under the rate condition specified in Assumption \ref{as.Rates} \eqref{ratei}, this theorem allows us to essentially revert to the scenario with sub-Gaussian errors. Condition \eqref{rateiii} is a slightly more restrictive version of the standard condition that $\sqrt{T}/(T\wedge p)=o(1)$ for inference in the factor regression model.\footnote{This condition is equivalently stated as $\sqrt{T}/p=o(1)$ in \cite{bai2006confidence,corradi2014testing} and many others.} Indeed, since $\varphi^*$ is of size $K$, it is reasonable to assume that $\|\varphi^*\|_2=O(1)$. Under this condition, \eqref{rateiii} corresponds to $\sqrt{T}/(T\wedge p)=o(1)$ up to logarithmic factors. As noted by a referee, the role of (iii) is to ensure that the error coming from the estimation of the factors is asymptotically negligible (see for instance \cite{bai2006confidence}). It may be possible to relax it using a more elaborate bootstrap scheme, see \cite{gonccalves2014bootstrapping}.  Additionally, it is worth noting that our proofs reveal that Assumption \ref{as.Rates} is stronger than necessary, and the validity of the test could be established under more complex but weaker rate conditions. However, for the sake of clarity, we present Assumption \ref{as.Rates} instead of a more intricate condition.
	
	We have the following theorem. 
	\begin{Theorem}\label{th}Let Assumptions \ref{as.nb},  \ref{as.moments}, \ref{as.tail}, \ref{as.mixing} and \ref{as.Rates} hold. For all $\alpha \in (0,1)$, we have 
		\begin{enumerate}[\textup{(}i\textup{)}] 
			\item\label{thi} If $\beta^*=0$, then $\P\left(T^{-1}\left\|\widehat{U}^\top \widetilde{Y}\right\|_\infty> \widehat{\lambda}_\alpha\right)\le \alpha+o(1).$
			\item\label{thii} If $\sqrt{\frac{\log(T\vee p)}{T\wedge p}}=o_P\left(T^{-1}\left\|U^\top U\beta^*\right\|_\infty\right) $, then $\P\left(T^{-1}\left\|\widehat{U}^\top \widetilde{Y}\right\|_\infty> \widehat{\lambda}_\alpha\right)\to 1$.
		\end{enumerate}
	\end{Theorem}
	The proof of Theorem \ref{th} can be found in Online Appendix \ref{OA-app.proof}. Statement \eqref{thi} means that the empirical size of the test tends to the nominal size. Statement \eqref{thii} shows that the test has asymptotic power equal to $1$ against sequences of alternatives such that $\sqrt{\frac{\log(T\vee p)}{T\wedge p}}=o_P\left(T^{-1}\left\|U^\top U\beta^*\right\|_\infty\right) $. As noted in \cite{lederer2021estimating}, such a condition is inevitable because the presence of the error $\varepsilon_t$ prevents us from distinguishing true $U\beta^*$ and $\varepsilon_t$ when $U\beta^*$ is too small. We discuss further this condition in Section \ref{OA-sec.rate-cond} of the Online Appendix.
	
\section{Monte Carlo simulations}\label{sec.sim}
	
In this section, we provide a Monte Carlo study shedding light on the finite sample performance of our proposed testing procedure.  We use the following model as our data-generating process (DGP):

\begin{equation*}\label{main.mcs}
\begin{aligned}
y_t&=f_t^\top \gamma^*+u_t^\top\beta^*+\varepsilon_t,\\
x_t&=Bf_t+u_t,\quad t=1\dots,T.
\end{aligned}
\end{equation*}

We generate samples with $T=\{200, 400\}$ observations, $p=\{T, 5T\}$ variables and $K=2$ factors.  The loadings $B=\{b_{jk}\}_{j\in[p],k\in[K]}$ are such that $b_{jk}\sim\mathcal{U}[-1,1]$. The factors are generated as $f_t=\rho_f f_{t-1}+ \tilde f_t$ for $t=2,\dots, T$, where $\tilde f_t$ are i.i.d. $\mathcal{N}\left(0,I_K\left(1-\rho_f^2\right)\right)$. The idiosyncratic components $\{u_t\}$ are such that $u_t=\rho_{u}u_{t-1}+\tilde u_t $ for $t=2,\dots, T$, where $\tilde u_t$ are i.i.d. $\mathcal{N}\left(0,\Sigma\left(1-\rho_{u}^2\right)\right)$, with $\Sigma_{ij}=c_u^{|i-j|},i,j\in[p]$., where $c_u$  reflects the amount of cross-sectional dependence. We also let $\varepsilon_t=\rho_{e} \varepsilon_{t-1}+\tilde \varepsilon_t $ for $t=2,\dots, T$, where $\tilde \varepsilon_t$ are i.i.d. $\mathcal{N}\left(0,\left(1-\rho_{e}^2\right)\right)$.

The parameters $\rho_f$, $\rho_{u}$, and $\rho_e$ control the level of time series dependence. The stationary distributions of $f_t$, $u_t$, $\varepsilon_t$ are, respectively, $\mathcal{N}(0,I_K)$, $\mathcal{N}(0,\Sigma)$ and $\mathcal{N}(0,1)$. We initialize $f_0$, $u_0$ and $\varepsilon_0$ as such.  We consider three dependency designs, where we vary the cross-sectional dependence via parameter $s$ and the time series dependence via parameters $(\rho_f, \rho_u, \rho_e)$ as follows:
\begin{itemize}
\item[]\textit{Design 1.} $c_u=\rho_f=\rho_{u}=\rho_e=0$, so that the data are i.i.d. across $j$ and $t$.
\item[]\textit{Design 2.} $c_u=0.1$, $\rho_f=0.6$, $\rho_{u}=0.1$ and $\rho_e=0$, which introduces cross-sectional dependence in the idiosyncratic shocks and time-series dependence in the factors and the idiosyncratic shocks.
\item[]\textit{Design 3.} $c_u=0.1$, $\rho_f=0.6$ and $\rho_{u}=\rho_e=0.1$, where, on top of the cross-sectional dependence, there is time series dependence in the factors, the idiosyncratic shocks and the error terms.
\end{itemize}
Our theory does not formally allow the third design, but we want to show that our test performs well even under weak serial correlation of $\{\varepsilon_t\}_t$.

\smallskip

Finally, we consider two cases for the target parameter $\beta^*$. For the first case we set $\beta^*=(1,0,\dots,0)^\top\times m$, where $m\in\{0,0.1,0.2,0.3,0.4\}$ controls the signal strength. This choice of $\beta^*$ corresponds to a \textit{sparse} design. For the second case, we consider $\beta^*=(1/\sqrt{p}, \dots, 1/\sqrt{p})^\top\times m$, which corresponds to \textit{dense} design. In both cases, we set $\gamma^*=(0.5,0.5)^\top$.  Note that the choice of $\beta^*$ ensures that for the dependence Design 1, the signal-to-noise ratio is the same for both sparse and dense cases.

\smallskip

We compute the rejection probabilities of our test at the significance levels $\alpha\in\{0.1,0.05, 0.01\}$ over $2000$ replications. To implement our test, we set $M=100$ and choose an equidistant grid of values for $\lambda$, using $L=1000$ bootstrap replications. The results are insensitive to the choice of $L$ and $M$ as long as they are sufficiently large, which is expected since their primary role is in the approximation of theoretical quantities. In our experience, $L=M=100$ already yields very precise results. The number of factors $K$ is estimated through the eigenvalue ratio estimator of \cite{ahn2013eigenvalue}.

\subsection{Main results}

The results for $T=200$ and sparse and dense alternatives are reported in Tables \ref{tab.rates_sparse}-\ref{tab.rates_dense}. In the Online Appendix, we present simulations under the same data-generating processes, but with the larger sample size $T=400$ (Tables \ref{OA-app.tab.rates_sparse} and \ref{OA-app.tab.rates_dense}). First, we observe that the empirical size is close to the nominal levels for all designs, indicating that both cross-sectional and serial dependence have little effect on the empirical size of our testing procedure. Notably, the results show consistent power when the alternative hypothesis is sparse (Table \ref{tab.rates_sparse}) across all data-generating processes (DGPs), suggesting that the dependency design does not significantly impact the power of our test. However, the power tends to slightly decrease in scenarios with large $p$ cases for both $T=200$ and $T=400$. Interestingly, in the case of dense alternatives (Table \ref{tab.rates_dense}), we see a significant drop (relative to sparse alternatives)  in power across all designs. This highlights that our testing procedure has low power when  $\beta^*\neq0$ is dense. An intuition for this result is as follows. Our test uses the LASSO estimator, which enforces sparsity. When $\beta^*$ is dense, the LASSO estimator often sets $\widehat{\beta}_{\widehat{\lambda}_\alpha}$ to $0$, leading to non-rejection of the null.

\begin{table}[ht]\setlength\extrarowheight{-4pt}
\centering
\begin{tabular}{lccccccc}
& & p/T = 200/200 & & & & p/T = 1000/200 & \\
\hline
$m$&   $\alpha =0.1$ &$\alpha = 0.05$ &$\alpha =0.01$ && $\alpha =0.1$ &$\alpha =0.05$ &$\alpha =0.01$ \\
\hline\\[-0.3cm]
\multicolumn{8}{c}{{\it Design 1}: $s=\rho_f=\rho_{u}=\rho_e=0$}\\
0.0 & 0.084 & 0.033 & 0.005 & & 0.090 & 0.040 & 0.005   \\
0.1 & 0.122 & 0.062 & 0.015 & & 0.108 & 0.049 & 0.010   \\
0.2 & 0.604 & 0.507 & 0.336 & & 0.274 & 0.184 & 0.082  \\
0.3 & 0.984 & 0.973 & 0.916 & & 0.704 & 0.614 & 0.418  \\
0.4 & 1.000 & 1.000 & 1.000 & & 0.958 & 0.927 & 0.833  \\
\multicolumn{8}{c}{{\it Design 2}: $s=0.1$, $\rho_f=0.6$, $\rho_{u}=0.1$ and $\rho_e=0$}\\
0.0 & 0.092 & 0.045 & 0.010 & & 0.090 & 0.040 & 0.004  \\
0.1 & 0.122 & 0.066 & 0.016 & & 0.104 & 0.048 & 0.009 \\
0.2 & 0.631 & 0.534 & 0.362 & & 0.274 & 0.184 & 0.080  \\
0.3 & 0.984 & 0.971 & 0.931 & & 0.731 & 0.633 & 0.435 \\
0.4 & 1.000 & 1.000 & 1.000 & & 0.961 & 0.931 & 0.853  \\
\multicolumn{8}{c}{{\it Design 3}: $s=0.1$, $\rho_f=0.6$ and $\rho_{u}=\rho_e=0.1$}\\
0.0 & 0.102 & 0.048 & 0.011 & & 0.099 & 0.044 & 0.004 \\
0.1 & 0.130 & 0.070 & 0.017 & & 0.108 & 0.054 & 0.009  \\
0.2 & 0.620 & 0.528 & 0.356 & & 0.273 & 0.186 & 0.084  \\
0.3 & 0.980 & 0.970 & 0.925 & & 0.723 & 0.631 & 0.429  \\
0.4 & 1.000 & 1.000 & 1.000 & & 0.962 & 0.933 & 0.849 \\
\hline\hline
\end{tabular}
\caption{Rejection probabilities for the three dependence designs we consider and sparse $\beta^*$. The data are generated with Gaussian variables. The sample size is $T=200$ while the number of regressors is $p\in\{200, 1000\}$.}
\label{tab.rates_sparse}
\end{table}

\begin{table}[ht]\setlength\extrarowheight{-4pt}
\centering
\begin{tabular}{lccccccc}
& & p/T = 200/200 & & & & p/T = 1000/200 & \\
\hline
$m$&   $\alpha =0.1$ &$\alpha = 0.05$ &$\alpha =0.01$ && $\alpha =0.1$ &$\alpha =0.05$ &$\alpha =0.01$ \\
\hline\\[-0.3cm]
\multicolumn{8}{c}{{\it Design 1}: $s=\rho_f=\rho_{u}=\rho_e=0$}\\
0.0 & 0.084 & 0.033 & 0.005 & & 0.090 & 0.040 & 0.005  \\
0.1 & 0.097 & 0.039 & 0.008 && 0.092 & 0.035 & 0.009  \\
0.2 & 0.114 & 0.060 & 0.007 & & 0.100 & 0.045 & 0.009  \\
0.3 & 0.140 & 0.074 & 0.011 & & 0.124 & 0.055 & 0.010  \\
0.4  & 0.176 & 0.096 & 0.017  & & 0.152 & 0.068 & 0.015 \\
\multicolumn{8}{c}{{\it Design 2}: $s=0.1$, $\rho_f=0.6$, $\rho_{u}=0.1$ and $\rho_e=0$}\\
0.0 & 0.084 & 0.040 & 0.006 & & 0.081 & 0.032 & 0.007 \\
0.1 & 0.085 & 0.041 & 0.009 & & 0.091 & 0.034 & 0.005  \\
0.2 & 0.108 & 0.051 & 0.012 & & 0.114 & 0.051 & 0.008 \\
0.3 & 0.155 & 0.072 & 0.015 & & 0.153 & 0.072 & 0.011  \\
0.4 & 0.210 & 0.109 & 0.023 & & 0.199 & 0.100 & 0.018  \\
\multicolumn{8}{c}{{\it Design 3}: $s=0.1$, $\rho_f=0.6$ and $\rho_{u}=\rho_e=0.1$}\\
0.0 & 0.094 & 0.046 & 0.005 & & 0.091 & 0.035 & 0.006  \\
0.1 & 0.100 & 0.046 & 0.010  & & 0.104 & 0.044 & 0.006 \\
0.2 & 0.121 & 0.060 & 0.012 & & 0.124 & 0.055 & 0.009  \\
0.3 & 0.171 & 0.087 & 0.018 & & 0.168 & 0.080 & 0.012  \\
0.4 & 0.228 & 0.116 & 0.025 & & 0.222 & 0.108 & 0.020 \\
\hline\hline
\end{tabular}
\caption{Rejection probabilities for the three dependence designs we consider and dense $\beta^*$. The data are generated with Gaussian variables. The sample size is $T=200$ while the number of regressors is $p\in\{200, 1000\}$.}
\label{tab.rates_dense}
\end{table}

\subsection{Heavy-tailed data, number of factors and lagged idiosyncratic shocks}\label{mc.heavy.k}

In this subsection, we consider the same data-generating process as in the main design Table \ref{tab.rates_sparse}, but we change elements of it to obtain results for heavy-tailed data and in cases when the number of factors is over- and under-estimated. For the heavy-tailed data scenarios, we generate factors, idiosyncratic shocks and regression errors from student-$t(5)$ distribution. This allows us to investigate the impact of the heavy tails on the proposed testing procedure. Note that our theoretical analysis imposes a condition for the existence of first $8$ finite moments, so that generating student-$t(5)$ errors goes beyond what our assumptions allow.  We generate heavy-tailed factors $f_t$, idiosyncratic components $u_t$ and the error terms $\epsilon_t$ by generating $\tilde{f}_t$, $\tilde{u}_t$ and $\tilde{\epsilon}_t$ from a student-$t(5)$ rather than a Gaussian distribution. Besides changing the distribution of the data, we also compute the performance of our test when the number of factors is either over-estimated or under-estimated. In this case, our data generating process is the same as for Table \ref{tab.rates_sparse}, but rather than estimating the number of factors $K$ using the eigenvalue ratio estimator, we set it to $K=5$ (over-estimated case) and $K=1$ (under-estimated case).

\smallskip

Results are reported in Tables \ref{OA-app.tab.rates_heavy}-\ref{OA-app.tab.rates_heavy2} and \ref{OA-app.tab.rates_k5}-\ref{OA-app.tab.rates2_k1} for heavy-tailed data and different number of factors, respectively. In the former case, we see a slight deterioration of our method compared to the Gaussian DGPs. This is not surprising since the residuals of the LASSO regression, as well as the factors, may be less accurately estimated in finite samples due to heavy tails, see, e.g., \cite{babii2020inference}. Next, we analyze the results of the case of the over-estimated number of factors. The performance is slightly affected compared to the case where we use the eigenvalue ratio estimator to determine the number of factors (see Table \ref{tab.rates_sparse}), but the differences are small. When comparing with the under-estimated case, we see that our testing procedure is over-sized. These results are in line with the literature on inference with factors models, see, e.g., \cite{moon2015linear}.

\smallskip

It is also interesting to investigate the performance of our testing procedure with lagged idiosyncratic shocks. In this case, we generate the data from the following model
\begin{equation}\label{laggedu}
\begin{aligned}
y_t&=f_t^\top \gamma^*+u_t^\top\beta_1^*+u_{t-1}^\top\beta_2^*+\varepsilon_t,\\
x_t&=Bf_t+u_t,\quad t\in[T],
\end{aligned}
\end{equation}
where all the elements are as in the main DGPs except that we add lagged idiosyncratic shock, set $\beta_1^* = \beta^*=(1,0,\dots,0)^\top\times m$ and all elements of $\beta_2^*=0$. The algorithm to compute the test for this model appears in the Online Appendix \ref{OA-algo-lags}. We report results for sparse and Gaussian DGPs which appear in Online Appendix Tables \ref{OA-app.tab.rates_laggedu}-\ref{OA-app.tab.rates2_laggedu}. Results show a slight deterioration of performance in terms of power, while the empirical size appears to be similar as in the main scenario Table \ref{tab.rates_sparse} (see also Table \ref{OA-app.tab.rates_sparse} in Online Appendix for a large $T$ comparison). The small decrease in power is due to an increase in the dimensionality of the regressors.

\section{Empirical applications}\label{sec.emp}
In this section, we present three empirical examples using well-established macroeconomic and financial datasets.

\smallskip

First, we examine the FRED-MD dataset, which includes 121 monthly macroeconomic series covering various sectors of the US economy. For further details, see \cite{mccracken2016fred}. We also study a quarterly variant of FRED-MD with 202 variables, named FRED-QD, for which results are reported in the Online Appendix. Second, we analyze a financial dataset comprising 100 representative anomalies from the literature, which we use to explain aggregate market returns and industry portfolio returns. For more information, we refer to \cite{dong2022anomalies}. Lastly, we investigate the network structure in asset returns using a finance dataset. Following the approach of \cite{fan2021bridging}, we consider a cross-section of monthly stock returns and a set of observable factors to study the relationships among financial firms.

\smallskip

We study these datasets for several reasons. First, FRED-MD and its quarterly variant FRED-QD are among the most widely studied high-dimensional macroeconomic datasets. Providing further insights into these datasets could be valuable for the extensive empirical macroeconomic literature. Second, the empirical asset pricing literature has proposed and examined many factors, known as anomalies, that explain asset prices. Our research offers new insights into applying long-short anomalies to explain the market and industry portfolio returns rather than individual asset returns. Third, applying our method to firm-level stock returns illustrates using a model with observed regressors, as detailed in Appendix Section \ref{app.modelw}. Overall, these diverse applications, using macroeconomic and financial datasets, demonstrate the versatility of our approach across different econometric settings, varying in sample sizes relative to the number of regressors. Lastly, we note that in all empirical applications, we select the number of factors using the eigenvalue ratio estimator of \cite{ahn2013eigenvalue}.

\subsection{Macroeconomic application}
In the macroeconomic application, we investigate the FRED-MD dataset over the sample period 1980 January to 2019 December containing $T=480$ observations. This is a commonly used dataset in various macroeconomic studies. In our analysis, we regress each of the 121 variables available in the dataset on common and idiosyncratic shocks. Specifically, we estimate the following model:
\begin{equation}\label{eq:model1}
\begin{aligned}
y_{t+1}&=f_t^\top \gamma^*+u_t^\top\beta^*+\varepsilon_{t+1},\\
x_t&=Bf_t+u_t,\quad t=1\dots,T,
\end{aligned}
\end{equation}
where $y_{t+1}$ represents one of the variables in the dataset at time $t+1$, and $x_t$ includes all the remaining regressors at time $t$. Note that we transform the original series using commonly applied transformations as suggested by \cite{mccracken2016fred}. To estimate $f_t$, we apply PCA using the eigenvalue ratio estimator to determine the number of common factors. We apply our procedure to test $H_0$. Results for each category of variable appear in Table \ref{tab.fredmdreject}.

\smallskip

First, results show that in most categories, we frequently reject the null hypothesis at the 10\%, 5\%, and even 1\% significance levels. This provides evidence of sparsity in the idiosyncratic shocks of the macroeconomic data. The categories where we reject the null most often are \textit{Output and Income}, \textit{Consumption, Orders, and Inventories}, \textit{Labor Market}, \textit{Interest and Exchange Rates}, and \textit{Prices}. For the remaining categories, the null hypothesis is not frequently rejected, suggesting that the sparse component is not important for those series. Interestingly, for the housing category, we never reject the null, while the other two categories for which we reject less frequently, i.e., \textit{Money and Credit} and \textit{Stock Market}, could be classified as financial rather than macroeconomic data. In addition to the results for the FRED-MD dataset, we also obtain similar rejection ratios for the FRED-QD dataset, which is a quarterly macroeconomic dataset similar to FRED-MD, but containing more variables and less observations. Results appear in Table \ref{OA-app.tab.fredqdreject}. The results for the FRED-QD dataset also show a similar pattern, providing evidence of sparsity in the idiosyncratic shocks.

\begin{table}[ht]
\centering
\begin{tabular}{rccc}
Category & 10\% & 5\% & 1\% \\
\hline
Output and Income (16) & 0.500 & 0.438 & 0.125 \\
Consumption, Orders, and Inventories (9) & 1.000 & 0.778 & 0.222 \\
Labor Market (31) & 0.774 & 0.677 & 0.419 \\
Housing (10) & 0.000 & 0.000 & 0.000 \\
Money and Credit (12) & 0.250 & 0.167 & 0.083 \\
Stock Market (3) & 0.333 & 0.000 & 0.000 \\
Interest and Exchange Rates (19) & 0.789 & 0.737 & 0.474 \\
Prices (20) & 0.800 & 0.650 & 0.450 \\
\hline\hline
\end{tabular}
\caption{Rejection ratios for each category of the FRED-MD dataset over the sample period 1980 January to 2019 December. In parentheses, we report the number of series per category. Data source: \cite{mccracken2016fred}. \label{tab.fredmdreject}}
\end{table}

\subsection{Finance application I}\label{finance-application-i}

In our second application, we test the sparse component of a financial dataset comprising a set of regressor variables shown to predict market excess returns (\citealp{dong2022anomalies}). Specifically, we examine whether the idiosyncratic sparse component is important in explaining market excess returns as well as returns of 49 industry portfolios when regressing on a representative sample of 100 long-short anomaly portfolio returns. The data spans from January 1970 to December 2017, therefore, the effective sample size is $T=575$. We estimate the same model as in \eqref{eq:model1}.

\smallskip

\cite{dong2022anomalies} have shown that these regressors lead to accurate forecasts of stock market excess return by employing various machine learning and forecast combination methods. For more details on the data and a full list of target variables, see Section  \ref{OA-app.sec.fin1} of the Online Appendix. Results are reported in Table \ref{tab:finapp1}.

\smallskip

Our results indicate that for more than 50\% of 49 industry returns, the idiosyncratic shocks are significant at the 10\% level, meaning we reject the null hypothesis at the 10\% level. For the aggregate market returns, the p-value is 0.042, providing evidence in favor of the sparse component. Additionally, we reject the null at the 1\% significance level for nine industry portfolios: \textit{Apparel} (Clths), \textit{Automobiles and Trucks} (Autos), \textit{Aircraft} (Aero), \textit{Rubber and Plastic Products} (Rubbr), \textit{Construction Materials} (BldMt), \textit{Real Estate} (RlEst), \textit{Printing and Publishing} (Books), \textit{Non-Metallic and Industrial Metal Mining} (Mines), and \textit{Business Supplies} (Paper). Furthermore, we analyze returns of 10 industry portfolios which are based on a broader classification compared to 49 industries. Results are presented in Table \ref{OA-app.tab:finapp1} in the Online  Appendix, which confirm a similar pattern. Therefore, we find strong evidence supporting the presence of sparse idiosyncratic components in equity returns for a variety of industries and the aggregate market.


\begin{table}
\centering\footnotesize
\begin{tabular}{cc|cc|cc|cc|cc}
\textbf{Market}$^{**}$ & 0.042 & \textbf{Clths}$^{***}$ & 0.001 & FabPr & 0.272 & \textbf{Oil}$^{**}$ & 0.014 & \textbf{Boxes}$^{**}$ &
0.015 \\
Agric & 0.184 & Hlth & 0.500 & \textbf{Mach}$^{**}$ & 0.012 & Util & 0.252 & \textbf{Trans}$^{**}$ &
0.024 \\
\textbf{Food}$^{**}$ & 0.012 & MedEq & 0.198 & \textbf{ElcEq}$^{**}$ & 0.034 & Telcm & 0.625 & Whlsl &
0.117 \\
Soda & 0.129 & Drugs & 0.286 & \textbf{Autos}$^{***}$  & 0.007 & \textbf{PerSv}$^{**}$  & 0.043 & \textbf{Rtail}$^{*}$  &
0.099 \\
Beer & 0.150 & \textbf{Chems}$^{**}$  & 0.014 & \textbf{Aero}$^{***}$  & 0.006 &  \textbf{BusSv}$^{*}$ & 0.051 & \textbf{Meals}$^{**}$  &
0.043 \\
Smoke & 0.130 & \textbf{Rubbr}$^{***}$  & 0.002 & \textbf{Ships}$^{**}$  & 0.034 & Hardw & 0.636 & \textbf{Banks}$^{**}$  &
0.011 \\
Toys & 0.205 & \textbf{Txtls}$^{**}$  & 0.023 & Guns & 0.173 & Softw & 0.206 & \textbf{Insur}$^{**}$  &
0.024 \\
\textbf{Fun}$^{*}$ & 0.057 & \textbf{BldMt}$^{***}$  & 0.004 & Gold & 0.504 & Chips & 0.646 & \textbf{RlEst}$^{***}$  &
0.003 \\
\textbf{Books}$^{***}$  & 0.001 & Cnstr & 0.123 & \textbf{Mines}$^{***}$  & 0.005 & LabEq & 0.618 & Fin &
0.195 \\
Hshld & 0.189 & Steel & 0.278 & Coal & 0.268 & \textbf{Paper}$^{***}$  & 0.001 & \textbf{Other}$^{**}$  &
0.010 \\
\hline\hline
\end{tabular}
\caption{p-Values of market and industry excess returns regressed on 100 long-short anomaly characteristics. Bold entries with $^{*}$, $^{**}$, and $^{***}$ indicate significance at 10\%, 5\% and 1\% significance level, respectively. \label{tab:finapp1}}
\end{table}

\subsection{Finance application II}\label{finance-application-ii}

In the third empirical application, we examine the sparse idiosyncratic components of individual stock returns using the dataset from \cite{jensen2023there}. Specifically, we use monthly stock return data for a sample of 728 firms with no missing data from January 1991 to December 2022, resulting in $T=384$ and  $p=727$ (the regressors $x_{it}$ are the returns of other firms). We test a model similar to \cite{fan2021bridging} where our test can be seen as a diagnostic check after the third step in the approach put forward by \cite{fan2021bridging}. However, our analysis diverges from \cite{fan2021bridging} by focusing on a balanced panel of firms. We select firms for which we have a complete time series of returns and observed regressors. Furthermore, in our industry or sector analyses, we group firms into 49 industries — the same as in finance I application —, whereas \cite{fan2021bridging} use a different firm classification.


\smallskip


Specifically, denote $y^{(i)}_{t}$ the stock excess return of firm $i$ at time $t$. We regress the firm's excess return on a set of observable factors — see Table \ref{OA-tab:finapp2.obs} for the list of factors — as well as common and idiosyncratic components stemming from all other returns. The model is thus an example of the extension of the main model with additional covariates, see Online Appendix \ref{app.modelw}. Denoting the returns of all firms except the $i^{th}$ firm as $x_{it} = (y_{jt})_{j\in[n]/i}$, where $n$ is the total number of firms, we report the rejection ratios $\beta^*$ by testing $\beta^*=0$ in the following model for each firm $i$:
\begin{equation}\label{model-firm-rets}
\begin{aligned}
y^{(i)}_{t}&= f_{t}^\top \gamma^*+ w_t^\top\delta^* + u_{it}^\top\beta^*+\varepsilon_{t},\\
x_{it}&=B f_{it}+ u_{it},\quad t=1\dots,T,
\end{aligned}
\end{equation}
where $w_t$ denotes the observable factors.

Results are reported in Table \ref{tab:finapp2}. First, we see that the rejection ratios (the proportion of firms $i$ for which we reject the null) are relatively low and only slightly above the nominal significance levels. This suggests that the sparse component is much less significant in individual stock returns compared to other applications we considered, i.e., we find weak evidence of the presence of sparse idiosyncratic shocks in firm-level returns. To gain further insight, we report results by grouping firms into 49 industries, as reported in Table \ref{OA-tab:finapp2.res.ind} in the Online Appendix. We find that for most industries, the rejection ratios are small. However, a few industries exhibit a larger proportion of rejections, notably \textit{Precious Metals} (Gold) and \textit{Communications} (Telcm). 

\begin{table}[ht]
\centering
\begin{tabular}{rrrr}
& 10 \% & 5 \% & 1 \% \\
Rejection rates & 0.130 & 0.082 & 0.038 \\
\hline\hline
\end{tabular}
\caption{Rejection rates for the firm-level financial returns dataset. \label{tab:finapp2}}
\end{table}

\section{Conclusion}\label{sec.ccl}

This paper proposes a new bootstrap test for the adequacy of the factor regression model against factor-augmented sparse alternatives. We establish the asymptotic validity of our test under time series dependence and polynomial tails. In a Monte Carlo study, we show that our procedure has excellent finite sample properties against sparse alternatives but low power against dense alternatives. We often reject the null when we apply our testing procedure to standard datasets in macroeconomics and finance. This suggests that sparsity is present - on top of a dense model - in several economic environments.

This message complements previous studies. Indeed, based on different approaches, \cite{giannone2021economic,kolesar2023fragility} found evidence against the presence of sparsity in several economic datasets, comparing only sparse and only dense models. Our analysis instead suggests that sparsity may still have a role to play on top of a dense component. These findings also constitute arguments in favor of using sparse plus dense models, which have recently gathered a lot of interest in the econometrics literature.

A potential limitation of our analysis is that we may reject $H_0$ because $\beta^*$ is nonzero but dense. In our Monte-Carlo simulations, we compare DGPs with sparse and dense $\beta^*$ with the same signal-to-noise ratio and find that our test has high power against the sparse alternative but low power against dense deviations from the null. This finding should mitigate the previous concern that rejection might be due to dense $ \beta^*\ne 0$. A complementary approach free of this concern is testing sparsity directly. \cite{kolesar2023fragility} takes this road by proposing a test for the null hypothesis of sparsity in a high-dimensional regression model (without factors). The limitations of their approach are that it is only valid when the number of variables is smaller than the sample size and that it does not reject the null when the regression coefficient is equal to $0$. This is a problem because, in our context, we would not conclude in favor of the existence of sparsity when $\beta^*=0$. For future research, it would be interesting to adapt the test of \cite{kolesar2023fragility} to the factor-augmented regression model and apply it to the datasets we consider in the present paper.
\appendix 
\renewcommand*{\thesection}{\Alph{section}}

\section*{Appendix: extension to additional regressors}

\section{A model with additional regressors}\label{app.modelw}
As in \cite{stock2002forecasting,bai2006confidence, lederer2021estimating}, in empirical applications we can augment the model with additional observed low-dimensional regressors $w_1,\dots,w_t\in\R^\ell$ (where $\ell$ is fixed with $T$).  We therefore can consider the alternative model.
	\begin{equation}\label{model-alternative}
		\begin{aligned}
			y_t&=f_t^\top \gamma^*+w_t^\top\delta^*+u_t^\top\beta^*+\varepsilon_t,\\
			x_t&=Bf_t+u_t,\quad t=1\dots,T,
		\end{aligned}
	\end{equation}
	Here, again, $\varepsilon_t\in \R$ represents a random error, $u_t$ is a $p$-dimensional random vector of idiosyncratic shocks, $f_t$ is a $K$-dimensional random vector of factors, and $B$ is a $p\times K$ random matrix of loadings. The parameters are $\gamma^*\in \R^K$, $\delta^*\in\R^\ell$, $\beta^*\in \R^p$. Note that here $w_t$ plays the role of an observed factor (with loading equal to $0$). This will be key to understanding the alternative testing procedure of Section \ref{subsec.test-alternative}.
	
	We focus on testing
	\begin{equation}\label{Htest-alternative}
		H_0:\beta^*=0\quad \text{against}\quad H_1:\beta^*\ne 0.
	\end{equation}
	To facilitate understanding, we again rewrite the model in matrix form as follows:
	\begin{equation*}
		\begin{aligned}
			Y&=F^\top \gamma^*+W\delta^*+ U^\top\beta^*+\mathcal{E},\\
			X&=BF+U,
		\end{aligned}
	\end{equation*}
	where $Y=(y_1,\dots,y_T)^\top$, $F=(f_1,\dots, f_T)^\top$ is a $T\times K$ matrix, $U=(u_1,\dots,u_T)^\top$, $W=(w_1,\dots,w_T)^\top$ and $X=(x_1,\dots,x_T)^\top$ are $T\times p$ matrices and $\mathcal{E}=(\varepsilon_1,\dots,\varepsilon_T)^\top.$\\
	
	\section{Testing procedure of the extended model} \label{subsec.test-alternative}
	Algorithm \ref{algo-alternative} present the test in this extended model. It is similar to Algorithm \ref{algo}. The only difference is that $\widehat{P}$ is now the projector on the columns of the $T\times (\widehat{K}+\ell)$ matrix $(\widehat{F}\ W)$ in Step 3. Essentially, $w_t$ is treated as an observed factor. 
	\begin{algorithm}
		\singlerule\vspace{0.1cm}
		\begin{algorithmic}
			\STATE \textbf{1.} Estimate $\widehat{K}$ by one of the available estimators of the number of factors.
			\STATE \textbf{2.} Let the columns of $\widehat{F}/\sqrt{T}$ be the eigenvectors corresponding to the leading $\widehat{K}$ eigenvalues of $XX^\top$.
			\STATE \textbf{3.} Compute $\widehat{U}=\left(I_T-\widehat{P}\right) X$ and $\widetilde{Y}=\left(I_T-\widehat{P}\right)Y$, where $\widehat{P}$ is the projector on the columns of the $T\times \left(\widehat{K}+\ell\right)$ matrix $\left(\widehat{F}\ W\right)$. Denote by $\widehat{u}_t$ the $T\times 1$ vector corresponding to the transpose of the $t^{th}$ row of $\widehat{U}$. 
			\STATE \textbf{4.} Calculate an approximation $\widehat{\lambda}_{\alpha,emp}$ of $\widehat{\lambda}_\alpha$ as follows:
			\STATE \begin{itemize}
				\item[\textbf{4a.}] Specify a grid $0<\lambda_1<\dots<\lambda_M<\bar{\lambda}$, with $\bar{\lambda}=2T^{-1}\left\|\widehat{U}^\top\widetilde{Y}\right\|_\infty$.
				\item[\textbf{4b.}] For $\lambda>0$, and $e\in\R^T$, let $\widehat{Q}\left(\lambda,e\right)=\left\|\frac2T \sum_{t=1}^T\widehat{u}_t\widehat{\varepsilon}_{\lambda,t} e_t,\right\|_\infty$, where $\widehat{\varepsilon}_{\lambda,t}=\widetilde{y}_t -\widehat{u}_t^\top\widehat{\beta}_{\lambda},\ t\in[T]$, for $		\widehat{\beta}_\lambda = \argmin_{\beta\in\R^p}\frac{1}{T}\left\|\widetilde{Y}-\widehat{U}\beta\right\|_2^2+\lambda\|\beta\|_1$. For $m\in[M]$, compute $\left\{\widehat{Q}\left(\lambda_m,e^{(\ell)}\right):\ \ell\in[L]\right\}$ for $L$ draws of $e\sim\mathcal{N}(0,I_T)$ and the corresponding empirical $(1-\alpha)$-quantile $\widehat{q}_{\alpha,emp}(\lambda_m)$ from them.
				\item[\textbf{4c.}] Let $\widehat{\lambda}_{\alpha,emp}=\widehat{q}_{\alpha,emp}(\lambda_{\widehat{m}})$,  with $\widehat{m}=\min\{m\in[M]:\ \widehat{q}_{\alpha,emp}(\lambda_{m'})\le \lambda_{m'}\text{ for all }m'\ge m\}.$
			\end{itemize}
			\STATE \textbf{5.} Reject $H_0$ when $2T^{-1}\left\|\widehat{U}^\top \widetilde{Y}\right\|_\infty>\widehat{\lambda}_{\alpha,emp}$.
		\end{algorithmic}
		\doublerule
		\caption{Conducting a test of level $\alpha\in(0,1)$ with additional regressors.\label{algo-alternative}}
	\end{algorithm}
	
	\section*{Supplementary material}
	
	\begin{description}
		
		\item[Online Appendix:] Additional empirical and simulation results, details on the data and the proof of Theorem \ref{th} (.pdf file). 
		\item[R package:] the R package \texttt{\textquotesingle FAS\textquotesingle}  that implements our test is available on CRAN: \url{https://cran.r-project.org/web/packages/FAS/index.html}.
	\end{description}
	
	\if1\blind
	{
		\section*{Acknowledgments}
		The authors thank Andrii Babii, Otilia Boldea, Peter Boswijk, Geert Dhaene, Art\={u}ras Juodis,  Frank Kleibergen, Lasse Heje Pedersen, seminar participants at Copenhagen Business School and BI Norwegian Business School, and conference attendants at the FinEML conference 2023 at Erasmus University (Rotterdam), the Leuven Statistics Days 2023, the Conference in Financial Econometrics (CFE) 2023 in Berlin, and the Netherlands Econometrics Study Group 2024 in Maastricht University. The authors also thank Michael Vogt for sharing his code for the procedure of \cite{lederer2021estimating}. Jad Beyhum undertook part of this work while employed by CREST, ENSAI (Rennes). Jad Beyhum gratefully acknowledges financial support from the Research Fund KU Leuven through the grant STG/23/014. Jonas Striaukas gratefully acknowledges the financial support from the European Commission, MSCA-2022-PF Individual Fellowship, Project 101103508. Project Acronym: MACROML. 
	} \fi

	\bibliographystyle{agsm}
	\bibliography{bibliography}

\end{document}


\def\spacingset#1{\renewcommand{\baselinestretch}%
		{#1}\small\normalsize} \spacingset{0}

	
	\if1\blind
	{
		\title{ Online appendix of ``Testing for sparse idiosyncratic components in factor-augmented regression models''}
		\author{Jad Beyhum\\\vspace{0.5em}
			Department of Economics, KU Leuven, Belgium\\\vspace{0.5em}
			Jonas Striaukas\hspace{.2cm}\\
			Department of Finance, Copenhagen Business School, Denmark.}
		\maketitle
	} \fi
	
	\if0\blind
	{
		\bigskip
		\bigskip
		\bigskip
		\begin{center}
			{\LARGE\bfPenalty }
		\end{center}
		\medskip
	} \fi

	\spacingset{1.5} 
	
	\renewcommand*{\thesection}{\Alph{section}}

	{\hypersetup{linkbordercolor=black,linkcolor = black,
			urlcolor  = blue,
			filecolor=black}
		\tableofcontents
	}
	\clearpage
	\setcounter{page}{1}
	\setcounter{section}{0}
	\setcounter{equation}{0}
	\setcounter{table}{0}
	\setcounter{figure}{0}
	\renewcommand{\theequation}{OA.\arabic{equation}}
	\renewcommand\thetable{OA.\arabic{table}}
	\renewcommand\thefigure{OA.\arabic{figure}}
	\renewcommand\thepage{OA. - \arabic{page}}
	\renewcommand\thealgorithm{OA.\arabic{algorithm}}

\section{Testing procedure of the model with lagged idiosyncratic terms}\label{subsec.lag}
Algorithm \ref{algo-lags} presents the test for the model with lagged idiosyncratic elements, see \eqref{P-laggedu}. This algorithm differs from the main procedure in Algorithm \ref{P-algo} by augmenting the lagged idiosyncratic terms into the sparse component. The procedure could be extended to accommodate observed factors, similar to Algorithm \ref{P-algo-alternative}. The algorithm with lagged idiosyncratic terms is outlined as follows and can be easily adapted in cases where there is more than one lag. 
	
\begin{algorithm}
	\singlerule\vspace{0.1cm}
	\begin{algorithmic}
		\STATE \textbf{1.} Estimate $\widehat{K}$ by one of the available estimators of the number of factors.
		\STATE \textbf{2.} Let the columns of $\widehat{F}/\sqrt{T}$ be the eigenvectors corresponding to the leading $\widehat{K}$ eigenvalues of $XX^\top$.
		\STATE \textbf{3.} Compute $\widehat{U}=\left(I_T-\widehat{P}\right) X$ and $\widetilde{Y}=\left(I_T-\widehat{P}\right)Y$, where $\widehat{P}=T^{-1} \widehat{F}\widehat{F}^\top$.
		\STATE \begin{itemize}
			\item[\textbf{3a.}] Denote by $\widehat{u}_t$ the $T\times 1$ vector corresponding to transpose of the $t^{th}$ row of $\widehat{U}$. 
			For all $t\in\{2,\dots T\}$, let $\widetilde{u}_t=\left(\widehat{u}_t^\top, \widehat{u}_{t-1}^\top\right)$.
		\end{itemize}
		\STATE \textbf{4.} Calculate an approximation $\widehat{\lambda}_{\alpha,emp}$ of $\widehat{\lambda}_\alpha$ as follows:
		\STATE \begin{itemize}
			\item[\textbf{4a.}] Specify a grid $0<\lambda_1<\dots<\lambda_M<\bar{\lambda}$, with $\bar{\lambda}=2(T-1)^{-1}\left\|\sum_{t=2}^T \widetilde{u}_t\widetilde{y}_t\right\|_\infty$.
			\item[\textbf{4b.}]  For $\lambda>0$, and $e\in\R^{\widetilde{T}}$, let $\widehat{Q}\left(\lambda,e\right)=\left\|\frac{2}{T-1} \sum_{t=2}^{T}\widetilde{u}_t\widehat{\varepsilon}_{\lambda,t} e_t,\right\|_\infty$, where $\widehat{\varepsilon}_{\lambda,t}=\widetilde{y}_t -\widetilde{u}_t^\top\widehat{\beta}_{\lambda},\ t\in\{2,\dots,T\}$, for $\widehat{\beta}_\lambda = \argmin_{\beta\in\R^p}\frac{1}{T-1}\sum_{t=2}^T
			\left(\widetilde{y}_t -\widetilde{u}_t^\top\widehat{\beta}_{\lambda}\right)^2+\lambda\|\beta\|_1$. For $m\in[M]$, compute $\left\{\widehat{Q}\left(\lambda_m,e^{(\ell)}\right):\ \ell\in[L]\right\}$ for $L$ draws of $e\sim\mathcal{N}(0,I_{T-1})$ and the corresponding empirical $(1-\alpha)$-quantile $\widehat{q}_{\alpha,emp}(\lambda_m)$ from them.
			\item[\textbf{4c.}] Let $\widehat{\lambda}_{\alpha,emp}=\widehat{q}_{\alpha,emp}(\lambda_{\widehat{m}})$,  with $\widehat{m}=\min\{m\in[M]:\ \widehat{q}_{\alpha,emp}(\lambda_{m'})\le \lambda_{m'}\text{ for all }m'\ge m\}.$
		\end{itemize}
		\STATE \textbf{5.} Reject $H_0$ when $2(T-1)^{-1}\left\|\sum_{t=2}^T \widetilde{u}_t\widetilde{y}_t\right\|_\infty>\widehat{\lambda}_{\alpha,emp}$.
	\end{algorithmic}
	\doublerule
	\caption{Conducting a test of level $\alpha\in(0,1)$ with lagged idiosyncratic terms.\label{algo-lags}}
\end{algorithm}

\section{Additional simulation results}\label{subsec.add.sim}
	
	\begin{table}[h!]\setlength\extrarowheight{-4pt}
		\begin{center}
			\begin{tabular}{lccccccc}
				& & p/T = 400/400 & & & & p/T = 2000/400 & \\
				\hline
				$m$&   $\alpha =0.1$ &$\alpha = 0.05$ &$\alpha =0.01$ && $\alpha =0.1$ &$\alpha =0.05$ &$\alpha =0.01$ \\
				\hline\\[-0.3cm]
				\multicolumn{8}{c}{{\it Design 1}: $s=\rho_f=\rho_{u}=\rho_e=0$}\\
				0.0 & 0.068 & 0.024 & 0.002 & & 0.072 & 0.026 & 0.005  \\
				0.1 & 0.085 & 0.038 & 0.005  & & 0.075 & 0.034 & 0.005 \\
				0.2 & 0.460 & 0.367 & 0.210 & & 0.170 & 0.101 & 0.034 \\
				0.3 & 0.949 & 0.926 & 0.847 & & 0.563 & 0.456 & 0.272  \\
				0.4 & 1.000 & 1.000 & 0.999 & & 0.898 & 0.856 & 0.734  \\
				\multicolumn{8}{c}{{\it Design 2}: $s=0.1$, $\rho_f=0.6$, $\rho_{u}=0.1$ and $\rho_e=0$}\\
				0.0 & 0.078 & 0.028 & 0.006 & & 0.070 & 0.027 & 0.005  \\
				0.1 & 0.098 & 0.046 & 0.011 & & 0.074 & 0.030 & 0.004  \\
				0.2 & 0.473 & 0.378 & 0.230 & & 0.174 & 0.101 & 0.034  \\
				0.3 & 0.961 & 0.942 & 0.873 & & 0.551 & 0.448 & 0.272  \\
				0.4 & 1.000 & 0.999 & 0.997 & & 0.900 & 0.850 & 0.720  \\
				\multicolumn{8}{c}{{\it Design 3}: $s=0.1$, $\rho_f=0.6$ and $\rho_{u}=\rho_e=0.1$}\\
				0.0 & 0.088 & 0.035 & 0.007  & & 0.084 & 0.032 & 0.005 \\
				0.1  & 0.112 & 0.052 & 0.014 & & 0.086 & 0.037 & 0.007  \\
				0.2 & 0.470 & 0.373 & 0.226 & & 0.185 & 0.102 & 0.035 \\
				0.3 & 0.961 & 0.935 & 0.870 & & 0.555 & 0.439 & 0.264  \\
				0.4 & 1.000 & 0.999 & 0.996 & & 0.895 & 0.843 & 0.721 \\ 
				\hline\hline
			\end{tabular}
		\end{center}
		\caption{Rejection probabilities for the three dependence designs we consider and sparse $\beta^*$. The data are generated with Gaussian variables. The sample size is $T=400$ while the number of regressors is $p\in\{400, 2000\}$.}
		\label{app.tab.rates_sparse}
	\end{table}

\begin{table}[h!]\setlength\extrarowheight{-4pt}
	\begin{center}
		\begin{tabular}{lccccccc}
			& & p/T = 400/400 & & & & p/T = 2000/400 & \\
			\hline
			$m$&   $\alpha =0.1$ &$\alpha = 0.05$ &$\alpha =0.01$ && $\alpha =0.1$ &$\alpha =0.05$ &$\alpha =0.01$ \\
			\hline\\[-0.3cm]
			\multicolumn{8}{c}{{\it Design 1}: $s=\rho_f=\rho_{u}=\rho_e=0$}\\
			0.0 & 0.068 & 0.024 & 0.002 & & 0.072 & 0.026 & 0.005  \\
			0.1  & 0.074 & 0.028 & 0.003 & & 0.070 & 0.030 & 0.005  \\
			0.2 & 0.086 & 0.032 & 0.004 & & 0.074 & 0.029 & 0.005  \\
			0.3 & 0.094 & 0.041 & 0.005 & & 0.084 & 0.034 & 0.004  \\
			0.4  & 0.104 & 0.052 & 0.006 & & 0.088 & 0.037 & 0.004  \\
			\multicolumn{8}{c}{{\it Design 2}: $s=0.1$, $\rho_f=0.6$, $\rho_{u}=0.1$ and $\rho_e=0$}\\
			0.0 & 0.073 & 0.035 & 0.005 & & 0.070 & 0.031 & 0.006 \\
			0.1 & 0.082 & 0.037 & 0.007 & & 0.070 & 0.032 & 0.008   \\
			0.2 & 0.092 & 0.041 & 0.009 & & 0.078 & 0.036 & 0.007   \\
			0.3  & 0.101 & 0.041 & 0.006 & & 0.080 & 0.036 & 0.008 \\
			0.4  & 0.112 & 0.048 & 0.007 & & 0.092 & 0.037 & 0.007 \\
			\multicolumn{8}{c}{{\it Design 3}: $s=0.1$, $\rho_f=0.6$ and $\rho_{u}=\rho_e=0.1$}\\
			0.0 & 0.090 & 0.043 & 0.006 & & 0.078 & 0.035 & 0.007  \\
			0.1 & 0.094 & 0.044 & 0.009  & & 0.083 & 0.038 & 0.008 \\
			0.2 & 0.110 & 0.046 & 0.008 & & 0.087 & 0.041 & 0.009  \\
			0.3 & 0.116 & 0.052 & 0.007 & & 0.094 & 0.046 & 0.009  \\
			0.4  & 0.131 & 0.056 & 0.007 & & 0.103 & 0.044 & 0.010 \\ 
			\hline\hline
		\end{tabular}
	\end{center}
	\caption{Rejection probabilities for the three dependence designs we consider and dense $\beta^*$. The data are generated  with Gaussian variables. The sample size is $T=400$ while the number of regressors is $p\in\{400, 2000\}$.}
	\label{app.tab.rates_dense}
\end{table}

\begin{table}[h!]\setlength\extrarowheight{-4pt}
	\begin{center}
		\begin{tabular}{lccccccc}
			& & p/T = 200/200 & & & & p/T = 1000/200 & \\
			\hline
			$m$&   $\alpha =0.1$ &$\alpha = 0.05$ &$\alpha =0.01$ && $\alpha =0.1$ &$\alpha =0.05$ &$\alpha =0.01$ \\
			\hline\\[-0.3cm]
			\multicolumn{8}{c}{{\it Design 1}: $s=\rho_f=\rho_{u}=\rho_e=0$}\\
			0.0 & 0.062 & 0.020 & 0.002 & & 0.054 & 0.018 & 0.002 \\
			0.1 & 0.082 & 0.034 & 0.005 & & 0.060 & 0.021 & 0.002  \\
			0.2 & 0.529 & 0.422 & 0.230  & & 0.200 & 0.116 & 0.035 \\
			0.3 & 0.936 & 0.886 & 0.766 & & 0.579 & 0.445 & 0.235  \\
			0.4 & 0.994 & 0.987 & 0.950  & & 0.872 & 0.785 & 0.585 \\
			\multicolumn{8}{c}{{\it Design 2}: $s=0.1$, $\rho_f=0.6$, $\rho_{u}=0.1$ and $\rho_e=0$}\\
			0.0 & 0.053 & 0.023 & 0.004 & & 0.050 & 0.014 & 0.002  \\
			0.1 & 0.090 & 0.040 & 0.006 & & 0.056 & 0.018 & 0.003  \\
			0.2 & 0.537 & 0.414 & 0.220 & & 0.196 & 0.114 & 0.040  \\
			0.3 & 0.942 & 0.894 & 0.761  & & 0.599 & 0.469 & 0.246 \\
			0.4 & 0.992 & 0.987 & 0.950 & & 0.878 & 0.801 & 0.594  \\
			\multicolumn{8}{c}{{\it Design 3}: $s=0.1$, $\rho_f=0.6$ and $\rho_{u}=\rho_e=0.1$}\\
			0.0 & 0.065 & 0.025 & 0.003  & & 0.060 & 0.018 & 0.002 \\
			0.1  & 0.096 & 0.048 & 0.006 & & 0.070 & 0.023 & 0.004  \\
			0.2  & 0.530 & 0.410 & 0.218 & & 0.210 & 0.114 & 0.040 \\
			0.3 & 0.936 & 0.892 & 0.755  & & 0.593 & 0.467 & 0.242 \\
			0.4 & 0.993 & 0.986 & 0.950 & & 0.876 & 0.799 & 0.590  \\ 
			\hline\hline
		\end{tabular}
	\end{center}
	\caption{Rejection probabilities for the three dependence designs we consider, sparse $\beta^*$ and with data generated with student-$t(5)$ variables. The sample size is $T=200$ while the number of regressors is $p\in\{200, 1000\}$.}
	\label{app.tab.rates_heavy}
\end{table}

\begin{table}[h!]\setlength\extrarowheight{-4pt}
	\begin{center}
		\begin{tabular}{lccccccc}
			& & p/T = 400/400 & & & & p/T = 2000/400 & \\
			\hline
			$m$&   $\alpha =0.1$ &$\alpha = 0.05$ &$\alpha =0.01$ && $\alpha =0.1$ &$\alpha =0.05$ &$\alpha =0.01$ \\
			\hline\\[-0.3cm]
			\multicolumn{8}{c}{{\it Design 1}: $s=\rho_f=\rho_{u}=\rho_e=0$}\\
			0.0 & 0.044 & 0.016 & 0.001 & & 0.037 & 0.010 & 0.001 \\
			0.1  & 0.056 & 0.024 & 0.004 & & 0.040 & 0.012 & 0.002  \\
			0.2 & 0.364 & 0.259 & 0.121 & & 0.118 & 0.055 & 0.015  \\
			0.3 & 0.880 & 0.822 & 0.632  & & 0.416 & 0.301 & 0.136 \\
			0.4 & 0.991 & 0.981 & 0.922 & & 0.761 & 0.643 & 0.418  \\
			\multicolumn{8}{c}{{\it Design 2}: $s=0.1$, $\rho_f=0.6$, $\rho_{u}=0.1$ and $\rho_e=0$}\\
			0.0 & 0.042 & 0.015 & 0.001 & & 0.031 & 0.009 & 0.002  \\
			0.1 & 0.054 & 0.021 & 0.004 & & 0.034 & 0.010 & 0.002 \\
			0.2 & 0.372 & 0.272 & 0.139 & & 0.104 & 0.049 & 0.014  \\
			0.3 & 0.890 & 0.829 & 0.641  & & 0.420 & 0.305 & 0.127 \\
			0.4 & 0.989 & 0.974 & 0.921  & & 0.760 & 0.657 & 0.410 \\
			\multicolumn{8}{c}{{\it Design 3}: $s=0.1$, $\rho_f=0.6$ and $\rho_{u}=\rho_e=0.1$}\\
			0.0 & 0.050 & 0.013 & 0.001  & & 0.039 & 0.011 & 0.002 \\
			0.1 & 0.056 & 0.021 & 0.004 & & 0.040 & 0.012 & 0.001  \\
			0.2 & 0.374 & 0.268 & 0.142  & & 0.114 & 0.054 & 0.014 \\
			0.3 & 0.883 & 0.820 & 0.633  & & 0.414 & 0.302 & 0.128 \\
			0.4 & 0.989 & 0.974 & 0.919 & & 0.753 & 0.656 & 0.404  \\ 
			\hline\hline
		\end{tabular}
	\end{center}
	\caption{Rejection probabilities for the three dependence designs we consider, sparse $\beta^*$ and with data generated with student-$t(5)$ variables. The sample size is $T=400$ while the number of regressors is $p\in\{400, 2000\}$.}
	\label{app.tab.rates_heavy2}
\end{table}

\begin{table}[h!]\setlength\extrarowheight{-4pt}
	\begin{center}
		\begin{tabular}{lccccccc}
			& & p/T = 200/200 & & & & p/T = 1000/200 & \\
			\hline
			$m$&   $\alpha =0.1$ &$\alpha = 0.05$ &$\alpha =0.01$ && $\alpha =0.1$ &$\alpha =0.05$ &$\alpha =0.01$ \\
			\hline\\[-0.3cm]
			\multicolumn{8}{c}{{\it Design 1}: $s=\rho_f=\rho_{u}=\rho_e=0$}\\
			0.0 & 0.084 & 0.036 & 0.006 & & 0.094 & 0.040 & 0.007  \\
			0.1  & 0.126 & 0.063 & 0.016  & & 0.106 & 0.052 & 0.011 \\
			0.2 & 0.594 & 0.494 & 0.322  & & 0.270 & 0.176 & 0.080 \\
			0.3 & 0.982 & 0.961 & 0.907  & & 0.669 & 0.580 & 0.391 \\
			0.4 & 1.000 & 1.000 & 0.999 & & 0.944 & 0.908 & 0.799  \\
			\multicolumn{8}{c}{{\it Design 2}: $s=0.1$, $\rho_f=0.6$, $\rho_{u}=0.1$ and $\rho_e=0$}\\
			0.0 & 0.092 & 0.046 & 0.008 & & 0.091 & 0.044 & 0.005  \\
			0.1 & 0.120 & 0.066 & 0.016 & & 0.111 & 0.055 & 0.009 \\
			0.2 & 0.609 & 0.518 & 0.348 & & 0.268 & 0.180 & 0.072  \\
			0.3 & 0.983 & 0.968 & 0.920 & & 0.704 & 0.613 & 0.408  \\
			0.4 & 1.000 & 1.000 & 0.999 & & 0.950 & 0.922 & 0.838  \\
			\multicolumn{8}{c}{{\it Design 3}: $s=0.1$, $\rho_f=0.6$ and $\rho_{u}=\rho_e=0.1$}\\
			0.0 & 0.091 & 0.041 & 0.006  & & 0.091 & 0.039 & 0.001 \\
			0.1 & 0.115 & 0.063 & 0.014 & & 0.109 & 0.052 & 0.005  \\
			0.2 & 0.607 & 0.517 & 0.347 & & 0.267 & 0.178 & 0.067  \\
			0.3 & 0.980 & 0.964 & 0.918  & & 0.700 & 0.611 & 0.405 \\
			0.4 & 0.997 & 0.999 & 0.997 & & 0.948 & 0.922 & 0.835  \\ 
			\hline\hline
		\end{tabular}
	\end{center}
	\caption{Rejection probabilities for the three dependence designs we consider, sparse $\beta^*$ and using $K = 5$ as the number of factors while the true number of factors is $K=2$. The data are generated with Gaussian variables. The sample size is $T=200$ while the number of regressors is $p\in\{200, 1000\}$.}
	\label{app.tab.rates_k5}
\end{table}

\begin{table}[h!]\setlength\extrarowheight{-4pt}
	\begin{center}
		\begin{tabular}{lccccccc}
			& & p/T = 400/400 & & & & p/T = 2000/400 & \\
			\hline
			$m$&   $\alpha =0.1$ &$\alpha = 0.05$ &$\alpha =0.01$ && $\alpha =0.1$ &$\alpha =0.05$ &$\alpha =0.01$ \\
			\hline\\[-0.3cm]
			\multicolumn{8}{c}{{\it Design 1}: $s=\rho_f=\rho_{u}=\rho_e=0$}\\
			0.0 & 0.079 & 0.028 & 0.002 & & 0.083 & 0.031 & 0.005  \\
			0.1 & 0.094 & 0.042 & 0.004  & & 0.083 & 0.034 & 0.005 \\
			0.2 & 0.451 & 0.362 & 0.212  & & 0.184 & 0.106 & 0.030 \\
			0.3 & 0.948 & 0.923 & 0.845 & & 0.548 & 0.456 & 0.268 \\
			0.4  & 1.000 & 1.000 & 0.998 & & 0.895 & 0.847 & 0.723  \\
			\multicolumn{8}{c}{{\it Design 2}: $s=0.1$, $\rho_f=0.6$, $\rho_{u}=0.1$ and $\rho_e=0$}\\
			0.0 & 0.080 & 0.035 & 0.008 & & 0.084 & 0.029 & 0.006  \\
			0.1 & 0.102 & 0.044 & 0.013 & & 0.088 & 0.032 & 0.005  \\
			0.2 & 0.467 & 0.374 & 0.233 & & 0.180 & 0.106 & 0.030  \\
			0.3 & 0.956 & 0.936 & 0.872  & & 0.545 & 0.442 & 0.262 \\
			0.4 & 1.000 & 0.999 & 0.998 & & 0.898 & 0.846 & 0.711  \\
			\multicolumn{8}{c}{{\it Design 3}: $s=0.1$, $\rho_f=0.6$ and $\rho_{u}=\rho_e=0.1$}\\
			0.0 & 0.077 & 0.034 & 0.007 & & 0.084 & 0.029 & 0.004  \\
			0.1 & 0.102 & 0.040 & 0.011 & & 0.088 & 0.032 & 0.003  \\
			0.2 & 0.465 & 0.371 & 0.231 & & 0.176 & 0.103 & 0.029 \\
			0.3 & 0.956 & 0.933 & 0.870  & & 0.540 & 0.438 & 0.259 \\
			0.4 & 0.995 & 0.996 & 0.996 & & 0.896 & 0.842 & 0.711  \\ 
			\hline\hline
		\end{tabular}
	\end{center}
	\caption{Rejection probabilities for the three dependence designs we consider, sparse $\beta^*$ and using $K = 5$ as the number of factors while the true number of factors is $K=2$. The data are generated with Gaussian variables. The sample size is $T=400$ while the number of regressors is $p\in\{400, 2000\}$.}
	\label{app.tab.rates2_k5}
\end{table}

\begin{table}[h!]\setlength\extrarowheight{-4pt}
	\begin{center}
		\begin{tabular}{lccccccc}
			& & p/T = 200/200 & & & & p/T = 1000/200 & \\
			\hline
			$m$&   $\alpha =0.1$ &$\alpha = 0.05$ &$\alpha =0.01$ && $\alpha =0.1$ &$\alpha =0.05$ &$\alpha =0.01$ \\
			\hline\\[-0.3cm]
			\multicolumn{8}{c}{{\it Design 1}: $s=\rho_f=\rho_{u}=\rho_e=0$}\\
			0.0 & 0.893 & 0.875 & 0.837  & & 0.843 & 0.812 & 0.743 \\
			0.1 & 0.895 & 0.876 & 0.837 & & 0.840 & 0.813 & 0.739  \\
			0.2 & 0.929 & 0.906 & 0.858 & & 0.853 & 0.824 & 0.740  \\
			0.3 & 0.987 & 0.973 & 0.931  & & 0.907 & 0.876 & 0.780 \\
			0.4  & 0.999 & 0.998 & 0.984 & & 0.969 & 0.934 & 0.847 \\
			\multicolumn{8}{c}{{\it Design 2}: $s=0.1$, $\rho_f=0.6$, $\rho_{u}=0.1$ and $\rho_e=0$}\\
			0.0 & 0.929 & 0.910 & 0.881  & & 0.877 & 0.856 & 0.800 \\
			0.1  & 0.928 & 0.913 & 0.879 & & 0.877 & 0.851 & 0.799  \\
			0.2 & 0.952 & 0.930 & 0.891  & & 0.882 & 0.858 & 0.810 \\
			0.3 & 0.986 & 0.969 & 0.928  & & 0.923 & 0.890 & 0.826 \\
			0.4 & 0.997 & 0.994 & 0.974  & & 0.964 & 0.936 & 0.858 \\
			\multicolumn{8}{c}{{\it Design 3}: $s=0.1$, $\rho_f=0.6$ and $\rho_{u}=\rho_e=0.1$}\\
			0.0 & 0.925 & 0.912 & 0.879 && 0.884 & 0.858 & 0.804   \\
			0.1 & 0.929 & 0.915 & 0.885 & & 0.879 & 0.850 & 0.804  \\
			0.2 & 0.951 & 0.925 & 0.892 & & 0.888 & 0.855 & 0.805  \\
			0.3 & 0.984 & 0.971 & 0.930 & & 0.918 & 0.885 & 0.821 \\
			0.4 & 0.997 & 0.994 & 0.976 & & 0.964 & 0.934 & 0.856   \\ 
			\hline\hline
		\end{tabular}
	\end{center}
	\caption{Rejection probabilities for the three dependence designs we consider, sparse $\beta^*$ and using $K = 1$ as the number of factors while the true number of factors is $K=2$. The data are generated  with Gaussian variables. The sample size is $T=200$ while the number of regressors is $p\in\{200, 1000\}$.}
	\label{app.tab.rates_k1}
\end{table}

\begin{table}[h!]\setlength\extrarowheight{-4pt}
	\begin{center}
		\begin{tabular}{lccccccc}
			& & p/T = 400/400 & & & & p/T = 2000/400 & \\
			\hline
			$m$&   $\alpha =0.1$ &$\alpha = 0.05$ &$\alpha =0.01$ && $\alpha =0.1$ &$\alpha =0.05$ &$\alpha =0.01$ \\
			\hline\\[-0.3cm]
			\multicolumn{8}{c}{{\it Design 1}: $s=\rho_f=\rho_{u}=\rho_e=0$}\\
			0.0 & 0.914 & 0.898 & 0.859 & & 0.867 & 0.839 & 0.777  \\
			0.1 & 0.917 & 0.895 & 0.859  & & 0.866 & 0.838 & 0.780 \\
			0.2 & 0.934 & 0.911 & 0.863  & & 0.868 & 0.843 & 0.780 \\
			0.3 & 0.973 & 0.959 & 0.906 & & 0.894 & 0.859 & 0.789  \\
			0.4 & 0.998 & 0.994 & 0.971 & & 0.943 & 0.909 & 0.824  \\
			\multicolumn{8}{c}{{\it Design 2}: $s=0.1$, $\rho_f=0.6$, $\rho_{u}=0.1$ and $\rho_e=0$}\\
			0.0 & 0.931 & 0.919 & 0.892 & & 0.890 & 0.869 & 0.829  \\
			0.1 & 0.935 & 0.921 & 0.892  & & 0.889 & 0.871 & 0.818 \\
			0.2 & 0.936 & 0.924 & 0.892 & & 0.889 & 0.869 & 0.820  \\
			0.3 & 0.978 & 0.961 & 0.919  & & 0.914 & 0.884 & 0.820 \\
			0.4 & 0.997 & 0.994 & 0.972 & & 0.946 & 0.914 & 0.840  \\
			\multicolumn{8}{c}{{\it Design 3}: $s=0.1$, $\rho_f=0.6$ and $\rho_{u}=\rho_e=0.1$}\\
			0.0 & 0.930 & 0.919 & 0.892 & & 0.890 & 0.866 & 0.829  \\
			0.1 & 0.933 & 0.922 & 0.891 & & 0.889 & 0.865 & 0.819  \\
			0.2 & 0.940 & 0.923 & 0.892 & & 0.891 & 0.869 & 0.815 \\
			0.3 & 0.977 & 0.961 & 0.920 & & 0.914 & 0.884 & 0.821  \\
			0.4 & 0.998 & 0.993 & 0.972 & & 0.945 & 0.914 & 0.838  \\ 
			\hline\hline
		\end{tabular}
	\end{center}
	\caption{Rejection probabilities for the three dependence designs we consider, sparse $\beta^*$ and using $K = 1$ as the number of factors while the true number of factors is $K=2$. The data are generated  with Gaussian variables. The sample size is $T=400$ while the number of regressors is $p\in\{400, 2000\}$.}
	\label{app.tab.rates2_k1}
\end{table}	
	
\begin{table}[h!]\setlength\extrarowheight{-4pt}
	\begin{center}
		\begin{tabular}{lccccccc}
			& & p/T = 200/200 & & & & p/T = 1000/200 & \\
			\hline
			$m$&   $\alpha =0.1$ &$\alpha = 0.05$ &$\alpha =0.01$ && $\alpha =0.1$ &$\alpha =0.05$ &$\alpha =0.01$ \\
			\hline\\[-0.3cm]
			\multicolumn{8}{c}{{\it Design 1}: $s=\rho_f=\rho_{u}=\rho_e=0$}\\
			0.0 &  0.084 & 0.038 & 0.007 &  &  0.073 & 0.032 & 0.003  \\
			0.1 &  0.115 & 0.063 & 0.015 &  &   0.078 & 0.035 & 0.004  \\
			0.2 & 0.537 & 0.442 & 0.290 &  &  0.216 & 0.135 & 0.052  \\
			0.3 &  0.973 & 0.958 & 0.896  &  &   0.633 & 0.530 & 0.346 \\
			0.4 &  1.000 & 1.000 & 0.999  &  &  0.933 & 0.902 & 0.794  \\
			\multicolumn{8}{c}{{\it Design 2}: $s=0.1$, $\rho_f=0.6$, $\rho_{u}=0.1$ and $\rho_e=0$}\\
			0.0 &  0.074 & 0.036 & 0.006 &  &  0.070 & 0.027 & 0.004  \\
			0.1  &  0.097 & 0.052 & 0.010 &  &  0.084 & 0.034 & 0.006 \\
			0.2 &  0.565 & 0.468 & 0.288 &  &  0.222 & 0.138 & 0.052  \\
			0.3 &  0.968 & 0.953 & 0.898 &  &  0.654 & 0.540 & 0.360  \\
			0.4 &  0.999 & 0.999 & 0.999 &  &  0.943 & 0.907 & 0.794  \\
			\multicolumn{8}{c}{{\it Design 3}: $s=0.1$, $\rho_f=0.6$ and $\rho_{u}=\rho_e=0.1$}\\
			0.0 &  0.089 & 0.042 & 0.010   &  &  0.084 & 0.030 & 0.004  \\
			0.1 &  0.112 & 0.059 & 0.014  &  &  0.093 & 0.038 & 0.008  \\
			0.2 &  0.571 & 0.460 & 0.282 &  &  0.226 & 0.135 & 0.055 \\
			0.3 &  0.969 & 0.951 & 0.895  &  &  0.647 & 0.535 & 0.355 \\
			0.4 &  0.999 & 0.999 & 0.999   &  &  0.935 & 0.897 & 0.792  \\
			\hline\hline
		\end{tabular}
	\end{center}
	\caption{Rejection probabilities for the three dependence designs we consider, sparse $\beta^*$ and lagged idiosyncratic terms. The data are generated with Gaussian variables. The sample size is $T=200$ while the number of regressors is $p\in\{200, 1000\}$.}
	\label{app.tab.rates_laggedu}
\end{table}

\begin{table}[h!]\setlength\extrarowheight{-4pt}
	\begin{center}
		\begin{tabular}{lccccccc}
			& & p/T = 400/400 & & & & p/T = 2000/400 & \\
			\hline
			$m$&   $\alpha =0.1$ &$\alpha = 0.05$ &$\alpha =0.01$ && $\alpha =0.1$ &$\alpha =0.05$ &$\alpha =0.01$ \\
			\hline\\[-0.3cm]
			\multicolumn{8}{c}{{\it Design 1}: $s=\rho_f=\rho_{u}=\rho_e=0$}\\
			0.0 &  0.070 & 0.028 & 0.005  &  &  0.061 & 0.027 & 0.005  \\
			0.1 &  0.081 & 0.038 & 0.011 &  & 0.064 & 0.028 & 0.004 \\
			0.2 & 0.400 & 0.320 & 0.184  &  &  0.127 & 0.066 & 0.016 \\
			0.3 &  0.932 & 0.903 & 0.823  &  &  0.467 & 0.373 & 0.206 \\
			0.4 &  1.000 & 0.999 & 0.995  &  &  0.862 & 0.812 & 0.664 \\
			\multicolumn{8}{c}{{\it Design 2}: $s=0.1$, $\rho_f=0.6$, $\rho_{u}=0.1$ and $\rho_e=0$}\\
			0.0 &  0.078 & 0.036 & 0.006  & &  0.059 & 0.020 & 0.001 \\
			0.1  &  0.088 & 0.042 & 0.008  & &  0.062 & 0.024 & 0.002 \\
			0.2 &  0.418 & 0.320 & 0.184  & & 0.130 & 0.070 & 0.021 \\
			0.3 &  0.936 & 0.905 & 0.823 & & 0.491 & 0.388 & 0.236   \\
			0.4 &  0.999 & 0.998 & 0.995  & & 0.878 & 0.826 & 0.693   \\
			\multicolumn{8}{c}{{\it Design 3}: $s=0.1$, $\rho_f=0.6$ and $\rho_{u}=\rho_e=0.1$}\\
			0.0 &  0.091 & 0.043 & 0.006 & &  0.068 & 0.028 & 0.002  \\
			0.1 &  0.106 & 0.048 & 0.007  & &  0.067 & 0.029 & 0.003  \\
			0.2 &  0.420 & 0.322 & 0.173 & & 0.137 & 0.077 & 0.024 \\
			0.3 &  0.931 & 0.904 & 0.812  & &  0.490 & 0.387 & 0.230 \\
			0.4 &  1.000 & 0.998 & 0.995  & &  0.875 & 0.823 & 0.682  \\
			\hline\hline
		\end{tabular}
	\end{center}
	\caption{Rejection probabilities for the three dependence designs we consider, sparse $\beta^*$, and lagged idiosyncratic terms. The data are generated with Gaussian variables. The sample size is $T=400$ while the number of regressors is $p\in\{400, 2000\}$.}
	\label{app.tab.rates2_laggedu}
\end{table}

\clearpage
	
\section{Additional empirical results}
		
\subsection{Macro application — FRED-QD}

\begin{table}[ht]
	\centering
	\begin{tabular}{rccc}
		Category & 10\% & 5\% & 1\% \\ 
		\hline
		NIPA (23)  & 0.478 & 0.435 & 0.261 \\
		Industrial Production (16) & 0.625 & 0.500 & 0.125 \\
		Employment and Unemployment (48)  & 0.688 & 0.562 & 0.375 \\
		Housing (10) & 0.400 & 0.200 & 0.200 \\
		Inventories, Orders, and Sales (7) & 0.286 & 0.286 & 0.286 \\
		Prices (36) & 0.417 & 0.306 & 0.194 \\
		Earnings and Productivity (13)  & 0.692 & 0.692 & 0.692 \\
		Interest Rates (15) & 0.400 & 0.400 & 0.133 \\
		Money and Credit (14) & 0.571 & 0.571 & 0.500 \\
		Household Balance Sheets (8)  & 0.250 & 0.125 &	0.125 \\
		Stock Markets (1) & 0.000 & 0.000 & 0.000 \\
		Exchange Rates (6) & 0.833 & 0.833 & 0.333 \\
		Other (2) & 0.000 & 0.000 & 0.000 \\
		Non-Household Balance Sheets (2) & 0.500 & 0.500 & 0.000 \\
		\hline\hline
	\end{tabular}
	\caption{Rejection ratios for each category of the FRED-QD dataset over the sample period 1959 Q3 to 2019 Q4. The sample size is $T=241$ while the number of regressors is $p=202$. In parentheses, we report the number of series per category. Data source: \cite{mccracken2021fred}. \label{app.tab.fredqdreject}}
\end{table}

\subsection{Finance I: additional details on data and results}\label{app.sec.fin1}

\begin{longtable}[ht]{rlll}
		& Variable name & Full name & Source \\ 
		\hline
		1 & Market & Market & \cite{dong2022anomalies} \\ 
		2 & Agric & Agriculture & \cite{french2024kenneth} \\ 
		3 & Food & Food Products & \cite{french2024kenneth} \\ 
		4 & Soda & Candy \& Soda & \cite{french2024kenneth} \\ 
		5 & Beer & Beer \& Liquor & \cite{french2024kenneth} \\ 
		6 & Smoke & Tobacco Products & \cite{french2024kenneth} \\ 
		7 & Toys & Recreation & \cite{french2024kenneth} \\ 
		8 & Fun & Entertainment & \cite{french2024kenneth} \\ 
		9 & Books & Printing and Publishing & \cite{french2024kenneth} \\ 
		10 & Hshld & Consumer Goods & \cite{french2024kenneth} \\ 
		11 & Clths & Apparel & \cite{french2024kenneth} \\ 
		12 & Hlth & Healthcare & \cite{french2024kenneth} \\ 
		13 & MedEq & Medical Equipment & \cite{french2024kenneth} \\ 
		14 & Drugs & Pharmaceutical Products & \cite{french2024kenneth} \\ 
		15 & Chems & Chemicals & \cite{french2024kenneth} \\ 
		16 & Rubbr & Rubber and Plastic Products & \cite{french2024kenneth} \\ 
		17 & Txtls & Textiles & \cite{french2024kenneth} \\ 
		18 & BldMt & Construction Materials & \cite{french2024kenneth} \\ 
		19 & Cnstr & Construction & \cite{french2024kenneth} \\ 
		20 & Steel & Steel Works Etc & \cite{french2024kenneth} \\ 
		21 & FabPr & Fabricated Products & \cite{french2024kenneth} \\ 
		22 & Mach & Machinery & \cite{french2024kenneth} \\ 
		23 & ElcEq & Electrical Equipment & \cite{french2024kenneth} \\ 
		24 & Autos & Automobiles and Trucks & \cite{french2024kenneth} \\ 
		25 & Aero & Aircraft & \cite{french2024kenneth} \\ 
		26 & Ships & Shipbuilding, Railroad Equipment & \cite{french2024kenneth} \\ 
		27 & Guns & Defense & \cite{french2024kenneth} \\ 
		28 & Gold & Precious Metals & \cite{french2024kenneth} \\ 
		29 & Mines & Non-Metallic and Industrial Metal Mining & \cite{french2024kenneth} \\ 
		30 & Coal & Coal & \cite{french2024kenneth} \\ 
		31 & Oil & Petroleum and Natural Gas & \cite{french2024kenneth} \\ 
		32 & Util & Utilities & \cite{french2024kenneth} \\ 
		33 & Telcm & Communication & \cite{french2024kenneth} \\ 
		34 & PerSv & Personal Services & \cite{french2024kenneth} \\ 
		35 & BusSv & Business Services & \cite{french2024kenneth} \\ 
		36 & Hardw & Computers & \cite{french2024kenneth} \\ 
		37 & Softw & Computer Software & \cite{french2024kenneth} \\ 
		38 & Chips & Electronic Equipment & \cite{french2024kenneth} \\ 
		39 & LabEq & Measuring and Control Equipment & \cite{french2024kenneth} \\ 
		40 & Paper & Business Supplies & \cite{french2024kenneth} \\ 
		41 & Boxes & Shipping Containers & \cite{french2024kenneth} \\ 
		42 & Trans & Transportation & \cite{french2024kenneth} \\ 
		43 & Whlsl & Wholesale & \cite{french2024kenneth} \\ 
		44 & Rtail & Retail & \cite{french2024kenneth} \\ 
		45 & Meals & Restaurants, Hotels, Motels & \cite{french2024kenneth} \\ 
		46 & Banks & Banking & \cite{french2024kenneth} \\ 
		47 & Insur & Insurance & \cite{french2024kenneth} \\ 
		48 & RlEst & Real Estate & \cite{french2024kenneth} \\ 
		49 & Fin & Trading & \cite{french2024kenneth} \\ 
		50 & Other & Other& \cite{french2024kenneth} \\ 
		\hline\hline
		\caption{A list of return portfolios for the results in reported in Table \ref{P-tab:finapp1}. We use average value weighted return. \label{app.tab:finapp1.target}}
\end{longtable}

\begin{table}[ht!]
	\centering\footnotesize
	\begin{tabular}{cc|cc|cc|cc|cc}
		\textbf{NoDur}$^{**}$ & 0.013 & \textbf{Durbl}$^{***}$ & 0.004 & \textbf{Manuf}$^{***}$ & 0.001 & \textbf{Enrgy}$^{**}$ & 0.016 & HiTec &
		0.533 \\
		Telcm & 0.625 & \textbf{Shops}$^{**}$ & 0.049 & Hlth & 0.265 & Utils & 0.252 & \textbf{Other}$^{***}$ &
		0.004 \\
		\hline\hline
	\end{tabular}
	\caption{P-values of industry excess returns (10 industry classification, see Table \ref{tab:finapp1.target2}) regressed on 100 characteristics. Bold entries with $*$, $**$, and $***$ indicate significance at the 10\%, 5\% and 1\% levels, respectively. \label{app.tab:finapp1}}
\end{table}

\begin{longtable}[ht]{rlll}
	& Variable name & Full name & Source \\ 
	\hline
	1 & NoDur & Consumer Nondurables & \cite{french2024kenneth} \\ 
	2 & Durbl & Consumer Durables & \cite{french2024kenneth} \\ 
	3 & Manuf & Manufacturing & \cite{french2024kenneth} \\ 
	4 &  Enrgy & Oil, Gas, and Coal & \cite{french2024kenneth} \\ 
	5 &  HiTec & Computers, Software, and Business/Electronic Equipment& \cite{french2024kenneth} \\ 
	6 &  Telcm & Telephone and Television Transmission & \cite{french2024kenneth} \\ 
	7 &  Shops & Wholesale, Retail, and Some Services  & \cite{french2024kenneth} \\ 
	8 &  Hlth & Healthcare, Medical Equipment, and Drugs & \cite{french2024kenneth} \\ 
	9 &  Utils & Utilities & \cite{french2024kenneth} \\ 
	10 & Other  & Other & \cite{french2024kenneth} \\ 
	\hline\hline
	\caption{A list of return portfolios for the results in reported in Table \ref{app.tab:finapp1}. We use average value weighted return. \label{tab:finapp1.target2}}
\end{longtable}

\subsection{Finance II: additional details on data}

 
 \begin{longtable}[ht]{rlll}
 	& Variable name & Factor & Source \\ 
 	\hline
 	1 & MKT  & Market & \cite{french2024kenneth} \\ 
	2 & SMB  & Small-minus-Big & \cite{french2024kenneth} \\ 
	3 & HML  & High-minus-Low & \cite{french2024kenneth} \\ 
	4 & CMA  & Conservative-minus-Aggressive & \cite{french2024kenneth} \\ 
	5 & RMW  & Robust-minus-Weak & \cite{french2024kenneth} \\ 
	6 & ni\_me& Earnings/price ratio & \cite{jensen2023there}\\
	7 & fcf\_me& Cash-flow/price ratio & \cite{jensen2023there}\\
	8 & div12m\_me& Dividend/price ratio & \cite{jensen2023there}\\
	9 & taccruals\_at & Accruals & \cite{jensen2023there}\\
	10 & beta\_60m & 60 Month CAPM Beta & \cite{jensen2023there}\\
	11 & turnover\_126d & Share turnover & \cite{jensen2023there}\\
	12 & rmax5\_rvol\_21d & Max Return to Volatility  & \cite{jensen2023there}\\
	13 & ivol\_ff3\_21d & Idiosync. vol. from the Fama-French 3-factor model  & \cite{jensen2023there}\\
	14 & ret\_3\_1 & Momentum 1-3 Months & \cite{jensen2023there}\\
	15 & ret\_60\_12 & Momentum 12-60 Months & \cite{jensen2023there}\\
 	\hline\hline
 	\caption{A list of observable factors used in Section \ref{P-finance-application-ii}.  \label{tab:finapp2.obs}}
 \end{longtable}

 \begin{longtable}[ht]{llllc}
		& 10 \% & 5 \% & 1 \% & Number of firms \\ 
		\hline
		Agric & 0.000 & 0.000 & 0.000 & 2 \\
		Food & 0.000 & 0.000 & 0.000 & 19 \\
		Soda & 0.000 & 0.000 & 0.000 & 4 \\
		Beer & 0.333 & 0.167 & 0.000 & 6 \\
		Smoke & 0.000 & 0.000 & 0.000 & 1 \\
		Toys & 0.000 & 0.000 & 0.000 & 5 \\
		Fun & 0.000 & 0.000 & 0.000 & 2 \\
		Books & 0.375 & 0.125 & 0.125 & 8 \\
		Hshld & 0.105 & 0.053 & 0.000 & 19 \\
		Clths & 0.000 & 0.000 & 0.000 & 9 \\
		Hlth & 0.100 & 0.100 & 0.100 & 10 \\
		MedEq & 0.053 & 0.000 & 0.000 & 19 \\
		Drugs & 0.053 & 0.053 & 0.000 & 19 \\
		Chems & 0.133 & 0.067 & 0.000 & 15 \\
		Rubbr & 0.000 & 0.000 & 0.000 & 3 \\
		Txtls & 0.000 & 0.000 & 0.000 & 5 \\
		BldMt & 0.000 & 0.000 & 0.000 & 15 \\
		Cnstr & 0.250 & 0.167 & 0.083 & 12 \\
		Steel & 0.222 & 0.111 & 0.111 & 9 \\
		FabPr & 0.000 & 0.000 & 0.000 & 1 \\
		Mach & 0.163 & 0.116 & 0.047 & 43 \\
		ElcEq & 0.167 & 0.167 & 0.083 & 12 \\
		Autos & 0.000 & 0.000 & 0.000 & 14 \\
		Aero & 0.083 & 0.083 & 0.000 & 12 \\
		Ships & 0.000 & 0.000 & 0.000 & 1 \\
		Guns & 0.000 & 0.000 & 0.000 & 4 \\
		Gold & 1.000 & 1.000 & 0.833 & 6 \\
		Mines & 0.000 & 0.000 & 0.000 & 2 \\ 
		Oil & 0.286 & 0.286 & 0.036 & 28 \\ 
		Util & 0.154 & 0.077 & 0.019 & 52 \\ 
		Telcm & 0.455 & 0.455 & 0.364 & 11 \\ 
		PerSv & 0.000 & 0.000 & 0.000 & 5 \\ 
		BusSv & 0.065 & 0.032 & 0.000 & 31 \\ 
		Hardw & 0.000 & 0.000 & 0.000 & 12 \\ 
		Softw & 0.100 & 0.050 & 0.000 & 20 \\
		Chips & 0.206 & 0.059 & 0.029 & 34 \\
		LabEq & 0.211 & 0.211 & 0.105 & 19 \\
		Paper & 0.000 & 0.000 & 0.000 & 11 \\
		Boxes & 0.000 & 0.000 & 0.000 & 3 \\
		Trans & 0.050 & 0.000 & 0.000 & 20 \\
		Whlsl & 0.179 & 0.143 & 0.071 & 28 \\
		Rtail & 0.138 & 0.069 & 0.069 & 29 \\
		Meals & 0.000 & 0.000 & 0.000 & 8 \\
		Banks & 0.096 & 0.027 & 0.014 & 73 \\
		Insur & 0.175 & 0.050 & 0.050 & 40 \\
		RlEst & 0.000 & 0.000 & 0.000 & 5 \\
		Fin & 0.077 & 0.077 & 0.000 & 13 \\
		Other & 0.000 & 0.000 & 0.000 & 2 \\
		\hline\hline
 	\caption{Rejection ratios for each industry of the returns of firms in the dataset of the application in Section  \ref{P-finance-application-ii} over the sample period January 1991 to December 2022, thus $T=384$ and $p=727$. In the Column \textit{Number of firms }, we report the number of firms per industry in our sample. Industry classification is based on 49 industry portfolios of \cite{french2024kenneth}, see Table \ref{app.tab:finapp1.target} for a full list of industries. Data source: \cite{jensen2023there}.  \label{tab:finapp2.res.ind}}
\end{longtable}

\section{On Theorem \ref{P-th}}\label{app.proof}This section contains material related to Theorem \ref{P-th}. In Section \ref{subsec.prem}, we define some useful mathematical objects. The proof of Theorem \ref{P-th} is given in Section \ref{subsec.proof} and makes use of results proved in later sections. Section \ref{subsec.distr} contains some auxiliary lemmas on distribution functions of random variables used in the proof of Theorem \ref{P-th}. Then, in Section \ref{subsec.prob}, we state and prove some lemmas on the probability of some events. Section \ref{subsec.seq}, contains results on some sequences introduced in the proof of Theorem \ref{P-th}.  Furthermore, Section \ref{subsec.fac} introduces results on the factors, the loadings and their estimators. Finally, Section \ref{subsec.pre} recalls pre-existing results on strong mixing sequences and high-dimensional Gaussian vectors. Finally, Section \ref{sec.rate-cond} discusses the rate condition in statement \eqref{P-thii} of Theorem \ref{P-th}. Our proofs borrow ideas and results from \cite{chernozukhov2013gaussian}, \cite{chernozhukov2015comparison}, \cite{lederer2021estimating}, \cite{fan2023latent} and \cite{fan2021bridging}. 

\subsection{Preliminaries}\label{subsec.prem}
We introduce some concepts which are latter useful in proving Theorem \ref{P-th}. 
	
First, we define (infeasible) estimators using the true number of factors.  These are all denoted using a ``check'' symbol. We let the columns of $\overline{F}/\sqrt{T}$ be the eigenvectors corresponding to the leading $K$ eigenvalues of $XX^\top$ and $\overline{B}=X^\top\overline{F} (\overline{F}^\top \overline{F})^{-1}=T^{-1} X^\top\overline{F}.$ Let also $\overline{P}=T^{-1} \overline{F}\left(\overline{F}^\top \overline{F}/T\right)^{-1}\overline{F}^\top=T^{-1} \overline{F}\overline{F}^\top$ be the projector on the vector space generated by the columns of $\overline{F}$. The estimator of $U$ is $\overline{U}=X-\overline{F}\overline{B}^\top =\left(I_T-\overline{P}\right) X$.  Similarly, we let $\overline{Y}=\left(I_T-\overline{P}\right)Y$ be the estimator of $Y-F\gamma^*$. The estimated loading $\overline{b}_{jk}$ corresponds to the $j^{th}$ element of the $k^{th}$ column of $\overline{B}$. Let also $\overline{b}_j=(\overline{b}_{j1},\dots,\overline{b}_{jK})^\top$. The second-step LASSO estimator using the true number of factors is then given by $$
\overline{\beta}_\lambda = \argmin_{\beta\in\R^p}\frac{1}{T}\left\|\overline{Y}-\overline{U}\beta\right\|_2^2+\lambda\|\beta\|_1,$$

For $t\in [T]$, we denote by $\overline{y}_t$ the $t^{th}$ element of $\overline{Y}$ and $\overline{u}_t$ as the $T\times 1$ vector corresponding to the $t^{th}$ row of $\overline{U}$.
For a given $\lambda>0$, let $\overline{\varepsilon}_{\lambda,t}=\overline{y}_t -\overline{u}_t^\top\overline{\beta}_{\lambda},\ t\in[T]$. The equivalent of $\widehat{Q}(\lambda,e)$ is then 
$$\overline{Q}(\lambda,e)=\left\|\frac2T \sum_{t=1}^T\overline{u}_t\overline{\varepsilon}_{\lambda,t} e_t,\right\|_\infty.$$
The estimator $\overline{q}_\alpha(\lambda)$ of $q_\alpha$ is the $(1-\alpha)$-quantile of the distribution of $\overline{Q}(\lambda,e)$ given $X$ and $Y$. Formally, $\overline{q}_\alpha(\lambda)=\inf\left\{q:\P_e(\overline{Q}(\lambda,e)\le q)\ge 1-\alpha\right\}$ (recall that $\P_e(\cdot)=\P(\cdot|X,Y)$). The analog of $\widehat{\lambda}_\alpha$ is given by
$$\label{choicelambda} \overline{\lambda}_\alpha =\inf\{\lambda>0\ :\  \overline{q}_\alpha(\lambda')\le \lambda' \text{ for all } \lambda'\ge \lambda\}.$$
Remark that, on the event $\{\widehat{K}=K\}$, we have $\widehat{F}=\overline{F}$, $\widehat{B}=\overline{B}$, $\widehat{U}=\overline{U}$, $\widetilde{Y}=\overline{Y}$, $\widehat{\beta}_\lambda=\overline{\beta}_\lambda$ and $\widehat{\lambda}_\alpha=\overline{\lambda}_\alpha$.

As in  \cite{lederer2021estimating}, we re-scale some quantities by multiplying them with $\sqrt{T}/2$. This re-scaling is convenient to apply some probabilistic results. For instance, we let $\overline{\Pi}(\mu,e)=\left\|\overline{W}(\mu,e)\right\|_\infty,$
where $$\overline{W}(\mu,e) =\left(\overline{W}_1(\mu,e), \dots, \overline{W}_p(\mu,e)\right)^\top,\ \text{with } \overline{W}_j(\mu,e)= \frac{1}{\sqrt{T}}\sum_{t=1}^T\overline{u}_{tj} \overline{\varepsilon}_{\frac{2}{\sqrt{T}}\mu,t} e_t .$$
Note that $\overline{\Pi}(\mu,e)=\frac{\sqrt{T}}{2}\overline{Q}(\lambda,e),$ for $\lambda =\frac{2}{\sqrt{T}}\mu.$
Similarly, for $\alpha\in(0,1)$, we define
\begin{align*}\overline{\pi}_\alpha(\mu)&=\inf\{q:\P_e(\overline{\Pi}(\mu,e)\le q)\ge 1-\alpha\};\\
	\overline{\mu}_\alpha &=\inf\{\mu>0\ :\  \overline{\pi}_\alpha(\mu')\le \mu' \text{ for all } \mu'\ge \mu\},	
\end{align*}
where $\overline{\mu}_\alpha =\frac{\sqrt{T}}{2}\overline{\lambda}_\alpha$. 
	
Next, to be able to compare $\overline{\Pi}(e)$ with population analogs, we define several additional quantities. Let $\Pi(e)=\left\|W(e)\right\|_\infty$, where 
$$W(e)=\left(W_1(e),\dots,W_p(e)\right)^\top,\ \text{with } W_j(e)=\frac{1}{\sqrt{T}}\sum_{t=1}^Tu_{tj} \varepsilon_te_t$$
and let $\mu_\alpha$ be the $(1-\alpha)-$quantile of $\Pi(e)$ conditionally on $(F,U,\mathcal{E})$.  Formally, $\mu_\alpha=\inf\{q:\P^*_e(\Pi(e)\le q)\ge 1-\alpha\}$, where $\P_e^*(\cdot)=\P(\cdot|F,U,\mathcal{E})$.
	
Moreover, we define $\Pi^*= \left\|W^*\right\|_\infty$, where $$W^*=\left(W_1^*,\dots,W_p^*\right)^\top,\ \text{with } W_j^*=\frac{1}{\sqrt{T}}\sum_{t=1}^Tu_{tj} \varepsilon_t,$$where $\mu_\alpha^*$ is the $(1-\alpha)$ quantile of $\Pi^*$. Finally, we also set $\Pi^G=\left\|G\right\|_\infty$ with $G$ a Gaussian vector with same covariance structure as $W^*$ and let $\mu_{\alpha}^G$ be the $(1-\alpha)$-quantile of $\Pi^G$. Auxiliary lemmas concerning the distributions of $\Pi(e)$, $\Pi^*$ and $\Pi^G$ can be found in Section \ref{subsec.distr}.
	
We also introduce the following useful quantities $$\Delta= \left\|\frac1T \sum_{t=1}^T u_tu_t^\top\varepsilon_t^2-\E\left[u_tu_t^\top\varepsilon_t^2\right]  \right\|_\infty\text{, }R(\mu,e)=\frac{1}{\sqrt{T}}\left\|\overline{W}(\mu,e)-W(e)\right\|_\infty,$$ for $\mu>0$ and the event $\mathcal{S}_{\mu}=\left\{\frac{2}{T}\left\|\overline{U}^\top (\overline{Y}-\overline{U}\beta^*)\right\|_\infty\le {\frac{2}{\sqrt{T}}\mu}\right\}.$ The above terms and events are controlled in Section \ref{subsec.prob}.
	
	The following sequences allow to bound some important terms in the proofs.
	\begin{align*}
		s_T^{(1)} &= \sqrt{\log(T)}\frac{p^{4/q}}{\sqrt{T}};\\
		s_T^{(2)} &= \sqrt{\log(T)}\left(\frac{1}{p}+\frac{p^{\frac2q}}{T}  +\frac{p^{\frac2q-\frac12}}{\sqrt{T}}\right)(\|\varphi^*\|_2\vee 1);\\
		s_T^{(3)} &= \sqrt{\log(T)}\left(1+\frac{p^{\frac{4}{q}}}{T} +\frac1p\right);\\
		s_T^{(4)}&= \sqrt{\log(T)} \left(\left(p^{\frac{4}{q}}T^{\frac{2}{q}-1}+\frac{T^{\frac{2}{q}}}{p}\right)+ \left(\frac1T+\frac{1}{p}\right)\left\|\varphi^*\right\|_2^2\right);\\
		s_T^{(5)} &=\sqrt{\log(T)} \frac{p^{\frac2q}}{\sqrt{T}};\\
		s_T^{(6)}&=\frac{2}{T^{1/4}}\sqrt{ \log(Tp)2\|\beta^*\|_1s_T^{(3)}};\\
		s_T^{(7)}&=\sqrt{\log(Tp)\frac{s_T^{(4)}}{T}};\\
		s_T^{(8)}&=K_1\left(\Delta\right)^{1/3}\left(1\vee 2\log(2p)\vee \log\left(1/\Delta\right)\right)^{1/3}\log(2p)^{1/3};\\
		s_T^{(9)}&= K_2\Big[\left(T^{\kappa-1/2}+T^{1-\frac{c}{2}(1-\frac{2}{h})}\right)\log(T)\log(p)+T^{-1/4}\log (p)^{3/2} \log(T)+ p^{\frac{2}{h}} T^{\frac{1}{h}-\frac12} \log(p)^2 \log(T)\\
		&\quad +\left( p\log(p)^{\frac{3}{2}h-4} \log(T)\log(Tp)\right)^{\frac{1}{h-2}}T^{-\frac14} +T^{\frac12-c\kappa} \left(p^{\frac{1}{h+1}} \sqrt{1\vee \log(p)}\right)\Big];\\
		s_T^{(10)}&=\frac{1}{T\vee p}+s_T^{(9)};\\
		s_T^{(11)}&= \kappa_2\left(\sqrt{2\log(2p)}+\sqrt{2\log(T\vee p)}\right);\\
		s_T^{(12)}&= s_T^{(6)}\left(1+s_T^{(11)}\right)+\frac{\sqrt{T}}{2}s_T^{(2)}+s_T^{(6)}\left(1+(1+s_T^{(6)})s_T^{(11)}+\frac{\sqrt{T}}{2}s_T^{(2)}\right)+s_T^{(7)};\\
		s_T^{(13)}&=s_T^{(6)}\left(1+s_T^{(11)}\right)+s_T^{(7)};\\
		s_T^{(14)}&=s_T^{(9)}+K_3s_T^{(12)}\sqrt{1\vee \log\left(2p/s_T^{(12)}\right)} +s_T^{(8)}+\frac3T,
	\end{align*}
	where $K_1,K_2$ and $K_3$ are constants introduced in Lemmas \ref{prop.LV2}, \ref{prop.LV1} and \ref{prop.LV3}, respectively. The constant $\kappa_2$ is defined in Assumption \ref{P-as.tail} and $h$ and $q$ are introduced in Assumptions \ref{P-as.tail} \eqref{P-tailiii} and \ref{P-as.mixing}. In Lemma \ref{lm.seq}, we show that these sequences all go to $0$ under Assumption \ref{P-as.Rates}. 
	
	Finally, we introduce the following events
	\begin{align*}
		\mathcal{S}_T^{(1)}&=\left\{\Delta \le s_T^{(1)}\right\};\\
		\mathcal{S}_T^{(2)} &=\left\{\left\|\frac{\overline{U}^\top (\overline{Y}-\overline{U}\beta^*)}{T}- \frac{U^\top\mathcal{E}}{T}\right\|_\infty\le s_T^{(2)}\right\};\\
		\mathcal{S}_T^{(3)}&=\left\{\max_{j\in[p]}\frac{1}{T}\sum_{t=1}^T \overline{u}_{tj}^2\le s_T^{(3)}\right\};\\
		\mathcal{S}_T^{(4)}&=\left\{\max_{j\in[p]}\frac{1}{T} \sum_{t=1}^T\left(\overline{u}_{tj}\widetilde{\varepsilon}_t+\widetilde{f}_t^\top\varphi^*- u_{tj}\varepsilon_t\right)^2\le s_T^{(4)}\right\};\\
		\mathcal{S}_T^{(5)}&= \left\{\left\|\frac{U^\top \mathcal{E}}{T}\right\|_\infty\le s_T^{(5)}\right\},
	\end{align*}
	where $\widetilde{\varepsilon}_t$ denotes the $t^{th}$ element of  $\left(I_T-\overline{P}\right)\mathcal{E}$ and $\widetilde{f}_t$ is the $K\times 1$ vector corresponding to the $t^{th}$ row of $\left(I_T-\overline{P}\right)F$.
	We show that the probabilities of these events go to $1$ with $T$ in Lemma \ref{lm.bound}. 
	
	\subsection{Proof of Theorem \ref{P-th}}\label{subsec.proof}
	\textit{Proof of \eqref{P-thi}.}
	We want to show that when $\beta^*=0$, we have 
	\begin{equation}\label{toshow206i1}\P\left(\left\|\frac{\widehat{U}^\top \widetilde{Y}}{T}\right\|_\infty> \widehat{\lambda}_\alpha\right)\le \alpha+o(1).\end{equation}
	First, we move from quantities depending on $\widehat{K}$ to quantities only depending on $K$. Notice that
	\begin{align*}
		&\P\left(\left\|\frac{\widehat{U}^\top \widetilde{Y}}{T}\right\|_\infty> \widehat{\lambda}_\alpha\right)\\&=\P\left(\left\{\left\|\frac{\widehat{U}^\top \widetilde{Y}}{T}\right\|_\infty> \widehat{\lambda}_\alpha\right\}\cap \left\{\widehat{K}=K\right\}\right)+  \P\left(\left\{\left\|\frac{\widehat{U}^\top \widetilde{Y}}{T}\right\|_\infty> \widehat{\lambda}_\alpha\right\}\cap \left\{\widehat{K}\ne K\right\}\right)\\
		&\le \P\left(\left\{\left\|\frac{\overline{U}^\top \overline{Y}}{T}\right\|_\infty> \overline{\lambda}_\alpha\right\}\cap \left\{\widehat{K}=K\right\}\right)+\P\left(\widehat{K}\ne K\right),
	\end{align*}
	where, we used the fact that, on the event $\{\widehat{K}=K\}$, we have $\widehat{U}=\overline{U}$, $\widetilde{Y}=\overline{Y}$ and $\widehat{\lambda}_\alpha=\overline{\lambda}_\alpha$.
	Now, using the formula $\P(C\cap D)=\P(C)+\P(D)-\P(C\cup D)$ (where $C$ and $D$ are two probabilistic events), we obtain
	\begin{align}
		\notag \P\left(\left\|\frac{\widehat{U}^\top \widetilde{Y}}{T}\right\|_\infty> \widehat{\lambda}_\alpha\right)&\le \P\left(\left\|\frac{\overline{U}^\top \overline{Y}}{T}\right\|_\infty> \overline{\lambda}_\alpha\right)+ \P(\widehat{K}=K) \\
		\notag &\quad - \P\left(\left\{\left\|\frac{\overline{U}^\top \overline{Y}}{T}\right\|_\infty> \overline{\lambda}_\alpha\right\}\cup \left\{\widehat{K}=K\right\}\right)+\P\left(\widehat{K}\ne K\right)\\
		\label{050324} &\le \P\left(\left\|\frac{\overline{U}^\top \overline{Y}}{T}\right\|_\infty> \overline{\lambda}_\alpha\right)+ \P\left(\widehat{K}\ne K\right)\\
		\label{0503242} &\le \P\left(\left\|\frac{\overline{U}^\top \overline{Y}}{T}\right\|_\infty> \overline{\lambda}_\alpha\right)+ o(1),
	\end{align}
	where, in equation \eqref{050324}, we used $\P(C\cup D)\ge \P(C)$ and, in equation \eqref{0503242}, we used Assumption \ref{P-as.nb}. This yields 
	\begin{align}
		\P\left(\left\|\frac{\widehat{U}^\top \widetilde{Y}}{T}\right\|_\infty> \widehat{\lambda}_\alpha\right)  &\le \P\left(\left\|\frac{\overline{U}^\top \overline{Y}}{T}\right\|_\infty>\overline{\lambda}_\alpha\right)+o(1)\\
		\notag &\le \P\left( \left\|\frac{U^\top \mathcal{E}}{T}\right\|_\infty+\left\|\frac{\overline{U}^\top \overline{Y}}{T}-\frac{U^\top \mathcal{E}}{T}\right\|_\infty> \overline{\lambda}_\alpha\right)+o(1)\\
		\notag &\le \P\left(\left\{\left\|\frac{U^\top \mathcal{E}}{T}\right\|_\infty>  \overline{\lambda}_\alpha -s_T^{(2)}\right\}\cap\mathcal{S}_T^{(2)} \right)+ \P\left((\mathcal{S}_T^{(2)})^c \right)+o(1)\\
		\label{toshow206i2} &\le \P\left(\left\|\frac{U^\top \mathcal{E}}{T}\right\|_\infty> \overline{\lambda}_\alpha - s_T^{(2)}\right)+o(1),
	\end{align}
	where in the last line we used Lemma \ref{lm.bound} (ii).
	
	Let us define
	$$\mathcal{T}_1=\mathcal{S}_{\mu^*_{\alpha+s_T^{(14)}}}\cap \mathcal{S}_T^{(1)}\cap \mathcal{S}_T^{(2)}\cap \mathcal{S}_T^{(3)}\cap\mathcal{S}_T^{(4)}.$$
	Note that, by Lemmas \ref{lm.bound} and \ref{lm.ps}, and the fact that $s_T^{(14)}\to 0$ by Lemma \ref{lm.seq} \eqref{lsiv}, \eqref{lsv} and \eqref{lsvi},  the event $\mathcal{T}_1$ has probability going to $1-\alpha$. 
	
	Hence, by \eqref{toshow206i2}, to show \eqref{toshow206i1}, it suffices to prove that,  on $\mathcal{T}_1$, we have \begin{equation}\label{toshow206i31}\overline{\lambda}_\alpha\ge \frac{2}{\sqrt{T}}\mu_{\alpha+s_T^{(14)}}^*+s_T^{(2)}.\end{equation}
	Indeed, in this case, we would have 
	\begin{align*}\P\left(\left\|\frac{\overline{U}^\top \overline{Y}}{T}\right\|_\infty> \overline{\lambda}_\alpha\right)&\le \P\left(\left\|\frac{U^\top \mathcal{E}}{T}\right\|_\infty>  \overline{\lambda}_\alpha - s_T^{(2)}\right)+o(1)\\
		&\le \P\left(\left\{\left\|\frac{U^\top \mathcal{E}}{T}\right\|_\infty>  \overline{\lambda}_\alpha - s_T^{(2)}\right\}\cap \mathcal{T}_1\right)+\P(\mathcal{T}_1^c)+o(1)\\
		&\le \P\left(\left\{\left\|\frac{U^\top \mathcal{E}}{T}\right\|_\infty>  \frac{2}{\sqrt{T}}\mu_{\alpha+s_T^{(14)}}^*\right\}\cap \mathcal{T}_1\right) +\P(\mathcal{T}_1^c)+o(1)\\
		&=0+\P(\mathcal{T}_1^c)+o(1) = \alpha+o(1),\end{align*}
	where, on the last line, we used $\mathcal{S}_{\mu^*_{\alpha+s_T^{(14)}}}\subset \mathcal{T}_1$.

	Let us therefore prove that, on $\mathcal{T}_1$, \eqref{toshow206i31} holds. To do so, we show that, on $\mathcal{T}_1$,
	\begin{equation}\label{toshow206i4}\P_e\left(\overline{\Pi}(\mu,e)> \mu\right) >\alpha \end{equation}
	for $\mu=(1+s_T^{(6)})\mu_{\alpha+s_T^{(14)}}^*+\frac{\sqrt{T}}{2}s_T^{(2)}>\mu_{\alpha+s_T^{(14)}}^*+\frac{\sqrt{T}}{2}s_T^{(2)}$, which implies that \eqref{toshow206i31} is true by definition of $\overline{\lambda}_\alpha$ and the fact that  $\overline{\mu}_\alpha =\frac{\sqrt{T}}{2}\overline{\lambda}_\alpha$.
	We have 
	\begin{align*}
		\P_e\left(\overline{\Pi}(\mu,e)> \mu\right)&\ge \P_e\left(\Pi(e)-R(\mu,e)> \mu\right)\\
		&\ge \P_e\left(\Pi(e)-R(\mu,e)> \mu, R(\mu,e)\le s_T^{(6)}\sqrt{\mu}+s_T^{(7)}\right)\\
		&\ge  \P_e\left(\Pi(e)> \mu+s_T^{(6)}\sqrt{\mu}+s_T^{(7)}\right)-\P_e\left( R(\mu,e)>s_T^{(6)}\sqrt{\mu}+s_T^{(7)}\right)\\
		&\ge \P_e\left(\Pi(e)> \mu+s_T^{(6)}\sqrt{\mu}+s_T^{(7)}\right)-\frac2T,
	\end{align*}
	where, on the last line, we used Lemma \ref{lm.lassodiff} and the facts that $ \mu\ge \mu_{\alpha+s_T^{(14)}}^*$ and we work on $\mathcal{S}^{(3)}_T\cap\mathcal{S}^{(4)}_T\cap \mathcal{S}_{\mu^*_{\alpha+s_T^{(14)}}}\subset \mathcal{T}_1$ to obtain that $\P_e\left( R(\mu,e)>s_T^{(6)}\sqrt{\mu}+s_T^{(7)}\right)\le \frac2T$.
	By Lemma \ref{prop.LV2}, we obtain
	\begin{equation}\label{eq2061}\P_e(\Pi(e)> \mu+s_T^{(6)}\sqrt{\mu}+s_T^{(7)})\ge \P(\Pi^G\ge\mu+s_T^{(6)}\sqrt{\mu}+s_T^{(7)}) -s_T^{(8)}-\frac2T.\end{equation}
	Since $\sqrt{\mu}\le (1+\mu)$, for $T$ large enough, it holds that
	\begin{align}
		\notag &\P\left((\Pi^G> \mu+s_T^{(6)}\sqrt{\mu}+s_T^{(7)}\right)\\
		\notag &\ge \P\left(\Pi^G> \mu+s_T^{(6)}(1+\mu)+s_T^{(7)}\right)\\
		\notag &\ge \P\left(\Pi^G> (1+s_T^{(6)})\mu_{\alpha+s_T^{(14)}}^*+\frac{\sqrt{T}}{2}s_T^{(2)}+s_T^{(6)}\left(1+(1+s_T^{(6)})\mu_{\alpha+s_T^{(14)}}^*+\frac{\sqrt{T}}{2}s_T^{(2)}\right)+s_T^{(7)}\right)\\
		\label{B4147}&\ge  \P\left(\Pi^G> \mu_{\alpha+s_T^{(14)}}^*+s_T^{(12)}\right)\\
		\notag&\ge \P\left(\Pi^G>\mu_{\alpha+s_T^{(14)}}^*)- \P(|\Pi^G-\mu_{\alpha+s_T^{(14)}}^*|\le s_T^{(12)}\right)\\
		\label{B3147}&\ge \P\left(\Pi^G>\mu_{\alpha+s_T^{(14)}}^*\right)-K_3s_T^{(12)}\sqrt{1\vee \log\left(2p/s_T^{(12)}\right)}\\
		\label{B1147}&\ge  \P\left(\Pi^*>\mu_{\alpha+s_T^{(14)}}^*\right)-s_T^{(9)}-K_3s_T^{(12)}\sqrt{1\vee \log\left(2p/s_T^{(12)}\right)}\\
		\notag &= \alpha+s_T^{(14)} -s_T^{(9)}-K_3s_T^{(12)}\sqrt{1\vee \log\left(2p/s_T^{(12)}\right)},
	\end{align}
	where, in \eqref{B4147}, we used  Lemma \ref{lm.G} and the fact that $s_T^{(14)}\to 0$ by Lemma \ref{lm.seq} \eqref{lsiv}, \eqref{lsv} and \eqref{lsvi}, to obtain that $\mu_{\alpha+s_T^{(14)}}^*\le s_T^{(11)}$ for $T$ large enough, in \eqref{B3147}, we leveraged Lemma \ref{prop.LV3} and \eqref{B1147} follows from Lemma \ref{prop.LV1}.
	This and \eqref{eq2061}, therefore yield
	$$\P_e\left(\Pi(e)> \mu\right) \ge  \alpha+s_T^{(14)} -s_T^{(9)}-K_3s_T^{(12)}\sqrt{1\vee \log\left(2p/s_T^{(12)}\right)} -s_T^{(8)}-\frac2T=\alpha+\frac1T>\alpha,$$
	by definition of $s_T^{(14)}$. This shows \eqref{toshow206i4} and therefore concludes the proof of \eqref{P-thi}. \\
	
	\noindent \textit{Proof of (ii).}
	We want to show that if $\sqrt{\frac{\log(T\vee p)}{T\wedge p}}=o_P\left(\left\|\frac{U^\top U\beta^*}{T}\right\|_\infty\right)$, we have 
	\begin{equation}\label{toshow206ii1}\P\left(\left\|\frac{\widehat{U}^\top \widehat{Y}}{T}\right\|_\infty> \widehat{\lambda}_\alpha\right)\to 1.\end{equation}
	As in the proof of (i), we first move from quantities depending on $\widehat{K}$ to quantities only depending on $K$. Notice that
	\begin{align}
		\notag &\P\left(\left\|\frac{\widehat{U}^\top \widetilde{Y}}{T}\right\|_\infty> \widehat{\lambda}_\alpha\right)\\
		\notag &\ge \P\left(\left\{\left\|\frac{\widehat{U}^\top \widehat{Y}}{T}\right\|_\infty> \widehat{\lambda}_\alpha\right\}\cap \left\{\widehat{K}=K\right\}\right)\\
		\notag &= \P\left(\left\{\left\|\frac{\overline{U}^\top \overline{Y}}{T}\right\|_\infty> \overline{\lambda}_\alpha\right\}\cap \left\{\widehat{K}=K\right\}\right)\\
		\label{0503243} &= \P\left(\left\|\frac{\overline{U}^\top \overline{Y}}{T}\right\|_\infty> \overline{\lambda}_\alpha\right) -  \P\left(\left\{\left\|\frac{\overline{U}^\top \overline{Y}}{T}\right\|_\infty> \overline{\lambda}_\alpha\right\}\cap \left\{\widehat{K}\ne K\right\}\right)\\
		\notag &\ge \P\left(\left\|\frac{\overline{U}^\top \overline{Y}}{T}\right\|_\infty> \overline{\lambda}_\alpha\right) -  \P\left(\widehat{K}\ne K\right)\\
		\label{0503245}&\ge \P\left(\left\|\frac{\overline{U}^\top \overline{Y}}{T}\right\|_\infty> \overline{\lambda}_\alpha\right)+o(1),
	\end{align}
	where, in \eqref{0503243}, we used the formula $\P(C)=\P(C\cap D)+\P(C\cap D^c)$, and \eqref{0503245} follows from Assumption \ref{P-as.nb}.
	Hence, it holds that
	\begin{align}
		\notag &\P\left(\left\|\frac{\widehat{U}^\top \widetilde{Y}}{T}\right\|_\infty> \widehat{\lambda}_\alpha\right)\\
		\notag &\ge  \P\left(\left\|\frac{\overline{U}^\top \overline{Y}}{T}\right\|_\infty> \overline{\lambda}_\alpha\right)\\
		\notag &\ge \P\left(\left\|\frac{U^\top U\beta^*}{T}\right\|_\infty-  \left\|\frac{U^\top \mathcal{E}}{T}\right\|_\infty-\left\|\frac{\overline{U}^\top (\overline{Y}-\overline{U}\beta^*)}{T}-\frac{U^\top \mathcal{E}}{T}\right\|_\infty> \overline{\lambda}_\alpha\right)\\
		\notag&\ge \P\left(\left\{\left\|\frac{U^\top U\beta^*}{T}\right\|_\infty>  \overline{\lambda}_\alpha + s_T^{(2)}+s_T^{(5)}\right\}\cap\mathcal{S}_T^{(2)}\cap\mathcal{S}_T^{(5)} \right)\\
		\notag& \ge \P\left(\left\|\frac{U^\top U\beta^*}{T}\right\|_\infty>  \overline{\lambda}_\alpha + s_T^{(2)}+s_T^{(5)}\right)- \P\left(\left(\mathcal{S}_T^{(2)}\cap\mathcal{S}_T^{(5)}\right)^c\right)\\
		\label{toshow206ii2} &= \P\left(\left\|\frac{U^\top U\beta^*}{T}\right\|_\infty>  \overline{\lambda}_\alpha + s_T^{(2)}+s_T^{(5)}\right)+o(1),
	\end{align}
	where, in the last line, we used Lemma \ref{lm.bound}. 
	
	Let us define 
	$$\mathcal{T}_2=\mathcal{S}_{\mu^*_{2s_T^{(10)}}}\cap \mathcal{S}_T^{(1)}\cap \mathcal{S}_T^{(2)}\cap \mathcal{S}_T^{(3)}\cap\mathcal{S}_T^{(4)}\cap \mathcal{S}_T^{(5)}.$$Note that, by Lemmas \ref{lm.bound} and \ref{lm.ps}, and the fact that $s_T^{(10)}=\frac{1}{T\vee p}+s_T^{(9)}\to 0$ by Lemma \ref{lm.seq} \eqref{lsv},  the event $\mathcal{T}_2$ has probability going to $1$. 
	
	Hence, by \eqref{toshow206ii2}, to show \eqref{toshow206ii1}, it suffices to prove that,  on $\mathcal{T}_2$, we have \begin{equation}\label{toshow206ii3}\overline{\lambda}_\alpha\le \frac{2}{\sqrt{T}}\mu_{2s_T^{(10)}}^*,\end{equation}
	for $T$ large enough.
	Indeed, in this case, we would have 
	\begin{align*}\P\left(\left\|\frac{\overline{U}^\top \overline{Y}}{T}\right\|_\infty\ge \overline{\lambda}_\alpha\right)&\ge  \P\left(\left\|\frac{U^\top U\beta^*}{T}\right\|_\infty>  \overline{\lambda}_\alpha + s_T^{(2)}+s_T^{(5)}\right)+o(1),\\
		&\ge \P\left(\left\{\left\|\frac{U^\top U\beta^*}{T}\right\|_\infty> \frac{2}{\sqrt{T}}\mu_{2s_T^{(10)}}^* + s_T^{(2)}+s_T^{(5)}\right\}\cap \mathcal{T}_2\right)+o(1)\\
		&\ge \P\left(\left\{\left\|\frac{U^\top U\beta^*}{T}\right\|_\infty>  \frac{2}{\sqrt{T}}s_T^{(11)} + s_T^{(2)}+s_T^{(5)}\right\}\cap \mathcal{T}_2\right)+o(1)\\
		&\ge \P\left(\left\|\frac{U^\top U\beta^*}{T}\right\|_\infty>  \frac{2}{\sqrt{T}}s_T^{(11)} + s_T^{(2)}+s_T^{(5)}\right)-\P(\mathcal{T}_2^c) +o(1)
		\to 1 ,\end{align*}
	where, in the third line, we used $\mu_{2s_T^{(10)}}^*\le s_T^{(11)} $ by Lemma \ref{lm.G} and, in the last line, we leveraged the facts $\frac{2}{\sqrt{T}}s_T^{(11)} + s_T^{(2)}+s_T^{(5)}=O_P\left(\sqrt{\frac{\log(T\vee p)}{T\wedge p}}\right)$ by Lemma \ref{lm.seq} \eqref{lsii} and that $\sqrt{\frac{\log(T\vee p)}{T\wedge p}}=o_P\left(\left\|\frac{U^\top U\beta^*}{T}\right\|_\infty\right)$ to obtain that $ \P\left(\left\|\frac{U^\top U\beta^*}{T}\right\|_\infty>  \frac{2}{\sqrt{T}}s_T^{(11)} + s_T^{(2)}+s_T^{(5)}\right)\to 1$.
	
	Let us therefore prove that, on $\mathcal{T}_2$, \eqref{toshow206ii3} holds for $T$ large enough. To do so, we show that, on $\mathcal{T}_2$, for $T$ large enough,
	\begin{equation}\label{toshow206ii4}\P_e\left(\overline{\Pi}(\mu_{2s_T^{(10)}}^*,e)> \mu_{2s_T^{(10)}}^*\right) \le \alpha, \end{equation}
	which implies \eqref{toshow206ii3} by definition of $\overline{\lambda}_\alpha$. 
	On $\mathcal{T}_2$, we have \begin{align*}
		\P_e\left(\overline{\Pi}(\mu_{2s_T^{(10)}}^*,e)> \mu_{2s_T^{(10)}}^*\right)&\le \P_e\left(\Pi(e)+R(\mu_{2s_T^{(10)}}^*,e)> \mu\right)\\
		&\le \P_e\left(\Pi(e)> \mu_{2s_T^{(10)}}^* -R(\mu_{2s_T^{(10)}}^*,e), R(\mu_{2s_T^{(10)}}^*,e)\le s_T^{(6)}\sqrt{\mu_{2s_T^{(10)}}^*}+s_T^{(7)}\right)\\
		&\quad +\P_e\left( R(\mu_{2s_T^{(10)}}^*,e)> s_T^{(6)}\sqrt{\mu_{2s_T^{(10)}}^*}+s_T^{(7)}\right)\\
		&\le  \P_e\left(\Pi(e)> \mu_{2s_T^{(10)}}^*-s_T^{(6)}\sqrt{\mu_{2s_T^{(10)}}^*}-s_T^{(7)}\right)+\frac2T,
	\end{align*}
	where, in the last line, we used Lemma \ref{lm.lassodiff}.
	By Lemma \ref{prop.LV2}, we obtain
	\begin{equation}\label{eq2061ii}\P_e\left(\Pi(e)>\mu_{2s_T^{(10)}}^*\right)\le \P\left(\Pi^G\ge \mu_{2s_T^{(10)}}^*-s_T^{(6)}\sqrt{\mu_{2s_T^{(10)}}^*}-s_T^{(7)}\right) +s_T^{(8)}+\frac2T.\end{equation}
	Since $\sqrt{\mu_{2s_T^{(10)}}^*}\le 1+\mu_{2s_T^{(10)}}^*$, it holds that\begin{align}
		\notag &\P\left(\Pi^G> \mu_{2s_T^{(10)}}^*-s_T^{(6)}\sqrt{\mu_{2s_T^{(10)}}^*}-s_T^{(7)}\right)\\
		\notag &\le \P\left(\Pi^G> \mu_{2s_T^{(10)}}^*-s_T^{(6)}(1+\mu_{2s_T^{(10)}}^*)-s_T^{(7)}\right)\\
		\label{B42147} &\le \P\left(\Pi^G>  \mu_{2s_T^{(10)}}^*-s_T^{(6)}(1+s_T^{(11)})-s_T^{(7)}\right)\\
		\notag &=  \P\left(\Pi^G>\mu_{2s_T^{(10)}}^*-s_T^{(13)}\right)\\
		\notag &\le  \P\left(\Pi^G>\mu_{2s_T^{(10)}}^*\right)+\P\left(|\Pi^G-\mu_{2s_T^{(10)}}^*|\le s_T^{(13)}\right)\\
		\label{B32147}&\le  \P\left(\Pi^G>\mu_{2s_T^{(10)}}^*\right)+K_3s_T^{(13)}\sqrt{1\vee \log\left(2p/s_T^{(13)}\right)}\\
		\label{B12147}&\le  \P\left(\Pi^*>\mu_{2s_T^{(10)}}^*\right)+s_T^{(9)}+K_3s_T^{(13)}\sqrt{1\vee \log\left(2p/s_T^{(13)}\right)}\\
		\notag&=2s_T^{(10)}+s_T^{(9)}+K_3s_T^{(13)}\sqrt{1\vee \log\left(2p/s_T^{(13)}\right)},
	\end{align}
	where, in \eqref{B42147},  we used Lemma \ref{lm.G} to obtain that $\mu_{2s_T^{(10)}}^*\le s_T^{(11)} $, in \eqref{B32147}, we leveraged Lemma \ref{prop.LV3} and \eqref{B12147} follows from Lemma \ref{prop.LV1}.
	This and \eqref{eq2061ii}, therefore yield
	$$\P_e\left(\Pi(e)> \mu_{2s_T^{(10)}}^*\right)\le 2s_T^{(10)}+s_T^{(9)}+K_3s_T^{(13)}\sqrt{1\vee \log\left(2p/s_T^{(13)}\right)} +s_T^{(8)}+\frac2T\le \alpha,$$
	for $T$ large enough by Lemma \ref{lm.seq} \eqref{lsiv}, \eqref{lsv}, \eqref{lsvi}.
	This shows that \eqref{toshow206ii4} holds and therefore concludes the proof of \eqref{P-thii}.
	
	\subsection{Auxiliary lemmas on distributions} \label{subsec.distr}
	
	\begin{Lemma}\label{prop.LV1} Under the assumptions of Theorem \ref{P-th}, it holds that
		$$ \sup_{z\in\R}\left|\P(\Pi^*\le z)-\P\left(\Pi^G\le z\right)\right|<s_T^{(9)} .$$
	\end{Lemma}
	\begin{Proof} The result is a direct consequence of Lemma \ref{lm.highdimclt} applied to $Z_t = u_t\varepsilon_t$ (and the constant $K_1$ used in the definition of $s_T^{(9)}$ is introduced in Lemma \ref{lm.highdimclt}). Condition \eqref{lhdi} of Lemma \ref{lm.highdimclt} is satisfied by Lemma \ref{lm.2exp} and Assumption \ref{P-as.tail} \eqref{P-tailiii}. Assumption \ref{P-as.mixing} implies that conditions \eqref{lhdii} and \eqref{lhdiii} hold. Concerning condition \eqref{lhdiv}, note that, by Assumption \ref{P-as.tail} \eqref{P-tailiv}, we have 
		\begin{align*}\E\left[\left(\frac{1}{\sqrt{T}} \sum_{t=1}^T u_t\varepsilon_t\right)\left(\frac{1}{\sqrt{T}} \sum_{t=1}^T u_t\varepsilon_t\right) ^\top \right]&=\frac{1}{T} \sum_{t=1}^T\sum_{s=1}^T\E\left[ u_t\varepsilon_tu_{s}^\top\varepsilon_{s} \right]= \E\left[ u_tu_t^\top\varepsilon_t^2\right].\end{align*}
		By Assumption \ref{P-as.tail} \eqref{P-tailii}, this implies that $$\sigma_{p}\left(\E\left[\left(\frac{1}{\sqrt{T}} \sum_{t=1}^T u_t\varepsilon_t\right)\left(\frac{1}{\sqrt{T}} \sum_{t=1}^T u_t\varepsilon_t\right) ^\top \right]\right)= \sigma_{p}\left(E\left[ u_tu_t^\top\varepsilon_t^2\right]\right)\ge \kappa_1>0,$$
		and therefore that condition \eqref{lhdiv} holds. 
	\end{Proof}
	\begin{Lemma}\label{prop.LV2} Let the assumptions of Theorem \ref{P-th} hold. On the event $\mathcal{S}_T^{(1)}$,
		$$\sup_{z\in\R}\left|\P_e(\Pi(e)\le z)-\P\left(\Pi^G\le z\right)\right|\le s_T^{(8)}.$$
	\end{Lemma}
	\begin{Proof} Conditionally on $U,\mathcal{E}$, $W(e)$ is a centered Gaussian vector with covariance matrix $T^{-1}\sum_{t=1}^T u_tu_t^\top\varepsilon_t^2 $. Moreover, $G$ is a centered Gaussian vector with covariance matrix $$\E\left[\left(\frac{1}{\sqrt{T}} \sum_{t=1}^T u_t\varepsilon_t\right)\left(\frac{1}{\sqrt{T}} \sum_{t=1}^T u_t\varepsilon_t\right) ^\top \right]=\E\left[\varepsilon_t^2u_tu_t^\top\right],$$
		see the proof of Lemma \ref{prop.LV1} for a justification of this equality. 
		Remark that, by Assumption \ref{P-as.tail} \eqref{P-tailii}, $$\kappa_2 > \E\left[u_{tj}u_{tj}^\top\varepsilon_t^2\right]> \kappa_1,$$
		for all $j\in[p]$.
		We can therefore apply Lemma \ref{lm.glv2} to get 
		$$\sup_{z\in\R}\left|\P_e(\Pi(e)\le z)-\P\left(\Pi^G\le z\right)\right|\le \pi(\Delta),$$
		where $\pi(\Delta)=K_1\Delta^{1/3}(1\vee \log(2p)\vee \log(1/\Delta))^{1/3}\log(2p)^{1/3}.$

	\end{Proof}
	\begin{Lemma}\label{prop.LV3}Under the assumptions of Theorem \ref{P-th}, there exists a constant $K_2>0$ such that, for all $z_1,z_2>0$, we have 
		$$\P\left(\left|\Pi^G-z_1\right|\le z_2\right)\le K_2z_2\sqrt{1\vee \log(2p/z_2)}.$$
	\end{Lemma}
	\begin{Proof}
		This is a direct consequence of Lemma \ref{lm.glv1} of which the conditions are satisfied by Assumption \ref{P-as.tail} (see the proofs of Lemmas \ref{prop.LV1} and \ref{prop.LV2} for more details).
	\end{Proof}
	
	\begin{Lemma}\label{lm.G}For every $\alpha>s_T^{(10)}$, we have
		$$\mu^{*}_\alpha\le s_T^{(11)}.$$
	\end{Lemma}
	\begin{Proof}
		Notice that, by Assumption \ref{P-as.tail} \eqref{P-tailiv},
		\begin{align*}\E\left[(W_j^*)^2\right]=\E\left[\left(\frac{1}{\sqrt{T}}\sum_{t=1}^Tu_{tj} \varepsilon_t\right)^2\right]= \E\left[u_{tj}^2 \varepsilon_t^2\right],
		\end{align*}
		which, by Assumption \ref{P-as.tail} \eqref{P-tailii} and \eqref{P-tailiv}, is bounded uniformly in $j$ and $t$ by $\kappa_2>0$.
		Using  Lemma 7 in \cite{chernozhukov2015comparison} and remark A.8 in \cite{lederer2021estimating}, we have, for every $r>0$, 
		$$\P\left(\|G\|_\infty \ge \E\left[\|G\|_\infty\right]+r\right)\le\exp\left(-\frac{r^2}{2\kappa_2}\right).$$
		Taking $r=\kappa_2\sqrt{2\log(T\vee p)}$, we get
		$$\P\left(\|G/\kappa_2\|_\infty \ge \E\left[\|G/\kappa_2\|_\infty\right]+\sqrt{2\log(T\vee p)}\right)\le\frac{1}{T\vee p}.$$
		By the Gaussian maximal inequality (see e.g., Exercise 2.17 in \cite{boucheron2013concentration}), it holds that $\E\left[\|G/\kappa_2\|_\infty\right]\le \sqrt{2\log(2p)}$, which yields
		$$\P\left(\|G\|_\infty \ge \kappa_2\left(\sqrt{2\log(2p)}+\sqrt{2\log(T\vee p)}\right)\right)\le\frac{1}{T\vee p},$$
		so that $\mu^{G}_\alpha\le \kappa_2 \left(\sqrt{2\log(2p)}+\sqrt{2\log(T\vee p)}\right)$ for $\alpha>1/(T\vee p)$ by definition of $\mu^G_\alpha$.
		Now, for $\alpha>s_T^{(10)}=(T\vee p)^{-1}+s_T^{(9)}$, by Lemma \ref{prop.LV1}, we have 
		$$\P\left(\Pi^*\ge\mu^G_{\alpha -s_T^{(9)}}\right)\le \P\left(\Pi^G\ge \mu^G_{\alpha -s_T^{(9)}}\right)+s_T^{(9)}\le \alpha- s_T^{(9)} +s_T^{(9)}=\alpha .$$
		Hence, we obtain $\mu^*_{\alpha}\le \mu^G_{\alpha -s_T^{(9)}}\le \kappa_2 \left(\sqrt{2\log(2p)}+\sqrt{2\log(T\vee p)}\right).$
	\end{Proof}

	\subsection{Auxiliary lemmas on probabilistic events} \label{subsec.prob}
	
	\begin{Lemma}\label{lm.bound} Under the assumptions of Theorem \ref{P-th}, it holds that \begin{enumerate}[\textup{(}i\textup{)}]  
			\item\label{lm.boundi}  $\P\left(\mathcal{S}_T^{(1)}\right)\to 1$;
			\item\label{lm.boundii}  $\P\left(\mathcal{S}_T^{(2)}\right)\to 1$;
			\item\label{lm.boundiii}  $\P\left(\mathcal{S}_T^{(3)}\right)\to 1$;
			\item\label{lm.boundiv}   $\P\left(\mathcal{S}_T^{(4)}\right)\to 1$;
			\item\label{lm.boundv}   $\P\left(\mathcal{S}_T^{(5)}\right)\to 1$.
		\end{enumerate}
	\end{Lemma}
	\begin{Proof} \newline
		Result \eqref{lm.boundi} follows directly from Lemma \ref{lm.lf2} \eqref{lf2v}; \eqref{lm.boundii} is a consequence of  Lemma \ref{fancomplex}; \eqref{lm.boundiii} comes from Lemmas \ref{lm.lf} \eqref{lfii} and Lemma \ref{lm.lf2} \eqref{lf2i} and the triangle inequality, \eqref{lm.boundiv} follows from Lemma \ref{20723} and \eqref{lm.boundv} is a direct consequence of Lemma \ref{lm.lf2} \eqref{lf2iii}. 
	\end{Proof}
	
	\begin{Lemma}\label{lm.lassodiff}Let the assumptions of Theorem \ref{P-th} hold. On the event $\mathcal{S}_T^{(3)}\cap \mathcal{S}_T^{(4)}\cap \mathcal{S}_{\mu}$,
		we have, for all $\mu'\ge \mu$,
		$$\P_e\left(R(\mu',e)\ge s_T^{(6)}\sqrt{ \mu'}+s_T^{(7)}\right) \le \frac2T.$$
	\end{Lemma}
	\begin{Proof} Take $\mu'\ge \mu$. Remember that $\overline{Y}=\left(I_T-\overline{P}\right)\left(X\beta+F\varphi^*+ \mathcal{E}\right)$. This yields that
		$$\overline{\varepsilon}_{\frac{2}{\sqrt{T}}\mu',t}=\widetilde{y}_t-\overline{u}_t^\top\overline{\beta}_{\frac{2}{\sqrt{T}}\mu'}=\overline{u}_t\left(\beta^*-\overline{\beta}_{\frac{2}{\sqrt{T}}\mu'}\right)+\widetilde{f}_t^\top\varphi^*+\widetilde{\varepsilon}_t,$$
		where we recall that $\widetilde{\varepsilon}_t$ is the $t^{th}$ element of  $\left(I_T-\overline{P}\right)\mathcal{E}$ and $\widetilde{f}_t$ is the $K\times 1$ vector corresponding to the $t^{th}$ row of $\left(I_T-\overline{P}\right)F$. This yields \begin{align}
			\notag &R(\mu',e)\\
			\notag &=\frac{1}{\sqrt{T}}\left\|\overline{W}(\mu',e)-W(e)\right\|_\infty\\
			\notag  &=\frac{1}{T}\max_{j\in[p]}\left|\sum_{t=1}^T\overline{u}_{tj} \overline{\varepsilon}_{\frac{2}{\sqrt{T}}\mu',t} e_t-  \sum_{t=1}^Tu_{tj} \varepsilon_te_t\right|\\
			\label{decomp6163}&\le \frac{1}{T}\max_{j\in[p]}\left|\sum_{t=1}^T\overline{u}_{tj}\overline{u}_t^\top\left(\beta^*-\overline{\beta}_{\frac{2}{\sqrt{T}}\mu'}\right) e_t \right|+\frac{1}{T}\max_{j\in[p]}\left| \sum_{t=1}^T\left(\overline{u}_{tj}\widetilde{\varepsilon}_t+\widetilde{f}_t^\top\varphi^*- u_{tj}\varepsilon_t\right) e_t\right| .
		\end{align}

		Now, we bound the two terms in \eqref{decomp6163}. We start with $\max_{j\in[p]}\left|\sum_{t=1}^T\overline{u}_{tj}\overline{u}_t^\top\left(\beta^*-\overline{\beta}_{\frac{2}{\sqrt{T}}\mu'}\right) e_t \right|$.  Remark that given $(F,U,\mathcal{E})$, we have
		\begin{align*}
			\frac{1}{T}\sum_{t=1}^T\overline{u}_{tj}\overline{u}_t^\top\left(\overline{\beta}_\lambda-\beta^*\right)e_t \sim\mathcal{N}\left(0, \frac{1}{T^2}\sum_{t=1}^T\left(\overline{u}_{tj}\overline{u}_{t}^\top\left(\overline{\beta}_\lambda-\beta^*\right)\right)^2\right)  
		\end{align*}					
		By the Gaussian tail bound (equation (2.10) in \cite{vershynin2018high}), for $z>0$, we obtain, for all $j\in[p]$ and $z>0$,
		\begin{equation}\label{0707232}\P_e^*\left(\left| \frac{1}{T}\sum_{t=1}^T\overline{u}_{tj}\overline{u}_t^\top\left(\overline{\beta}_\lambda-\beta^*\right)e_t \right|>z\right) \le 2\exp\left(-\frac{z^2}{\frac{1}{T^2}\sum_{t=1}^T\left(\overline{u}_{tj}\overline{u}^\top_t\left(\overline{\beta}_\lambda-\beta^*\right)\right)^2}\right).\end{equation}
		Next, let $\lambda= {\frac{2}{\sqrt{T}}\mu}$ and $\lambda'= {\frac{2}{\sqrt{T}}\mu'}$. By definition of $\overline{\beta}_{\lambda'}$, it holds that 
		$$\frac{1}{T}\left\|\overline{Y}-\overline{U}\overline{\beta}_{\lambda'}\right\|^2_2+ \lambda' \left\|\overline{\beta}_\lambda\right\|_1\le\frac{1}{T}\left\|\overline{Y}-\overline{U}\beta^*\right\|^2_2+ \lambda \|\beta^*\|_1.$$
		This yields
		\begin{align}\notag &\frac{1}{T}\left\|\overline{U}(\beta^*-\overline{\beta}_{\lambda'})\right\|^2_2\\
			\notag& \le \frac2T \left(\overline{Y}-\overline{U}\beta^*\right)^\top\overline{U}\left(\overline{\beta}_{\lambda'}-\beta^*\right)+ \lambda' \left(\|\beta^*\|_1-\left\|\overline{\beta}_{\lambda'}\right\|_1\right)\\
			\notag &\le \frac2T \left\|\overline{U}^\top \left(\overline{Y}-\overline{U}\beta^*\right)\right\|_\infty\left\|\overline{\beta}_{\lambda'}-\beta^*\right\|_1+ \lambda'\left(\|\beta^*\|_1- \left\|\overline{\beta}_{\lambda'}\right\|_1\right)\\
			\notag &\le \lambda' \|\overline{\beta}_{\lambda'}-\beta^*\|_1+ \lambda' \left(\|\beta^*\|_1-\left\|\overline{\beta}_{\lambda'}\right\|_1\right)\\
			\label{ineqlasso} &\le 2  \lambda'\|\beta^*\|_1.
		\end{align}
		where we used Hölder's inequality and the fact that we work on $\mathcal{S}_\mu$.
		Moreover, we have 
		\begin{equation}\label{0707231}
			\begin{aligned}
				\frac{1}{T^2}\sum_{t=1}^T(\overline{u}_{tj}\overline{u}^\top_t(\overline{\beta}_\lambda-\beta^*))^2&\le \frac{1}{T}\sum_{t=1}^T \overline{u}_{tj}^2 \frac{1}{T}\left\|\overline{U}\left(\beta^*-\overline{\beta}_\lambda\right)\right\|^2_2\\
				&\le  s_T^{(3)}2\lambda'\|\beta^*\|_1,
			\end{aligned}
		\end{equation}
		by \eqref{ineqlasso} and because we work on $\mathcal{S}_T^{(3)}$. Recall that $s_T^{(6)}= 2\sqrt{ \log(Tp)\|\beta^*\|_1s_T^{(3)}T^{-1/2}}$. Using \eqref{0707232}, \eqref{0707231} and the union bound, we get
		\begin{align}\notag &\P_e^*\left(\frac{1}{T}\max_{j\in[p]}\left|\sum_{t=1}^T\overline{u}_{tj}\overline{u}_t^\top\left(\beta^*-\overline{\beta}_{\frac{2}{\sqrt{T}}\mu'}\right) e_t \right|_\infty>s_T^{(6)}\sqrt{\mu'}\right)\\
			\notag &\le p \max_{j\in[p]}\P_e^*\left(\left| \frac{1}{T}\sum_{t=1}^T\overline{u}_{tj}\overline{u}_t^\top\left(\overline{\beta}_\lambda-\beta^*\right)e_t \right|>s_T^{(6)}\sqrt{\mu'}\right)\\
			\label{boundr1}&\le  \exp\left(-\frac{(s_T^{(6)})^2\mu'}{2\lambda'\|\beta^*\|_1s_T^{(3)}} +\log(p)\right)=T^{-1}.
		\end{align}
		
		Let us now bound the term
		$\max_{j\in[p]}\left| \sum_{t=1}^T\left(\overline{u}_{tj}\widetilde{\varepsilon}_t+\widetilde{f}_t^\top\varphi^*- u_{tj}\varepsilon_t\right) e_t\right|$. Conditional on $(F,U,\mathcal{E})$, we have
		\begin{align*}
			\frac1T\sum_{t=1}^T\left(\overline{u}_{tj}\widetilde{\varepsilon}_t+\widetilde{f}_t^\top\varphi^*- u_{tj}\varepsilon_t\right)  e_t &\sim \mathcal{N}\left(0, \frac{1}{T^2}\sum_{t=1}^T\left(\overline{u}_{tj}\widetilde{\varepsilon}_t+\widetilde{f}_t^\top\varphi^*- u_{tj}\varepsilon_t\right)^2\right).
		\end{align*}
		Since we work on $\mathcal{S}_T^{(4)}$, by the Gaussian tail bound, this yields, for all $j\in[p]$ and $z>0$,
		$$\P_e^*\left(\left|\frac1T\sum_{t=1}^T\left(\overline{u}_{tj}\widetilde{\varepsilon}_t+\widetilde{f}_t^\top\varphi^*- u_{tj}\varepsilon_t\right)  e_t\right|>z \right)\le \exp\left(-\frac{Tz^2}{s_T^{(4)}}\right).$$
		Recall that $s_T^{(7)}=\sqrt{\log(Tp) T^{-1}s_T^{(4)}}$. Using the union bound, we get
		\begin{align}\notag &\P_e^*\left(\max_{j\in[p]}\left|\frac1T\sum_{t=1}^T\left(\overline{u}_{tj}\widetilde{\varepsilon}_t+\widetilde{f}_t^\top\varphi^*- u_{tj}\varepsilon_t\right) e_t\right|>s_T^{(7)}\right)\\
			\notag &\le p \max_{j\in[p]}\P_e\left(\left|\frac1T\sum_{t=1}^T\left(\overline{u}_{tj}\widetilde{\varepsilon}_t+\widetilde{f}_t^\top\varphi^*- u_{tj}\varepsilon_t\right) e_t\right|>s_T^{(7)}\right)\\
			\label{boundr2}&\le p\exp\left(-\frac{\left(s_T^{(7)}\right)^2}{T^{-1}s_T^{(4)}}\right)=T^{-1}.
		\end{align}
		Using the pigeonhole principle, \eqref{decomp6163}, \eqref{boundr1} and \eqref{boundr2}, we get 
		$\P_e^*\left(R(\mu',e)\ge s_T^{(6)}\sqrt{ \mu'}+s_T^{(7)}\right) \le 2T^{-1},$
		which yields $\P_e\left(R(\mu',e)\ge s_T^{(6)}\sqrt{ \mu'}+s_T^{(7)}\right) \le 2T^{-1},$ integrating over the distribution of $(F,U,\mathcal{E})$.
		
	\end{Proof}
	
	\begin{Lemma}\label{lm.ps}Under the assumptions of Theorem \ref{P-th}, we have 
		$$\sup_{\alpha'\in(0,1)}\left|\P\left(\mathcal{S}_{\mu_{\alpha'}^*} \right)-(1-\alpha')\right|=o(1).$$
	\end{Lemma}
	\begin{Proof} Let us first bound $\P\left(\mathcal{S}_{\mu_{\alpha'}^*}\right)$ from above. For $\alpha'\in(0,1)$, we have \begin{align}
			\notag \P\left(\mathcal{S}_{\mu_{\alpha'}^*}\right)&=\P\left(2\left\|\frac{\overline{U}^\top (\overline{Y}-\overline{U}\beta^*)}{T}\right\|_\infty\le \frac{2}{\sqrt{T}}\mu_{\alpha'}^*\right)\\
			\notag &\le \P\left(\left\|\frac{U^\top \mathcal{E}}{T}\right\|_\infty-\left\|\frac{\overline{U}^\top (\overline{Y}-\overline{U}\beta^*)}{T}-\frac{U^\top \mathcal{E}}{T}\right\|_\infty\le  \frac{1}{\sqrt{T}}\mu_{\alpha'}^*\right) \\
			\notag &\le \P\left(\left\{\left\|\frac{U^\top \mathcal{E}}{T}\right\|_\infty-\left\|\frac{\overline{U}^\top (\overline{Y}-\overline{U}\beta^*)}{T}-\frac{U^\top \mathcal{E}}{T}\right\|_\infty\le  \frac{1}{\sqrt{T}}\mu_{\alpha'}^*\right\}\cap\mathcal{S}_T^{(2)}\right) + \P\left( \left(\mathcal{S}_T^{(2)}\right)^c\right)
			\\
			\label{ps1} &\le \P\left(\left\|\frac{U^\top \mathcal{E}}{T}\right\|_\infty\le  \frac{1}{\sqrt{T}}\mu_{\alpha'}^*+s_T^{(2)}\right) +\P\left( \left(\mathcal{S}_T^{(2)}\right)^c\right).
		\end{align}

		Now,  we have 
		\begin{align}\notag \P\left(\left\|\frac{U^\top \mathcal{E}}{T}\right\|_\infty\le  \frac{1}{\sqrt{T}}\mu_{\alpha'}^*+s_T^{(2)}\right)&=\P\left(\Pi^*\le  \mu_{\alpha'}^*+\sqrt{T}s_T^{(2)}\right)
			\\
			\notag & \le \P\left(\Pi^G\le  \mu_{\alpha'}^*+\sqrt{T}s_T^{(2)}\right)+s_T^{(9)}\\
			\notag &\le \P\left(\Pi^G\le  \mu_{\alpha'}^*\right)+\P\left(\left|\Pi^G-  \mu_{\alpha'}^*\right|\le \sqrt{T}s_T^{(2)}\right)+s_T^{(9)}\\
			\notag &\le \P\left(\Pi^*\le  \mu_{\alpha'}^*\right)+\P\left(\left|\Pi^G-  \mu_{\alpha'}^*\right|\le \sqrt{T}s_T^{(2)}\right)+2s_T^{(9)}\\
			\label{ps2}&\le 1-\alpha'+\P\left(\left|\Pi^G-  \mu_{\alpha'}^*\right|\le \sqrt{T}s_T^{(2)}\right)+2s_T^{(9)},
		\end{align}
		where we used Lemma \ref{prop.LV1} in the second and fourth lines.
		By Lemma \ref{prop.LV3}, we have $$\P\left(\left|\Pi^G-  \mu_{\alpha'}^*\right|\le s_T^{(2)}\right)\le K_4\sqrt{T}s_T^{(2)}\sqrt{1\vee \log\left(\frac{2p}{\sqrt{T}s_T^{(2)}}\right)}.$$
		Combining this, \eqref{ps1} and \eqref{ps2}, we get \begin{equation}\label{ps3}\P\left(\mathcal{S}_{\mu_{\alpha'}^*}\right)\le 1-\alpha'+ \P\left( \left(\mathcal{S}_T^{(2)}\right)^c\right)+K_4\sqrt{T}s_T^{(2)}\sqrt{1\vee \log\left(\frac{2p}{\sqrt{T}s_T^{(2)}}\right)}+2s_T^{(9)}.\end{equation}
		
		By a similar reasoning, we can show that \begin{equation}\label{ps4}\P\left(\mathcal{S}_{\mu_{\alpha'}^*}\right)\ge 1-\alpha'- \P\left( \left(\mathcal{S}_T^{(2)}\right)^c\right)-K_4\sqrt{T}s_T^{(2)}\sqrt{1\vee \log\left(\frac{2p}{\sqrt{T}s_T^{(2)}}\right)}-2s_T^{(9)}.\end{equation}
		Since $\sqrt{T}s_T^{(2)}\sqrt{1\vee \log\left(\frac{2p}{\sqrt{T}s_T^{(2)}}\right)}\to 0$, $s_T^{(8)}\to 0$, $s_T^{(9)}\to 0$ by Lemma \ref{lm.seq} \eqref{lsiii}, \eqref{lsiv}, \eqref{lsv} and $\P\left( \left(\mathcal{S}_T^{(2)}\right)^c\right)\to0 $ by  Lemma \ref{lm.bound} \eqref{lm.boundii}, \eqref{ps3} and \eqref{ps4} yield the result
	\end{Proof}

	\subsection{Auxiliary lemma on sequences}\label{subsec.seq}
	\begin{Lemma} \label{lm.seq} Under Assumption \ref{P-as.Rates}, we have \begin{enumerate}[\textup{(}i\textup{)}]  \item\label{lsi} $s_T^{(12)}\sqrt{1\vee \log\left(2p/s_T^{(12)}\right)}=o(1)$;  \item\label{lsii} $2T^{-1/2}s_T^{(11)}+s_T^{(2)}+s_T^{(5)}=O\left(\sqrt{\frac{\log(T\vee p)}{(T\wedge p)}}p^\frac2q\right)\left(\|\varphi^*\|_2\vee 1\right)$;  \item\label{lsiii} $\sqrt{T}s_T^{(2)}\sqrt{1\vee \log\left(\frac{2p}{\sqrt{T}s_T^{(2)}}\right)}=o(1)$;  \item\label{lsiv} $s_T^{(8)}=o_P(1)$;\item\label{lsv} $s_T^{(9)}=o(1)$; \item\label{lsvi}$s_T^{(13)}\sqrt{1\vee \log\left(2p/s_T^{(13)}\right)}=o(1).$
		\end{enumerate}
	\end{Lemma}
	\begin{Proof}
		\newline \noindent \textit{Proof of \eqref{lsi}.} By Assumption \ref{P-as.Rates} \eqref{P-ratei}, we have $s_T^{(3)}=O\left(\sqrt{\log(T)}\right)$, so that \begin{equation}\label{seq1}s_T^{(6)}=O\left(\left(\frac{\log(T\vee p)^{2}}{\sqrt{T}}\|\beta^*\|_1\right)^{1/2}\right).\end{equation} Since $s_T^{(11)}=O\left(\sqrt{\log(T\vee p)}\right)$, this yields
		\begin{equation}\label{seq2} s_T^{(6)}s_T^{(11)}=O\left(\left(\frac{\log(T\vee p)^{3}}{\sqrt{T}}\|\beta^*\|_1\right)^{1/2}\right).\end{equation}
		We also have \begin{equation}\label{seq3} \left(s_T^{(6)}\right)^2s_T^{(11)}=O\left(\left(\frac{\log(T\vee p)^{3}}{\sqrt{T}}\|\beta^*\|_1\right)^{1/2}\right),\end{equation}
		because $s_T^{(6)}=o(1)$ by \eqref{seq1} and Assumption \ref{P-as.Rates} \eqref{P-rateii}. Next, it holds that 
		\begin{equation}\label{seq4}s_T^{(2)}=O\left(\frac{\sqrt{\log(T)}p^{\frac2q}}{T\wedge p}(\|\varphi^*\|_2\vee1)\right),\end{equation}
		so that \begin{equation}\label{seq5}s_T^{(6)}s_T^{(2)}=o\left(s_T^{(2)}\right),\end{equation}
		since $s_T^{(6)}=o(1)$ by \eqref{seq1} and Assumption \ref{P-as.Rates} \eqref{P-rateii}. Moreover, it holds that $$s_T^{(4)}=O\left(\sqrt{\log(T)}\frac{p^\frac4q T^\frac2q}{T\wedge p}\left(\|\varphi^*\|_2^2\vee 1\right)\right)$$
		and, therefore, \begin{equation}\label{seq6} s_T^{(7)}=O\left(\sqrt{\frac{\log(T\vee p)^2p^\frac4q T^\frac2q}{T(T\wedge p)}}\left(\|\varphi^*\|_2\vee 1\right)\right).\end{equation} Recall that
		$$ s_T^{(12)}= 2s_T^{(6)}+2s_T^{(6)}s_T^{(11)}+\left(s_T^{(6)}\right)^2s_T^{(11)}+\frac{\sqrt{T}}{2}s_T^{(2)}+\frac{\sqrt{T}}{2}s_T^{(6)}s_T^{(2)}+s_T^{(7)}.$$
		By \eqref{seq1}, \eqref{seq2}, \eqref{seq3}, \eqref{seq4}, \eqref{seq5}, \eqref{seq6}, we obtain 
		\begin{align}\notag  s_T^{(12)}&=O\left(\left(\frac{\log(T\vee p)^{3}}{\sqrt{T}}\|\beta^*\|_1\right)^{1/2}+ \left(\frac{\log(T\vee p)^{3/2}\sqrt{T}}{(T\wedge p)}+\sqrt{\frac{\log(T\vee p)^2p^\frac4q T^\frac2q}{T(T\wedge p)}}\right)(\|\varphi^*\|_2\vee1)\right)\\
			\label{seq7}&=O\left(\left(\frac{\log(T\vee p)^{3}}{\sqrt{T}}\|\beta^*\|_1\right)^{1/2}+ \frac{\log(T\vee p)^{3/2}\sqrt{T}}{(T\wedge p)}\left(1+\sqrt{\frac{p^\frac4q T^\frac2q}{T}}\right)(\|\varphi^*\|_2\vee1)\right)  .\end{align}
		Additionally, we have $(T(T\wedge p))^{-1/2}=o_P\left(s_T^{(7)}\right)=O\left(s_T^{(12)}\right)$ so that $\log\left(2p/s_T^{(12)}\right)=O\left(\log(2p)+\log\left(\sqrt{T(T\wedge p)}\right)\right)=O(\log(T\vee p))$. This and \eqref{seq7} imply
		\begin{align*}&s_T^{(12)}\sqrt{1\vee \log\left(2p/s_T^{(12)}\right)}\\
			&=O\left(\left(\frac{\log(T\vee p)^{5}}{\sqrt{T}}\|\beta^*\|_1\right)^{1/2}+\frac{\log(T\vee p)^{5/2}\sqrt{T}}{(T\wedge p)}\left(1+\sqrt{\frac{p^\frac4q T^\frac2q}{T}}\right)(\|\varphi^*\|_2\vee1)\right)=o(1),\end{align*}
		by Assumption \ref{P-as.Rates} and the fact that $q\ge 8$.\\
		
		\noindent \textit{Proof of \eqref{lsii}.} The result follows directly from Assumption \ref{P-as.Rates} and \eqref{seq4}.\\
		
		\noindent \textit{Proof of \eqref{lsiii}.} We have $(T\wedge p)^{-1}=o_P\left(\sqrt{T}s_T^{(2)}\right)$, hence $\log\left(\frac{2p}{\sqrt{T}s_T^{(2)}}\right)=O(\log(T\vee p))$, so that 
		\begin{align*}&\sqrt{T}s_T^{(2)}\sqrt{1\vee \log\left(\frac{2p}{\sqrt{T}s_T^{(2)}}\right)}\\
			&=O\left(\log(T\vee p)^{\frac32}\sqrt{T}\left(\frac{1}{p}+\frac{p^{\frac2q}}{T}  +\frac{p^{\frac2q-\frac12}}{\sqrt{T}}\right)(\|\varphi^*\|_2\vee 1) \right)=o(1)\end{align*}
		by the definition of $s_T^{(2)}$ and Assumption \ref{P-as.Rates}.
		
		\noindent \textit{Proof of \eqref{lsiv}.} We have $\Delta=o_P(1)$ by the facts that $\P(\mathcal{S}_T^{(1)})\to 1$ and that $s_T^{(1)}=o(1)$ by Assumption \ref{P-as.Rates}. This yields the result using that $\lim\limits_{x\to 0^+} x\log(x)\to 0$.\\
		
		\noindent \textit{Proof of \eqref{lsv}.} Recall that
		\begin{align*}
			s_T^{(9)} &=  K_1\Big[\left(T^{\kappa-1/2}+T^{1-\frac{c}{2}(1-\frac{2}{h})}\right)\log(T)\log(p)+T^{-1/4}\log (p)^{3/2} \log(T)\\
			& \quad + p^{\frac{2}{h}} T^{\frac{1}{h}-\frac12} \log(p)^2 \log(T)\\
			&\quad +\left( p\log(p)^{\frac{3}{2}h-4} \log(T)\log(Tp)\right)^{\frac{1}{h-2}}T^{-\frac14} +T^{\frac12-c\kappa} \left(p^{\frac{1}{h+1}} \sqrt{1\vee \log(p)}\right)\Big]\\
			&=O_P\Big( \left(T^{\kappa-1/2}+T^{1-\frac{c}{2}(1-\frac{2}{h})}\right)\log(T)\log(T^{r})+T^{-1/4}\log (T^{r})^{3/2} \log(T)\\
			&\quad+T^{\frac{2r}{h}+\frac{1}{h}-\frac12}  \log(T^r)^2 \log(T)\\
			&\quad +\left(\log(T^r)^{\frac{3}{2}h-4} \log(T)\log(T^{r+1})\right)^{\frac{1}{h-2}}T^{\frac{r}{h-2}-\frac14} +T^{\frac{r}{h+1}+\frac12-c\kappa}  \sqrt{1\vee \log(T^r)}\Big).
		\end{align*}
		By Assumption \ref{P-as.mixing}, we have $\kappa-1/2<0$ and $1-\frac{c}{2}(1-\frac{2}{h})<0$, so that $$\left(T^{\kappa-1/2}+T^{1-\frac{c}{2}(1-\frac{2}{h})}\right)\log(T)\log(T^{r})\to 0.$$ As $r$ is finite, $T^{-1/4}\log (T^{r})^{3/2} \log(T)\to 0$.  Moreover, since $r <\frac{h}{4} -\frac12$, we have $$T^{\frac{2r}{h}+\frac{1}{h}-\frac12}  \log(T^r)^2 \log(T)\to 0$$ and $$\left(\log(T^r)^{\frac{3}{2}h-4} \log(T)\log(T^{r+1})\right)^{\frac{1}{h-2}}T^{\frac{r}{h-2}-\frac14}\to 0$$ Additionally, it holds that $T^{\frac{r}{h+1}+\frac12-c\kappa}  \sqrt{1\vee \log(T^r)}\to 0$ because $r< (h+1)(c\kappa -\frac12)$. All of this yields \eqref{lsv}.\\
		
		\noindent \textit{Proof of \eqref{lsvi}.} The proof is similar to that of \eqref{lsi} and therefore omitted.
	\end{Proof}

	\subsection{Auxiliary lemmas on factors and loadings} \label{subsec.fac}
	In this Section, we prove useful results on the factors, the factor loadings and their estimators. Let $H=T^{-1}V\overline{F}^\top FB^\top B$, where $V$ is the $K\times K$ matrix corresponding the $K$ largest eigenvalues of $T^{-1}XX^\top$. Recall that the estimated loadings are $\overline{B}= X^\top \overline{F}\left(\overline{F}^\top \overline{F}\right)^{-1}=T^{-1}X^\top \overline{F} $ and $\overline{b}_j$ and $b_j$ are the $K\times 1$ vectors corresponding to the $j^{th}$ rows of $\overline{B}$ and $B$, respectively.

	\begin{Lemma}\label{lm.lf2} Under the assumptions of Theorem \ref{P-th}, the following holds:
		\begin{enumerate}[\textup{(}i\textup{)}]  
			\item\label{lf2i}$\max_{j\in[p]}\left|\sum_{t=1}^T u_{tj}^2\right|=O_P(T)$;
			\item\label{lf2ii}$\max_{j\in[p],k\in[K]}\left|\sum_{t=1}^T u_{tj}f_{tk}\right|=O_P\left(p^{\frac{2}{q}}\sqrt{T}\right)$;
			\item\label{lf2iii}$\left\|U^\top \mathcal{E}\right\|_\infty=O_P\left(p^{\frac{2}{q}}\sqrt{T}\right)$;
			\item\label{lf2iv}$\max_{j\in[p],k\in[K]}\left|\sum_{t=1}^T  u_{tj}\left(\sum_{\ell=1}^pu_{t\ell}b_{\ell k}\right)\right|=O_P\left(\sqrt{T}p^{\frac{2}{q}+\frac12}+T\right)$;
			\item\label{lf2v}  $\left\|\frac1T \sum_{t=1}^T u_tu_t^\top\varepsilon_t^2-\E\left[u_tu_t^\top\varepsilon_t^2\right]  \right\|_\infty=O_P\left(\frac{p^{4/q}}{\sqrt{T}}\right);$
			\item\label{lf2vi}$\| \mathcal{E}\|_2=O_P\left(\sqrt{T}\right)$;
			\item\label{lf2vibis}$\| F\|_2=O_P\left(\sqrt{T}\right)$;
			\item\label{lf2vibisii}$\left\| \frac1T F^\top F-I_K\right\|_2=O_P\left(\frac{1}{\sqrt{T}}\right)$;
			\item\label{lf2vii}$\| U\|_2=O_P\left(\sqrt{Tp}\right)$;
			\item\label{lf2viii}$\left\|F^\top \mathcal{E}\right\|_2=O_P\left(\sqrt{T}\right)$;
			\item\label{lf2x} $\|UB\|_2=O_P\left(\sqrt{Tp}\right)$;
			\item\label{lf2ix}$\left\|F^\top U\right\|_2=O_P\left(\sqrt{Tp}\right)$;
			\item\label{lf2ixbis}$\left\|\mathcal{E}^\top U\right\|_2=O_P\left(\sqrt{Tp}\right)$;
			\item\label{lf2xi} $\left\|F^\top UB\right\|_2^2=O_P\left(Tp\right)$;
			\item\label{lf2xii} $\left\|\mathcal{E}^\top UB\right\|_2^2=O_P\left(Tp\right)$.
		\end{enumerate}
	\end{Lemma}
	\begin{Proof} In this proof, we will often apply Lemmas \ref{lm.2exp}, \ref{lm.nagaev}  and \ref{lm.nagaevsup} to some specific processes. Following the arguments of the proof of Lemma \ref{prop.LV1}, it can be checked that the conditions of these Lemmas hold for these processes under the assumptions of Theorem \ref{P-th}.\\
		
		\noindent\textit{Proof of \eqref{lfi}.} We apply Lemmas \ref{lm.2exp} and \ref{lm.nagaevsup} to $Z_t=\left(u_{tj}^2-\E\left[u_{tj}^2\right]\right)_{j=1}^p$ and get
		\begin{equation}\label{187}\max_{j\in[p]}\left|\frac1T\sum_{t=1}^T u_{tj}^2-\E\left[u_{tj}^2\right]\right|=O_P\left(\frac{p^{\frac{2}{q}}}{\sqrt{T}}\right).\end{equation} By the triangle inequality, we obtain
		\begin{align*}\max_{j\in[p]}\left|\sum_{t=1}^T u_{tj}^2\right|&\le T \max_{j\in[p]}\left|\frac1T\sum_{t=1}^T u_{tj}^2-\E[u_{tj}^2]\right| + T \max_{j\in[p]}\E\left[u_{tj}^2\right]\\
			&=O_P\left(T+\sqrt{T}p^{\frac{2}{q}}\right)=O_P(T),\end{align*}
		where we used $\max_{j\in[p]}\E[u_{tj}^2]\le \|\Sigma\|_\infty =O(1)$ by Assumption \ref{P-as.tail} \eqref{P-tailii} and the fact that $p^{\frac{2}{q}}/\sqrt{T}\to 0$ by Assumption \ref{P-as.Rates} \eqref{P-ratei}.\\
		
		\noindent\textit{Proof of \eqref{lf2ii}, \eqref{lf2iii}.} We apply Lemmas \ref{lm.2exp} and \ref{lm.nagaevsup} to 
		\begin{align*}Z_t&=((u_{tj}f_{tk})_{j=1}^p)_{k=1}^K;\\
			Z_t&=(u_{tj}\varepsilon_t)_{j=1}^p,
		\end{align*}
		and obtain \eqref{lf2ii}, \eqref{lf2iii}.  \\
		
		\noindent\textit{Proof of \eqref{lf2iv}.} We apply Lemmas \ref{lm.2exp} and  \ref{lm.nagaevsup} to 
		\begin{align*}
			Z_t&=\left(\left(u_{tj}\left(p^{-1/2}\sum_{\ell=1}^pu_{t\ell}b_{\ell k}\right)-\E\left[u_{tj}\left(p^{-1/2}\sum_{\ell=1}^pu_{t\ell}b_{\ell k}\right)\right]\right)_{j=1}^p\right)_{k=1}^K,
		\end{align*}
		and obtain 
		\begin{equation}\label{2407soir}\max_{j\in[p],k\in[K]}\left| \sum_{t=1}^T\left(u_{tj}\left(p^{-1/2}\sum_{\ell=1}^pu_{t\ell}b_{\ell k}\right)-\E\left[u_{tj}\left(p^{-1/2}\sum_{\ell=1}^pu_{t\ell}b_{\ell k}\right)\right]\right)\right| =O_P\left(p^{\frac{2}{q}}\sqrt{T}\right).\end{equation}
		Next, by Assumption \ref{P-as.tail} \eqref{P-tailv}, we have
		\begin{align}
			\label{2407soir2} \max_{j\in[p],k\in[K]}\left| \sum_{t=1}^T\E\left[u_{tj}\left(\sum_{\ell=1}^pu_{t\ell}b_{\ell k}\right)\right]\right|   =O(T).
		\end{align}
		By the triangle inequality and equations \eqref{2407soir} and \eqref{2407soir2}, we obtain
		\begin{align*}
			&\max_{j\in[p],k\in[K]}\left|\sum_{t=1}^T  u_{tj}\left(\sum_{\ell=1}^pu_{t\ell}b_{\ell k}\right)\right|\\
			&\le \sqrt{p} \max_{j\in[p],k\in[K]}\left| \sum_{t=1}^T\left(u_{tj}\left(p^{-1/2}\sum_{\ell=1}^pu_{t\ell}b_{\ell k}\right)-\E\left[u_{tj}\left(p^{-1/2}\sum_{\ell=1}^pu_{t\ell}b_{\ell k}\right)\right]\right)\right|\\
			&\quad + \max_{j\in[p],k\in[K]}\left| \sum_{t=1}^T\E\left[u_{tj}\left(\sum_{\ell=1}^pu_{t\ell}b_{\ell k}\right)\right]\right|   =O_P\left(\sqrt{T}p^{\frac{2}{q}+\frac12}+T\right).\\
		\end{align*}
		
		\noindent\textit{Proof of \eqref{lf2v}.} The result directly follows from the application of Lemmas \ref{lm.2exp} and \ref{lm.nagaevsup} to  $Z_t= u_tu_t^\top\varepsilon_t^2 -\E\left[u_tu_t^\top\varepsilon_t^2\right]$.\\
		
		\noindent\textit{Proof of \eqref{lf2vi}.} The result follows from applying Lemmas \ref{lm.2exp} and \ref{lm.nagaevsup} to $Z_t=\varepsilon_t^2-\E\left[\varepsilon_t^2\right] $ and using the triangle inequality.\\
		
		\noindent\textit{Proof of \eqref{lf2vibis}.}  To obtain this statement, we apply Lemmas \ref{lm.2exp} and \ref{lm.nagaevsup} to $Z_t=f_{tk}^2-\E[f_{tk}^2]$, sum over $k$ and use the triangle inequality, noticing that $\E[f_{tk}^2]=1$ by Assumption \ref{P-as.moments} \eqref{P-f4i}. \\
		
		\noindent\textit{Proof of \eqref{lf2vibisii}.}  Statement \eqref{lf2vibisii} follows from the application of Lemmas \ref{lm.2exp} and \ref{lm.nagaevsup} to $Z_t=f_{tk}f_{t\ell}-\E[f_{tk}f_{t\ell}]$, summing over $k,\ell$ and using the fact that $\E[f_{t}f_t^\top]=I_K$ by Assumption \ref{P-as.moments} \eqref{P-f4ii} from the main text. \\
		
		\noindent\textit{Proof of \eqref{lf2vii}.} We use the fact that $\E\left[\|U\|_2^2\right] =\E\left[\sum_{t=1}^T\sum_{j=1}^p u_{tj}^2\right]=O(Tp)$ by Assumption \ref{P-as.tail} \eqref{P-tailii} and Markov's inequality. \\
		
		\noindent\textit{Proof of \eqref{lf2viii}.} We apply Lemmas \ref{lm.2exp} and \ref{lm.nagaevsup} to $Z_t=\varepsilon_tf_{tk}$ and obtain $\sum_{t=1}^T\varepsilon_tf_{tk}=O_P\left(\sqrt{T}\right)$. This yields \eqref{lf2viii}, by $\left\|F^\top \mathcal{E}\right\|_2=\sqrt{\sum_{k=1}^K\left(\sum_{t=1}^T\varepsilon_tf_{tk}\right)^2}=O_P\left(\sqrt{T}\right).$ \\
		
		\noindent\textit{Proof of \eqref{lf2x}, \eqref{lf2ix} and \eqref{lf2ixbis}.} We apply Lemmas \ref{lm.2exp} and \ref{lm.nagaevsup} to 
		\begin{align*}Z_t&=\left(p^{-1/2}\sum_{\ell=1}^pu_{t\ell}b_{\ell k}\right)^2-\E\left[\left(p^{-1/2}\sum_{\ell=1}^pu_{t\ell}b_{\ell k}\right)^2\right]
		\end{align*}
		and obtain 
		\begin{align}
			\label{2507}\max_{k\in[K]}\left|\sum_{t=1}^T \left(\left(p^{-1/2}\sum_{\ell=1}^pu_{t\ell}b_{\ell k}\right)^2-\E\left[\left(p^{-1/2}\sum_{\ell=1}^pu_{t\ell}b_{\ell k}\right)^2\right]\right)\right| &=O_P\left(\sqrt{T}\right).
		\end{align}
		Note that, by Assumption \ref{P-as.tail} \eqref{P-tailiii}, 
		\begin{align}
			\label{24072}\max_{k\in[K]}\E\left[\left(p^{-1/2}\sum_{\ell=1}^pu_{t\ell}b_{\ell k}\right)^2\right]=O(1).
		\end{align}
		Then, we obtain the result using the triangle inequality and equations \eqref{2407} and \eqref{24072}:
		\begin{align*}
			\|UB\|_2^2&= p\sum_{t=1}^T \sum_{k=1}^K \left(p^{-1/2}\sum_{\ell=1}^pu_{t\ell}b_{\ell k}\right)^2\\
			&\le K p \max_{k\in[K]}\left|\sum_{t=1}^T\left( \left(p^{-1/2}\sum_{\ell=1}^pu_{t\ell}b_{\ell k}\right)^2-\E\left[\left(p^{-1/2}\sum_{\ell=1}^pu_{t\ell}b_{\ell k}\right)^2\right)\right]\right|\\
			&\quad + KTp\max_{k\in[K]}\E\left[\left(p^{-1/2}\sum_{\ell=1}^pu_{t\ell}b_{\ell k}\right)^2\right]=O_P\left(Tp\right).
		\end{align*}
		The proofs of  \eqref{lf2ix} and \eqref{lf2ixbis} are similar and therefore omitted.\\
		
		\noindent\textit{Proof of \eqref{lf2xi}, \eqref{lf2xii}.} We apply Lemmas \ref{lm.2exp} and \ref{lm.nagaevsup} to 
		\begin{align*}
			Z_t&=f_{t}\left(p^{-1/2}\sum_{\ell=1}^pu_{t\ell}b_{\ell k}\right);\\
			Z_t&=\varepsilon_{t}\left(p^{-1/2}\sum_{\ell=1}^pu_{t\ell}b_{\ell k}\right)
		\end{align*}
		and obtain the result by summing over $k$.
	\end{Proof}
	
	\begin{Lemma}\label{lm.lf}  Under the assumptions of Theorem \ref{P-th}, the following holds:
		\begin{enumerate}[\textup{(}i\textup{)}]  
			\item\label{lfi} $\left\|\overline{F}-FH^\top\right\|_2^2=O_P\left(\frac{T}{p}+1\right)$;
			\item\label{lfiii} $\left\|H^\top H-I_K\right\|_2^2 =O_P\left(\frac1T+\frac1p\right)$;
			\item\label{lfiv} $\max_{j\in[p]}\left\|\overline{b}_j -Hb_j\right\|_2=O_P\left(\frac{1}{\sqrt{p}}+\frac{p^{\frac{2}{q}}}{\sqrt{T}}\right)$;
			\item\label{lfv} $\left\|V^{-1}\right\|_2=O_P\left(\frac1p\right)$;
			\item\label{lfvi} $\left\|\overline{U}-U\right\|_\infty=o_P\left(\frac{p^{\frac{2}{q}}}{T^{1/2-1/q}} + \frac{T^{1/q}}{\sqrt{p}}\right)$;
			\item\label{lfii} $\max_{j\in[p]}\sum_{t=1}^T |\overline{u}_{tj}-u_{tj}|^2=O_P\left(p^{4/q} +\frac{T}{p}\right)$;
		\end{enumerate}
	\end{Lemma}
	\begin{Proof}
		The results \eqref{lfi} to \eqref{lfvi} follow from Lemmas S.9 and S.11 and Theorem 2 in \cite{fan2021bridging}. Assumption 2 in \cite{fan2021bridging} is satisfied by our Assumption \ref{P-as.tail}. Assumption 3 in \cite{fan2021bridging} corresponds to our Assumption \ref{P-as.moments}. 
		
		To show \eqref{lfii}, note that, for all $t\in[T]$ and $j\in[p]$, it holds that
		\begin{align*} \overline{u}_{tj}-u_{tj}&=x_{tj}^\top -\overline{b}_j^\top \overline{f}_t-u_{tj}\\
			&= b_j^\top f_t -\overline{b}_j^\top \overline{f}_t\\
			&=b_j^\top \left(I_K-H^\top H\right)f_t -\left(\overline{b}_j -Hb_j \right)^\top\overline{f}_t- b_j^\top H^\top \left(\overline{f}_t-Hf_t\right).
		\end{align*}
		This yields 
		\begin{align*}\max_{j\in[p]}\sum_{t=1}^T |\overline{u}_{tj}-u_{tj}|^2&\le K\|B\|_\infty^2\left\|H^\top H-I_K\right\|_2^2\|F\|_2^2\\
			&\quad +\max_{j\in[p]}\left\|\overline{b}_j -Hb_j\right\|_2^2T \\
			&\quad +  K\|B\|_\infty^2\|H\|_2^2\left\|\overline{F}-FH^\top\right\|_2^2=O_P\left(p^{4/q} +\frac{T}{p}\right),
		\end{align*}
		where we used \eqref{lfi}, \eqref{lfiii}, \eqref{lfiv}, Lemma \ref{lm.lf2} \eqref{lf2vibis} and Assumption \ref{P-as.moments} \eqref{P-f4iv}.
	\end{Proof}
	
	\begin{Lemma}\label{lm.bai}Under the assumptions of Theorem \ref{P-th}, it holds that $$\overline{F}-FH^\top = \frac1T FB^\top U^\top \overline{F}V^{-1}+\frac1T  UBF^\top \overline{F}V^{-1}+\frac1T UU^\top  \overline{F}V^{-1}.$$
	\end{Lemma}
	\begin{Proof}
		Recall that $H=\frac1T \overline{V}\overline{F}^\top FB^\top B$ and $\overline{F}V=\frac1T XX^\top\overline{F}$. As a result, we have
		\begin{align*}
			\overline{F}V& = T^{-1}XX^\top\overline{F}\\
			&=  \frac1T (FB^\top  + U)(FB^\top  + U)^\top\overline{F}\\
			&= \frac1T FB^\top B F^\top\overline{F}+\frac1T  FB^\top U^\top \overline{F}+ \frac1T UBF^\top \overline{F}+T^{-1}UU^\top  \overline{F}.
		\end{align*}
		Multiplying both sides by $V^{-1}$, we get the result.
	\end{Proof}
	\begin{Lemma}\label{lm.complex} 
		Under the assumptions of Theorem \ref{P-th}, we have 
		$$\left\|\left(\overline{F}-FH^\top\right)^\top \mathcal{E}\right\|_2=O_P\left(\sqrt{\frac{T}{p}} +1 \right).$$
	\end{Lemma}
	\begin{Proof}
		By Lemma \ref{lm.bai}, we have \begin{equation}\label{bai1}\left\|(\overline{F}-FH^\top)^\top \mathcal{E}\right\|_2 \le J_1+J_2+J_3,\end{equation}
		where 
		\begin{align*}
			J_1&=  \frac1T \left\|\mathcal{E}^\top FB^\top U^\top \overline{F}V^{-1}\right\|_2;\\
			J_2&=  \frac1T \left\|\mathcal{E}^\top UBF^\top \overline{F}V^{-1}\right\|_2;\\
			J_3&=  \frac1T \left\|\mathcal{E}^\top UU^\top  \overline{F}V^{-1}\right\|_2.
		\end{align*}
		We have 
		\begin{align}
			\notag J_1&\le \frac1T \left\|\mathcal{E}^\top F \right\|_2\left(\|UB\|_2 \left\|\overline{F}-FH^\top\right\|_2+\|H\|_2\left\|B^\top U^\top F\right\|_2\right)\|V^{-1}\|_2\\
			\label{bai2}&=O_P\left(\frac{1}{T} \sqrt{T}\left(\sqrt{Tp}\sqrt{\frac{T}{p} +1}+\sqrt{Tp}\right)\frac1p\right)=O_P\left( \frac{\sqrt{T}}{p}\frac{1}{\sqrt{p}}\right),
		\end{align}
		by Lemmas \ref{lm.lf} \eqref{lfi}, \eqref{lfiii}, \eqref{lfv} and \ref{lm.lf2} \eqref{lf2viii}, \eqref{lf2x}, \eqref{lf2xi}.
		Moreover, it holds that
		\begin{align}
			\notag J_2&\le \frac1T \|\mathcal{E}^\top UB\|_2\left\| F\right\|_2\left\|\overline{F}\right\|_2\|V^{-1}\|_2\\
			\label{bai3}&=O_P\left(\frac{1}{T}\sqrt{T} \sqrt{Tp}\sqrt{T}\frac1p\right)=O_P\left(\sqrt{\frac{T}{p}}\right),
		\end{align}
		by Lemmas \ref{lm.lf} \eqref{lfv} and \ref{lm.lf2} \eqref{lf2vibis}, \eqref{lf2xii} and the fact that $\left\|\overline{F}\right\|_2=\sqrt{KT}$.
		We also have 
		\begin{align}
			\notag J_3&\le \frac1T \left\|\mathcal{E}^\top U\right\|_2\left(\|U\|_2 \left\|\overline{F}-FH^\top\right\|_2+\left\|U^\top F\right\|_2\right)\|V^{-1}\|_2\\
			\notag &=O_P\left(\frac{1}{T}\sqrt{Tp}\left(\sqrt{Tp}\sqrt{\frac{T}{p}+1}+ \sqrt{Tp}\right)\frac1p\right)\\
			\label{bai4}&=O_P\left(\sqrt{\frac{T}{p}} + \frac{1}{\sqrt{p}} +1\right),
		\end{align}
		where we used Lemmas \ref{lm.lf} \eqref{lfi}, \eqref{lfv} and \ref{lm.lf2} \eqref{lf2vii}, \eqref{lf2ix}, \eqref{lf2ixbis}.
		We obtain the result by \eqref{bai1}, \eqref{bai2}, \eqref{bai3} and \eqref{bai4}.
	\end{Proof}
	\begin{Lemma}\label{20723} Under the assumptions of Theorem \ref{P-th}, we have 
		$$\max_{j\in[p]} \sum_{t=1}^T\left(\overline{u}_{tj}\widetilde{\varepsilon}_t+\widetilde{f}_t^\top\varphi^*- u_{tj}\varepsilon_t\right)^2=O_P\left(\left(p^{\frac{4}{q}}T^{\frac{2}{q}}+\frac{T^{\frac{2}{q}+1}}{p}\right)+ \left(1+\frac{T}{p}\right)\left\|\varphi^*\right\|_2^2\right).$$
	\end{Lemma}
	\begin{Proof}
		First, notice that, by the triangle inequality,
		\begin{align}
			\notag &\sqrt{ \sum_{t=1}^T\left(\overline{u}_{tj}\widetilde{\varepsilon}_t+\widetilde{f}_t^\top\varphi^*- u_{tj}\varepsilon_t\right)^2}\\
			\notag &=\sqrt{ \sum_{t=1}^T\left(\overline{u}_{tj}\left(\widetilde{\varepsilon}_t-\varepsilon_t\right)+\widetilde{f}_t^\top\varphi^*+ \left(\overline{u}_{tj}- u_{tj}\right)\varepsilon_t\right)^2}\\
			\label{2071}&\le \sqrt{ \sum_{t=1}^T\left(\overline{u}_{tj}\left(\widetilde{\varepsilon}_t-\varepsilon_t\right)\right)^2}+
			\sqrt{ \sum_{t=1}^T\left(\widetilde{f}_t^\top\varphi^*\right)^2} +\sqrt{ \sum_{t=1}^T\left( \left(\overline{u}_{tj}- u_{tj}\right)\varepsilon_t\right)^2}.
		\end{align}
		We first bound the term 
		$\sum_{t=1}^T\left(\overline{u}_{tj}\left(\widetilde{\varepsilon}_t-\varepsilon_t\right)\right)^2$. Remark that 
		\begin{equation}\label{2072}\sum_{t=1}^T\left(\overline{u}_{tj}\left(\widetilde{\varepsilon}_t-\varepsilon_t\right)\right)^2\le \left\|\overline{U}\right\|_\infty^2 \left\| \left(I_T-\overline{P}\right)\mathcal{E} - \mathcal{E}\right\|_2^2=\left\|\overline{U}\right\|_\infty^2 \left\| \overline{P}\mathcal{E}\right\|_2^2.\end{equation}
		Now, using Lemma \ref{lm.nagaev} and the tail bound in Assumption \ref{P-as.tail} \eqref{P-tailiii}, we obtain $\left\|U\right\|_\infty=O_P\left((Tp)^{1/q}\right)$. Combining this with Lemma \ref{lm.lf} \eqref{lfvi} and $\left\|\overline{U}\right\|_\infty\le \left\|\overline{U}-U\right\|_\infty+\left\|U\right\|_\infty$, we get \begin{equation}\label{2073}\left\|\overline{U}\right\|_\infty^2=O_P\left((Tp)^{2/q}\right).\end{equation} Next, recall that $\overline{P}=T^{-1} \overline{F}\overline{F}^\top\mathcal{E}$ and $\left\|\overline{F}\right\|_2=\sqrt{KT}$. This yields
		\begin{align}\notag \left\| \overline{P}\mathcal{E}\right\|_2&\le \frac1T\left\|\overline{F}\right\|_2\left\| \left(\overline{F}-FH^\top\right)^\top\mathcal{E}\right\|_2+\frac1T\left\|\overline{F}\right\|_2\left\| H\right\|_2\left\|F^\top\mathcal{E}\right\|_2\\
			\label{2074} &=\frac{1}{\sqrt{T}} O_P\left(\sqrt{\frac{T}{p}} +1\right)= O_P\left(\frac{1}{\sqrt{T}}+\frac{1}{\sqrt{p}}\right),
		\end{align}
		by Lemmas \ref{lm.lf} \eqref{lfiii}, \ref{lm.lf2} \eqref{lf2viii} and \ref{lm.complex}. Thanks to \eqref{2072}, \eqref{2073} and \eqref{2074}, we obtain 
		\begin{equation}\label{2075} \sum_{t=1}^T\left(\overline{u}_{tj}\left(\widetilde{\varepsilon}_t-\varepsilon_t\right)\right)^2=O_P\left((Tp)^{\frac{2}{q}}\left(\frac{1}{T}+\frac{1}{p}\right)\right).
		\end{equation}
		Let us now bound the term $ \sum_{t=1}^T\left(\widetilde{f}_t^\top\varphi^*\right)^2$. We have 
		\begin{equation}\label{2076}
			\sum_{t=1}^T\left(\widetilde{f}_t^\top\varphi^*\right)^2= \left\|\left(I_T- \overline{P}\right) F\varphi^*\right\|_2^2
			\le \left\|\left(I_T- \overline{P}\right) F\right\|_2^2\left\|\varphi^*\right\|_2^2.
		\end{equation}
		Next, notice that 
		\begin{align}
			\notag \left\|\left(I_T- \overline{P}\right) F\right\|_2&= \left\|\left(I_T-\frac1T\overline{F}\overline{F}^\top\right) F\right\|_2\\
			\notag &\le \left\|\frac1T\left(\overline{F}-FH^\top\right)\left(FH^\top\right)^\top F\right\|_2+ \left\| \frac1TFH^\top \left(\overline{F}-FH^\top\right)^\top F\right\|_2\\
			\label{2077}&\quad + \left\|\left(I_T- \frac1TFH^\top\left(FH^\top\right)^\top \right)F\right\|_2.
		\end{align}
		Then, notice that 
		\begin{align}
			\notag &\left\|\frac1T\left(\overline{F}-FH^\top\right)\left(FH^\top\right)^\top F\right\|_2+ \left\| \frac1TFH^\top \left(\overline{F}-FH^\top\right)^\top F\right\|_2\\
			\label{2078} &\le \frac2T \left\|\overline{F}-FH^\top\right\|_2 \left\|F\right\|_2^2 \left\| H\right\|_2=O_P\left(\sqrt{\frac{T}{p}}+1\right),
		\end{align}
		by Lemmas \ref{lm.lf} \eqref{lfi}, \eqref{lfiii} and \ref{lm.lf2} \eqref{lf2vibis}.
		Moreover, we have
		\begin{align}
			\notag &  \left\|\left(I_T- \frac1TFH^\top\left(FH^\top\right)^\top \right)F\right\|_2\\
			\notag &\le \left\|\left(I_T- \frac1TFF^\top \right)F\right\|_2+  \left\| \frac1TF\left(H^\top H-I_K\right)F^\top F\right\|_2\\
			\label{2079}&\le \left\|F\right\|_2\left\|I_K- \frac1T F^\top F \right\|_2+  \frac1T\left\| F\right\|_2\left\|H^\top H-I_K\right\|_2\left\|F\right\|_2^2= O_P\left(1+\sqrt{\frac{T}{p}}\right),
		\end{align}
		by Lemmas \ref{lm.lf2} \eqref{lf2vibis}, \eqref{lf2vibisii} and \ref{lm.lf} \eqref{lfiii}. Combining \eqref{2076}, \eqref{2077}, \eqref{2078} and \eqref{2079}, we get 
		\begin{equation}\label{20710} \sum_{t=1}^T\left(\widetilde{f}_t^\top\varphi^*\right)^2= O_P\left(1+\frac{T}{p}\right)\left\|\varphi^*\right\|_2^2. \end{equation}
		Finally, we bound $\sum_{t=1}^T\left(\left(\overline{u}_{tj}- u_{tj}\right)\varepsilon_t\right)^2$. Notice that 
		\begin{equation}\label{decomp6162}
			\max_{j\in[p]}\sum_{t=1}^T\left((\overline{u}_{tj}-u_{tj})\varepsilon_t\right)^2 \le  \|\mathcal{E}\|_\infty^2\max_{j\in[p]} \sum_{t=1}^T(\overline{u}_{tj}-u_{tj})^2
		\end{equation}
		Next, using Lemma \ref{lm.nagaev} and the tail bound in Assumption \ref{P-as.tail} \eqref{P-tailiii}, we have $ \|\mathcal{E}\|_\infty^2=O_P\left(T^{\frac{2}{q}}\right).$ This, Lemma \ref{lm.lf} \eqref{lfii} and equation
		\eqref{decomp6162} yield that 
		\begin{equation}\label{20711}
			\max_{j\in[p]}\sum_{t=1}^T\left(\left(\overline{u}_{tj}- u_{tj}\right)\varepsilon_t\right)^2=O_P\left(T^{\frac{2}{q}}p^{\frac{4}{q}} +\frac{T^{1+\frac{2}{q}}}{p}\right).
		\end{equation}
		Combining \eqref{2071}, \eqref{2075}, \eqref{20710} and \eqref{20711}, we obtain the result.
	\end{Proof}
	\begin{Lemma}\label{fancomplex} Under the assumptions of Theorem \ref{P-th}, we have 
		$$\left\|\overline{U}^\top \left(\overline{Y}-\overline{U}\beta^*\right)-U^\top \mathcal{E}\right\|_{\infty}=(\|\varphi^*\|_2\vee 1)O_P\left(\frac{T}{p}+p^{\frac2q}  +\sqrt{T}p^{\frac2q-\frac12}\right).$$
	\end{Lemma}
	\begin{Proof}
		In all this proof, we work on the event $\mathcal{E}_{\sigma}=\{\sigma_{p}\left(H^\top H\right)\ge 1/2\}$ which has probability going to $1$ by Lemma \ref{lm.lf} \eqref{lfiii}. Note that, on $\mathcal{E}_\sigma$, we have \begin{equation}\label{2307}\left\|\left(H^{\top}\right)^{-1}\right\|_2\le \sqrt{K}\left\|\left(H^{\top}\right)^{-1}\right\|_{op} \le \sqrt{K}\sigma_{p}\left(H^\top H\right)^{-1/2}\le \sqrt{2K}.\end{equation} Recall that $\overline{Y}=\left(I_T-\overline{P}\right)(X\beta^* +F\varphi^* +\mathcal{E})$. This yields
		\begin{equation}\label{key1}\begin{aligned}\left\|\overline{U}^\top \left(\overline{Y}-\overline{U}\beta^*\right)- U^\top\mathcal{E}\right\|_\infty &\le \left\|\overline{U}^\top (F\varphi^*+\mathcal{E})- U^\top\mathcal{E}\right\|_\infty\\
				&\le \left\|\overline{U}^\top F\varphi^*\right\|_\infty +\left\|\left(\overline{U}- U\right)^\top\mathcal{E}\right\|_\infty.
			\end{aligned}
		\end{equation}
		Let us first bound $ \left\|\overline{U}^\top F\varphi^*\right\|_\infty$. Since $\overline{U}^\top\overline{F}=0$ and $H^\top$ is invertible on the event $\mathcal{E}_{\sigma}$, it holds that
		\begin{equation}\label{key2}\left\|\overline{U}^\top F\varphi^*\right\|_\infty\le \left\|\left(\overline{U}-U\right)^\top \left(FH^\top -\overline{F}\right)\left(H^{\top}\right)^{-1}\varphi^*\right\|_\infty +\left\|U^\top \left(FH^\top -\overline{F}\right)\left(H^{\top}\right)^{-1}\varphi^*\right\|_\infty.
		\end{equation}
		We now bound the first term on the right-hand side of \eqref{key2}. By the inequality of Cauchy-Schwartz, we have 
		\begin{align}
			\notag &\left\|\left(\overline{U}-U\right)^\top \left(FH^\top -\overline{F}\right)\left(H^{\top}\right)^{-1}\varphi\right\|_\infty\\
			\notag &=\max_{j\in[p]}\left|\left(\left(\overline{U}-U\right)^\top \left(FH^\top -\overline{F}\right)\left(H^{\top}\right)^{-1}\varphi^*\right)_j\right|\\
			\notag &\le \left(\max_{j\in[p]}\sum_{t=1}^T \left|\overline{u}_{tj}-u_{tj}\right|^2\right)^{1/2}\left\|\overline{F}-FH^\top\right\|_2\left\|\left(H^{\top}\right)^{-1}\right\|_2\|\varphi^*\|_2\\
			\label{key3} &=\|\varphi^*\|_2O_P\left(\sqrt{p^{\frac{4}{q}}+\frac{T}{p}}\sqrt{\frac{T}{p}+1}\right)=\|\varphi^*\|_2O_P\left(\frac{T}{p} + p^{\frac{2}{q}}+p^{\frac{2}{q}-\frac12}\sqrt{T}\right),
		\end{align}
		where we used Lemma \ref{lm.lf} \eqref{lfi}, \eqref{lfiii}, \eqref{lfii} and equation \eqref{2307}.
		Next, we control the second term on the right-hand side of \eqref{key2}. By Lemma \ref{lm.bai}, it holds that 
		
		\begin{equation}\label{key4}\left\|U^\top \left(FH^\top -\overline{F}\right)\left(H^{\top}\right)^{-1}\varphi^*\right\|_\infty\le J_{1}+J_{2}+ J_{3},\end{equation}
		where 
		\begin{align*}J_{1}&= \frac1T\left\| U^\top FB^\top U^\top \overline{F}V^{-1}\left(H^{\top}\right)^{-1}\varphi^*\right\|_\infty;\\
			J_2&= \frac1T\left\| U^\top UBF^\top \overline{F}V^{-1}\left(H^{\top}\right)^{-1}\varphi^*\right\|_\infty;\\
			J_3&= \frac1T\left\| U^\top UU^\top  \overline{F}\left(H^{\top}\right)^{-1}\varphi^*\right\|_\infty.
		\end{align*}
		Remark that
		\begin{align}
			\notag \left\|B^\top U^\top \overline{F}\right\|_2&\le\left\|B^\top U^\top\right\|_2 \left\|\overline{F}-FH^\top\right\|_2+\|H\|_2\left\|B^\top U^\top F\right\|_2\\
			\label{160723}&=O_P\left(\sqrt{Tp}\sqrt{\frac{T}{p}+1} +\sqrt{Tp} \right)=O_P(T+\sqrt{Tp}),
		\end{align}
		by Lemmas \ref{lm.lf2} \eqref{lf2x}, \eqref{lf2xi} and \ref{lm.lf} \eqref{lfi}, \eqref{lfiii}.
		By the inequality of Cauchy-Schwartz, this yields
		\begin{align}
			\notag J_{1}
			\notag &=\frac1T \max_{j\in[p]}\left|\left(U^\top FB^\top U^\top \overline{F}V^{-1}\left(H^{\top}\right)^{-1}\varphi^*\right)_j\right|\\
			\notag &=\frac1T \max_{j\in[p]}\left|\sum_{k=1}^K\left(\sum_{t=1}^T u_{tj} f_{tk} \right) \left(B^\top U^\top \overline{F}V^{-1}\left(H^{\top}\right)^{-1}\varphi^*\right)_k\right|\\
			\notag &\le \frac1T \left(\max_{j\in[p]}\left|\sum_{k=1}^K\left(\sum_{t=1}^T u_{tj} f_{tk} \right)^2\right|\right)^{1/2} \left\|B^\top U^\top \overline{F}\right\|_2 \left\|V^{-1}\right\|_2\left\|(H^{-1})^\top\right\|_2\|\varphi^*\|_2\\
			\notag &\le \frac1T  \sqrt{K} \max_{j\in[p]}\left|\sum_{t=1}^T u_{tj} f_{tk} \right| \left\|B^\top U^\top \overline{F}\right\|_2 \left\|V^{-1}\right\|_2\left\|(H^{-1})^\top\right\|_2\|\varphi^*\|_2\\
			\label{key5} &=O_P\left(\frac{1}{Tp}\left(T+\sqrt{Tp} \right)\sqrt{T}p^{\frac{2}{q}} \right)\|\varphi^*\|_2=O_P\left(\sqrt{T}p^{\frac{2}{q}-1}+p^{\frac{2}{q}-\frac12}\right)\|\varphi^*\|_2,
		\end{align}
		where we used Lemmas  \ref{lm.lf2} \eqref{lf2ii} and \ref{lm.lf} \eqref{lfiii}, \eqref{lfv} and equations \eqref{160723} and \eqref{2307}.
		Then, notice that, by Lemma \ref{lm.lf} \eqref{lfi} and \eqref{lfiii}, we have 
		\begin{equation}\label{2407}\left\|F^\top \overline{F}\right\|_2\le \|F\|_2 \left\|\overline{F}-FH^\top\right\|_2+  \|F\|_2^2 \|H\|_2=O_P(T).\end{equation}
		This allows to bound $J_{2}$. Indeed, by the inequality of Cauchy-Schwartz, it holds that 
		\begin{align}
			J_{2}\notag &=\frac1T \max_{j\in[p]}\left|\left(U^\top UBF^\top \overline{F}V^{-1}\left(H^{\top}\right)^{-1}\varphi^*\right)_j\right|\\
			\notag &= \frac1T \max_{j\in[p]}\left|\sum_{k=1}^K\sum_{t=1}^T  u_{tj}\left(\sum_{\ell=1}^pu_{t\ell}b_{\ell k}\right) \left(F^\top \overline{F}V^{-1}\left(H^{\top}\right)^{-1}\varphi^*\right)_k\right|\\
			\notag &\le \frac1T \max_{j\in[p]} \sqrt{\sum_{k=1}^K\left(\sum_{t=1}^T  u_{tj}\left(\sum_{\ell=1}^pu_{t\ell}b_{\ell k}\right)\right)^2 } \left\| F^\top \overline{F}V^{-1}\left(H^{\top}\right)^{-1}\varphi^*\right\|_2\\
			\notag &\le \frac{1}{T}\sqrt{K} \max_{j\in[p],k\in[K]}\left|\sum_{t=1}^T  u_{tj}\left(\sum_{\ell=1}^pu_{t\ell}b_{\ell k}\right)\right| \left\|F^\top \overline{F}\right\|_2   \left\|V^{-1}\right\|_2\left\|(H^{-1})^\top\right\|_2\|\varphi^*\|_2\\
			\label{key6}&=O_P\left(\frac{1}{Tp}T\left(T+p^{\frac{2}{q}+\frac12}\sqrt{T} \right)\right)\|\varphi^*\|_2=O_P\left(\frac{T}{p}+\sqrt{T} p^{\frac{2}{q}-\frac12}\right)\|\varphi^*\|_2,
		\end{align}
		by Lemmas \ref{lm.lf2} \eqref{lf2iv}, \eqref{lf2vibis} and \ref{lm.lf} \eqref{lfiii}, \eqref{lfv} and equations \eqref{2307} and \eqref{2407}.
		Finally note that 
		\begin{align}\notag \left\|U^\top \overline{F}\right\|_2 &\le \left\|U^\top F\right\|_2 \left\|H^\top\right\|_2+ \|U\|_2 \left\|\overline{F}-FH^\top\right\|_2\\
			\label{24072}& =O_P\left(\sqrt{T}+\sqrt{T}\sqrt{\frac{T}{p}+1}\right)=O_P\left(\sqrt{T}+ \frac{T}{\sqrt{p}}\right),\end{align}
		by Lemmas \ref{lm.lf} \eqref{lfi}, \eqref{lfiii} and \ref{lm.lf2} \eqref{lf2vii}, \eqref{lf2ix}. Thanks to this, we can bound $J_3$. Indeed, by the inequality of Cauchy-Schwartz, we have 
		\begin{align}
			\notag J_{3}
			\notag &=\frac1T \max_{j\in[p]}\left|\left(U^\top UU^\top  \overline{F}\left(H^{\top}\right)^{-1}\varphi^*\right)_j\right|\\
			\notag &= \frac1T \max_{j\in[p]}\left|\sum_{\ell=1}^p\left(\sum_{t=1}^T  u_{tj}u_{t\ell}\right) \left(U^\top  \overline{F}\left(H^{\top}\right)^{-1}\varphi^*\right)_{\ell}\right|\\
			&\le \frac{1}{T}  \max_{j\in[p]}\sqrt{\sum_{\ell=1}^p\left(\sum_{t=1}^T  u_{tj}u_{t\ell}\right)^2} \left\|U^\top \overline{F}\right\|_2   \left\|V^{-1}\right\|_2\left\|(H^{-1})^\top\right\|_2\|\varphi^*\|_2 \\
			\notag &\le \frac{1}{T} \sqrt{p} \max_{j\in[p]}\left|\sum_{t=1}^T  u_{tj}^2\right|\left\|U^\top \overline{F}\right\|_2   \left\|V^{-1}\right\|_2\left\|(H^{-1})^\top\right\|_2\|\varphi^*\|_2 \\
			\label{key7}&=O_P\left(\frac{1}{Tp}T\sqrt{p} \left( \sqrt{T} +\frac{T}{\sqrt{p}}   \right)\right)\|\varphi^*\|_2=O_P\left(\sqrt{\frac{T}{p}}+ \frac{T}{p}\right)\|\varphi^*\|_2,
		\end{align}
		where we used Lemmas \ref{lm.lf2} \eqref{lf2i} and \ref{lm.lf} \eqref{lfv} and equations \eqref{2307} and \eqref{24072}.
		Then, \eqref{key2}, \eqref{key3}, \eqref{key4}, \eqref{key5}, \eqref{key6}, \eqref{key7} imply that 
		\begin{equation}\label{key8}\left\|\overline{U}^\top F\varphi^*\right\|_\infty=O_P\left(\frac{T}{p}+p^{\frac{2}{q}} +\sqrt{T}p^{\frac{2}{q}-\frac12} \right)\|\varphi^*\|_2.\end{equation}
		
		Let us now bound the second term on the right-hand side of \eqref{key1}, that is $\left\|\left(\overline{U}- U\right)^\top\mathcal{E}\right\|_\infty.$
		Note that \begin{align*} \overline{U}^\top-U^\top&=X^\top -\overline{B}\overline{F}^\top-U^\top\\
			&= BF^\top -\overline{B}\overline{F}^\top\\
			&= B\left(I_K-H^\top H\right)F^\top -\left(\overline{B}-BH^\top \right)\overline{F}^\top - BH^\top \left(\overline{F}-FH\right)^\top.
		\end{align*}
		This yields
		\begin{equation}\label{key9}
			\left\|\left(\overline{U}- U\right)^\top\mathcal{E}\right\|_\infty \le K_1+K_2+K_3,\end{equation} 
		where
		\begin{align*}K_1&=\left\|B\left(I_K-H^\top H\right)F^\top\mathcal{E} \right\|_\infty;\\
			K_2&= \left\|\left(\overline{B}-BH^\top \right)\overline{F}^\top\mathcal{E}  \right\|_\infty;\\
			K_3&= \left\|BH^\top \left(\overline{F}-FH\right)^\top\mathcal{E}\right\|_\infty.
		\end{align*}
		By the inequality of Cauchy-Schwartz, Lemmas \ref{lm.lf2} \eqref{lf2viii} and \ref{lm.lf} \eqref{lfiii} and Assumption \ref{P-as.moments} \eqref{P-f4iv}, it holds that
		\begin{align}\notag K_1 &=\max_{j\in[p]} \left|\sum_{k=1}^K b_{jk}\left(\left(I_K-H^\top H\right)F^\top\mathcal{E}\right)_{k} \right|\\
			\notag &\le  \sqrt{K}\|B\|_\infty \left\|I_K-H^\top H\right\|_2\left\|F^\top \mathcal{E}\right\|_2\\
			\label{key10}&=O_P\left(\sqrt{\frac{1}{T}+\frac{1}{p}} \sqrt{T}\right)= O_P\left(1+\sqrt{\frac{T}{p}}\right).
		\end{align}
		Next, we have
		\begin{align}
			\notag K_2&=\max_{j\in[p]} \left|\sum_{k=1}^K \left(\overline{b}_{j}-Hb_j\right)_k\left(\overline{F}^\top\mathcal{E}\right)_{k} \right|\\
			\notag &\le  \max_{j\in[p]}\left\|\overline{b}_j -Hb_j\right\|_2 \left\|\overline{F}^\top\mathcal{E}\right\|_2\\
			\notag &\le\max_{j\in[p]}\left\|\overline{b}_j -Hb_j\right\|_2 \left(\left\|\left(\overline{F}-FH^\top\right)^\top\mathcal{E}\right\|_2+\|H\|_2\left\| F^\top \mathcal{E}\right\|_2\right)\\
			\notag &= O_P\left(\left(\frac{1}{\sqrt{p}}+\frac{p^\frac{2}{q}}{\sqrt{T}}\right) \left(\sqrt{T} +\sqrt{\frac{T}{p}}\right)\right)\\
			\label{key11} &=O_P\left(\sqrt{\frac{T}{p}} +p^\frac{2}{q}\right).
		\end{align}
		where we used the inequality of Cauchy-Schwarz, Lemmas \ref{lm.lf2} \eqref{lf2viii}, \ref{lm.lf} \eqref{lfiii}, \eqref{lfiv}, and \ref{lm.complex}.
		Finally, by the inequality of Cauchy-Schwartz, Lemmas \ref{lm.lf2} \eqref{lfiii} and \ref{lm.complex} and Assumption \ref{P-as.moments} \eqref{P-f4iv}, it holds that 
		\begin{align}\notag K_3&=\max_{j\in[p]} \left|\sum_{k=1}^Kb_{jk}\left(H^\top \left(\overline{F}-FH\right)^\top\mathcal{E}\right)_{k} \right|\\
			\notag &\le  \sqrt{K}\left\|B\right\|_\infty\left\|\left(\overline{F}-FH^\top\right)^\top\mathcal{E}\right\|_2 \left\|H\right\|_2\\
			\label{key12}&=O_P\left(\sqrt{\frac{T}{p}}+1\right).
		\end{align}
		Combining \eqref{key9}, \eqref{key10}, \eqref{key11} and \eqref{key12} yields
		\begin{align}
			\label{key13}\frac1T\left\|\left(\overline{U}- U\right)^\top\mathcal{E}\right\|_\infty=O_P\left(\sqrt{\frac{T}{p}}+p^{\frac2q}\right).
		\end{align}
		We obtain the result of the lemma by \eqref{key1}, \eqref{key8} and \eqref{key13}. 
	\end{Proof}
	
	\subsection{Pre-existing results on strong mixing sequences and high-dimensional Gaussian vectors}\label{subsec.pre}
	In this section, we state some useful lemmas and reformulate some results of \cite{fan2021bridging} and \cite{lederer2021estimating} that we use to prove Theorem \ref{P-th}.
	\subsubsection{Results on variables with polynomial tails}
	The following result is a direct consequence of the inequality of Cauchy-Schwarz. This lemma allows to show that products of variables in $u_{tj},f_{tk},\varepsilon_t$, $p^{-1/2} \sum_{j=1}^pb_ju_{tj}$ have polynomial tails.
	\begin{Lemma}\label{lm.2exp}
		Let $Z_1$ and $Z_2$ be random variables such that $\vertiii{Z_1}_q \le C$ and  $\vertiii{Z_2}_q\le C$,
		for some constants $C,q>0$. Then, we have $\vertiii{Z_1Z_2}_{q/2}\le C^2$.
	\end{Lemma}
	The next lemma serves to bound the sup norm of some variables.
	\begin{Lemma}\label{lm.nagaev}
		Let $Z$ be a mean-zero $p$-dimensional random vector. Assume that there exist constants $C,h>0$ such that we have 
		$\vertiii{Z}_{h}\le C.$ Then, it holds that $\|Z\|_\infty=O_P\left(p^{1/h} \right).$
		
	\end{Lemma}
	
	\begin{Proof} For all $j\in[p]$ and $z>0$, we have $\P(|Z_j|\le z)= \P(|Z_j|^h\le z^h)\le \E[|Z_j|^h]/z^h$ by Markov's inequality. By the union bound, we obtain $\P(|Z|_\infty\ge z)\le p \max_{j\in[p]} \P(|Z_j|_\infty\ge z)\le p \left( \E[|Z_j|^h]/z^h\right)$. Taking $z\propto p^{1/h} $ we obtain the result.
	\end{Proof}
	
	\subsubsection{Results on strong mixing sequences}
	
	The next Lemma is a direct consequence of Lemma S.4 and Remark 4 in \cite{fan2021bridging}.
	\begin{Lemma}\label{lm.nagaevsup}
		Let $S_T=\sum_{t=1}^T Z_t$, where $\{Z_t\}_{t}$ is a sequence of mean-zero $p$-dimensional random vectors such that 
		\begin{enumerate}[\textup{(}i\textup{)}]  
			\item\label{lnagi}  There exist constants $C_1,\xi>0$ and $h\ge 2$ such that, for all $t\in[T]$, we have 
			$$\vertiii{Z_t}_{h+\xi}\le C_1;$$
			\item\label{lnagii} Theres exist constants $C_2,c>0$ such that the strong mixing coefficients of the sequence $\{Z_t\}_{t}$ satisfy $\tilde\alpha(t)\le C_2 t^{-c}$ for all $t\in\mathbb{Z}_+$;
			\item\label{lnagiii} $c>\frac{(h-1)(h+\xi)}{\xi}$.
		\end{enumerate}
		Then, it holds that $\|S_T\|_\infty=O_P\left(p^{1/h} \sqrt{T}\right).$
		
	\end{Lemma}

	The last result of this subsection is a direct consequence of the high-dimensional central limit theorem for strong mixing sequences due to Theorem S.7 (a) in \cite{fan2021bridging}.
	\begin{Lemma}\label{lm.highdimclt}
		Let $S_T=n^{-1/2}\sum_{t=1}^T Z_t$, where $\{Z_t\}_{t}$ is a sequence of mean-zero $p$-dimensional random vectors, such that 
		\begin{enumerate}[\textup{(}i\textup{)}] 
			\item\label{lhdi} There exist $C_1,\xi>0$ and $h\ge 4$ such that, for all $t\in[T]$, $j\in[p]$ and $z>0$, we have 
			$\vertiii{Z_{t}}_{h+\xi}\le C_1 ;$
			\item\label{lhdii}  There exist constants $C_2,c>0$ such that the strong mixing coefficients of the sequence $\{Z_{t}\}_{t}$ satisfy $\tilde \alpha(t)\le C_2t^{-c}$ for all $t\ge 2$;
			\item\label{lhdiii} $c >\left[ \left(\frac{h+\xi}{\xi}\right)\left(\frac{h}{2}-1\right)\right]\vee \left(\frac{2}{1-\frac{2}{h}}\right)$ and $\kappa= \left(\frac{\frac{1}{2} +\frac{h}{4(h+1)}}{c+\frac{h}{2(h+1)}}\right)<\frac12$
			\item\label{lhdiv}  There exists $\sigma_*>0$ such that $\sigma_{p}(\Sigma) \ge \sigma_*^2$, where $\Sigma=\E\left[S_TS_T^\top\right]$;
		\end{enumerate} Let also $G\sim\mathcal{N}(0,\Sigma)$. Then, there exists a constant $K_2>0$ such that, for all $z\ge0$, we have
		\begin{align*}
			&\sup_{z\in\R_+}\left|\P\left(\left\|S_T\right\|_\infty\le z\right)-\P(|G|_\infty \le z) \right|\\
			&\le  K_2\Big[\left(T^{\kappa-1/2}+T^{1-\frac{c}{2}(1-\frac{2}{h})}\right)\log(T)\log(p)+T^{-1/4}\log (p)^{3/2} \log(T)+ p^{\frac{2}{h}} T^{\frac{1}{h}-\frac12} \log(p)^2 \log(T)\\
			&\quad +\left( p\log(p)^{\frac{3}{2}h-4} \log(T)\log(Tp)\right)^{\frac{1}{h-2}}T^{-\frac14} +T^{\frac12-c\kappa} \left(p^{\frac{1}{h+1}} \sqrt{1\vee \log(p)}\right)\Big].
		\end{align*}
		
	\end{Lemma}
	\subsubsection{Results on high-dimensional Gaussian vectors}
	The following two lemmas are direct consequences of Lemmas A.4 and A.5 and Remark A.8 in \cite{lederer2021estimating}. (Note that the lemmas in \cite{lederer2021estimating} themselves follow from results in \cite{chernozukhov2013gaussian} and \cite{chernozhukov2015comparison}.)
	\begin{Lemma}\label{lm.glv1}
		Let $G:=(G_1,\dots,G_p)^\top$ be a mean zero $p$-dimensional Gaussian vector. Suppose that there exist constants $c_3,C_3$ such that $c_3\le \E[G_j^2]\le C_3$ for all $j\in[p]$, then, for every $z,\delta>0$, we have 
		$$\P\left(\left|\left\|G\right\|_\infty -z\right|\le\delta\right)\le C\delta\sqrt{1\vee \log(2p/\delta)},$$
		where $C>0$ depends only on $c_3,C_3$. 
	\end{Lemma}
	
	\begin{Lemma}\label{lm.glv2}
		Let $G:=(G_1,\dots,G_p)^\top$ and $G':=(G'_1,\dots,G_p')^\top$ be two mean zero $p$-dimensional Gaussian vectors with respective covariance matrices $\Sigma^{G}$ and $\Sigma^{G'}$. Define $\delta= \left\|\Sigma^{G}-\Sigma^{G'}\right\|_\infty$. 
		Suppose that there exist constants $c_3,C_3$ such that $c_3\le \E[G_j^2]\le C_3$ for all $j\in[p]$. Then, there exists a constant $C>0$ depending only on $c_3,C_3$ such that 
		$$\sup_{z\in\R}\left|\P\left(\left\|G\right\|_\infty\le z\right)-\P\left(\left\|G'\right\|_\infty\le z\right)\right|\le C\delta^{1/3}(1\vee 2\log(2p)\vee \log(1/\delta))^{1/3} (\log(2p))^{1/3}.$$
	\end{Lemma}
	
	\subsection{On the rate condition in statement \eqref{P-thii} of Theorem \ref{P-th}.}\label{sec.rate-cond}
	\subsubsection{Discussion}
	The power analysis in Theorem \ref{P-th} contains the rate conditon \begin{equation}\label{rate.th.suff}\sqrt{\frac{\log(T\vee p)}{T\wedge p}}=o_P\left(\frac1T\left\|U^\top U\beta^*\right\|_\infty\right),\end{equation} which we analyze in this subsection. First, we state the following Lemma, which contains sufficient conditions in terms of nonrandom quantities for \eqref{rate.th.suff} to hold (concerning \eqref{suff.thi}) and not hold (concerning \eqref{suff.thii}).
			\begin{Lemma}\label{lm.suff.th}Let the assumptions of Theorem \ref{P-th} hold. We have 
		\begin{enumerate}[\textup{(}i\textup{)}] 
					\item\label{suff.thi} If $\sqrt{\frac{\log(T\vee p)}{T\wedge p}}+ \frac{p^{\frac{4}{q}}}{\sqrt{T}}\|\beta^*\|_1=o\left(\|\Sigma \beta^*\|_\infty\right)$, then \eqref{rate.th.suff} holds. 
					\item\label{suff.thii} If $\|\Sigma \beta^*\|_\infty+\frac{p^{\frac{4}{q}}}{\sqrt{T}}\|\beta^*\|_1=o\left(\sqrt{\frac{\log(T\vee p)}{T\wedge p}}\right)$, then \eqref{rate.th.suff} does not hold. 
		\end{enumerate}

		\end{Lemma}
		Lemma  \ref{lm.suff.th} is proved in Section \ref{sec.proof.lm.suff.th}. This allows us to give examples of sequences of $\beta^*$ and $\Sigma$ such that \eqref{rate.th.suff} holds or does not hold under our assumptions.\\

		\noindent \textbf{Example where \eqref{rate.th.suff} holds.} Let $\beta^*=(b,0,\dots,0)$ and $\Sigma_{11}\ge\underline{\sigma}$, where $b\ne 0$ and $\underline{\sigma}>0$ are constants that do not depend on $T$. Assume also that $p=o(T^{\frac{q}{8}})$. Then, we have $\|\Sigma \beta^*\|_\infty\ge \Sigma_{11} |b|\ge \underline{\sigma}|b|$ and, therefore, by the fact that $p=o(T^{\frac{q}{8}})$, we obtain
		$$\sqrt{\frac{\log(T\vee p)}{T\wedge p}}+ \frac{p^{\frac{4}{q}}}{\sqrt{T}}\|\beta^*\|_1=\sqrt{\frac{\log(T\vee p)}{T\wedge p}}+ \frac{p^{\frac{4}{q}}}{\sqrt{T}}|b|=o(\underline{\sigma}|b|)= o\left(\|\Sigma \beta^*\|_\infty\right).$$
		By Lemma \ref{lm.suff.th} \eqref{suff.thi}, this shows that \eqref{rate.th.suff} holds. \\
		
\noindent \textbf{Example where \eqref{rate.th.suff} does not hold.} Let $\beta^*=\left(\left(\frac{p^{\frac{4}{q}}}{\sqrt{T}}\right)^{-1}\frac{1}{T\wedge p},0,\dots,0\right)$ and $\max_{j\in[p]} |\Sigma_{1j}|=O(1)$. Assume also that $p=o(T^{\frac{q}{8}})$. Then, we have $$\|\Sigma \beta^*\|_\infty\le \left(\max_{j\in[p]} |\Sigma_{1j}|\right)\left(\frac{p^{\frac{4}{q}}}{\sqrt{T}}\right)^{-1}\frac{1}{T\wedge p}$$ and, therefore,  we obtain
		$$\|\Sigma \beta^*\|_\infty+\frac{p^{\frac{4}{q}}}{\sqrt{T}}\|\beta^*\|_1\le  \left(\max_{j\in[p]} |\Sigma_{1j}|\right)\left(\frac{p^{\frac{4}{q}}}{\sqrt{T}}\right)^{-1}\frac{1}{T\wedge p}+\frac{1}{T\wedge p} = o\left(\sqrt{\frac{\log(T\vee p)}{T\wedge p}}\right).$$
		By Lemma \ref{lm.suff.th} \eqref{suff.thii}, this shows that \eqref{rate.th.suff} does not hold. 	

		\subsubsection{Proof of Lemma \ref{lm.suff.th}} \label{sec.proof.lm.suff.th}	
		
		First, we apply Lemmas \ref{lm.2exp} and \ref{lm.nagaevsup} to $Z_t=\left(\left(u_{tj}u_{tk}-\E\left[u_{tj}u_{tk}\right]\right)_{j=1}^p\right)_{k=1}^p$ (it can be shown that the conditions of these lemmas hold following the arguments of the proof of Lemma \ref{prop.LV1}). These Lemmas yield
		\begin{equation}\label{70624}\left\|\frac{U^\top U}{T}-\Sigma\right\|_\infty\le \max_{j,k\in[p]}\left|\frac1T\sum_{t=1}^T u_{tj}u_{tk}-\E\left[u_{tj}u_{tk}\right]\right|=O_P\left(\frac{p^{\frac{4}{q}}}{\sqrt{T}}\right).\end{equation}
		
		Next, we show \eqref{suff.thi}. By the triangle inequality, we have 
	$$\left\|\Sigma\beta^*\right\|_\infty\le  \left\|\left(\frac{U^\top U}{T}-\Sigma\right)\beta^*\right\|_\infty +\left\|\frac{U^\top U}{T}\beta^*\right\|_\infty .$$
	By Hölder's inequality applied to each component of the vector $\left(\frac{U^\top U}{T}-\Sigma\right)\beta^*$, it holds that 
	$$ \left\|\left(\frac{U^\top U}{T}-\Sigma\right)\beta^*\right\|_\infty\le \left\|\frac{U^\top U}{T}-\Sigma\right\|_\infty\|\beta^*\|_1.$$
	This leads to $$\left\|\frac{U^\top U}{T}\beta^*\right\|_\infty \ge  \|\Sigma \beta^*\|_\infty -\left\|\frac{U^\top U}{T}-\Sigma\right\|_\infty\|\beta^*\|_1.$$
		Hence, by \eqref{70624} and since, under \eqref{suff.thi}, it holds that
		 $\sqrt{\frac{\log(T\vee p)}{T\wedge p}}+ \frac{p^{\frac{4}{q}}}{\sqrt{T}}\|\beta^*\|_1=o\left(\|\Sigma \beta^*\|_\infty\right)$, we have 
		\begin{align*}\sqrt{\frac{\log(T\vee p)}{T\wedge p}}&=o\left(\|\Sigma \beta^*\|_\infty- \frac{p^{\frac{4}{q}}}{\sqrt{T}}\|\beta^*\|_1\right) \\
		&=o_P\left(\|\Sigma \beta^*\|_\infty -\left\|\frac{U^\top U}{T}-\Sigma\right\|_\infty\|\beta^*\|_1\right)\\
		&=o_P\left(T^{-1}\left\|U^\top U\beta^*\right\|_\infty\right),\end{align*}
		which proves \eqref{suff.thi}.

Finally, we show \eqref{suff.thii}. By a reasoning similar to that of the proof of \eqref{suff.thi}, we have 
	\begin{align*}\left\|\frac{U^\top U}{T}\beta^*\right\|_\infty &\le  \left\|\left(\frac{U^\top U}{T}-\Sigma\right)\beta^*\right\|_\infty +\left\|\Sigma\beta^*\right\|_\infty\\
	& \le  \left\|\frac{U^\top U}{T}-\Sigma\right\|_\infty\|\beta^*\|_1 +\left\|\Sigma\beta^*\right\|_\infty.
	\end{align*}
Hence, by \eqref{70624} and since, under \eqref{suff.thii}, it holds that
		 $\|\Sigma \beta^*\|_\infty+\frac{p^{\frac{4}{q}}}{\sqrt{T}}\|\beta^*\|_1=o\left(\sqrt{\frac{\log(T\vee p)}{T\wedge p}}\right)$, we have
		\begin{align*}\left\|\frac{U^\top U}{T}\beta^*\right\|_\infty&\le  \left\|\frac{U^\top U}{T}-\Sigma\right\|_\infty\|\beta^*\|_1 +\left\|\Sigma\beta^*\right\|_\infty \\
		&=O_P\left(\|\Sigma \beta^*\|_\infty+\frac{p^{\frac{4}{q}}}{\sqrt{T}}\|\beta^*\|_1\right)\\
		&=o_P\left(\sqrt{\frac{\log(T\vee p)}{T\wedge p}}\right),\end{align*}
		which proves \eqref{suff.thii} (i.e., that \eqref{rate.th.suff} cannot hold).

	\bibliographystyle{agsm}
	\bibliography{bibliography}